\documentclass[fleqn,usenatbib]{mnras}

\usepackage{newtxtext,newtxmath}

\usepackage[T1]{fontenc}
\usepackage{ae,aecompl}
\usepackage{hyperref}

\usepackage[usenames]{xcolor}
\usepackage{color, soul}

\usepackage{graphicx}	
\usepackage{amsmath}	

\usepackage{amssymb}	
\usepackage{times}
\usepackage{pifont}
\usepackage{adjustbox}

\newcommand{\cmark}{\ding{51}}%
\newcommand{\xmark}{\ding{55}}%

\title[Observational Signatures of Circumbinary Discs I]{Observational Signatures of Circumbinary Discs --- I: Kinematics}

\author[J. Calcino et al.]{
Josh Calcino$^{1}$\thanks{Contact e-mail: \href{mailto:jcalcino@lanl.gov}{jcalcino@lanl.gov}},
Daniel J. Price$^{2}$,
Christophe Pinte$^{2,3}$,
Himanshi Garg$^{2}$,
Brodie J. Norfolk$^{4}$,
\newauthor
Valentin Christiaens$^{2,5}$,
Hui Li$^{1}$,
Richard Teague$^{6}$
\\
$^{1}$Theoretical Division, Los Alamos National Laboratory, Los Alamos, NM 87545, USA\\
$^{2}$School of Physics and Astronomy, Monash University, Vic 3800, Australia\\
$^{3}$Univ. Grenoble Alpes, CNRS, IPAG, F-38000 Grenoble, France\\
$^{4}$Centre for Astrophysics and Supercomputing (CAS), Swinburne University of Technology, Hawthorn, Victoria 3122, Australia\\
$^{5}$Space sciences, Technologies \& Astrophysics Research (STAR) Institute, Universit\'e de Li\`ege, All\'ee du Six Ao\^ut 19c, B-4000 Sart Tilman, Belgium\\
$^{6}$Center for Astrophysics | Harvard \& Smithsonian, 60 Garden Street, Cambridge, MA 02138, USA\\
}
\date{Accepted XXX. Received YYY; in original form ZZZ}

\pubyear{2021}

\begin{document}
\label{firstpage}
\pagerange{\pageref{firstpage}--\pageref{lastpage}}
\maketitle

\begin{abstract}
We present five morphological and kinematic criteria to aid in asserting the binary nature of a protoplanetary disc, based on 3D hydrodynamical simulations of circumbinary discs post-processed with Monte Carlo radiative transfer. We find that circumbinary discs may be identified by i) a central cavity, ii) spiral arms both in and outside of their central cavities, iii) non-localised perturbations in their iso-velocity curves, iv) asymmetry between the lines of maximum speed of the blue and red-shifted wings and v) asymmetry between the area of the blue and red-shifted wings. We provide quantitative metrics for the last two criteria that can be used, in conjunction with the morphological criteria, to signal whether a protoplanetary disc is likely to be a circumbinary disc.

\end{abstract}




\section{Introduction}

Recent observations of protoplanetary discs, from optical and near infrared, to centimetre wavelengths, have revealed an abundance of substructures such as, spiral arms, rings, gaps, and cavities \citep[e.g.][]{dong2018c, FLong2018, andrews2018, norfolk2021, vandermarel2021}. 
Discerning a single origin for these structures has remained a challenge. More often than not they have been attributed to the interaction of companions, of planetary or stellar mass, with the gas and dust content in the disc \citep[e.g. see][]{dong2015b, calcino2019, baruteau2019, calcino2020, veronesi2020}. Point-like features seen with direct imaging provide the most compelling evidence of companions. In all but a few cases (e.g. PDS~70 \cite{keppler2018}, HD~142527 \citet{biller2012,lacour2016}, and HD~100453 \citet{benisty2017, Rosotti2020, gonzalez2020}) convincing evidence is lacking. Emission from the central star and scattering from the protoplanetary disc make such detections difficult. 

One may try to infer the existence of perturbing bodies in protoplanetary discs by matching scattered light and/or continuum observations of these discs \citep[e.g.][]{dipierro2014, dong2015b, dipierro2018, calcino2019, baruteau2019, calcino2020, veronesi2020}, but models are degenerate.
For example, \cite{calcino2019} explained the substructures of IRS~48 using a stellar mass companion, while \cite{vandermarel2013} and \cite{zhus2014} argued for a planet. Hence one cannot always rule out models.

A more robust way is to use kinematics \citep[e.g.][]{pinte2018, teague2018,pinte2019, pinte2020, calcino2022}. The idea is to detect planets influencing the surrounding disc material by observing rotational line transitions from species such as carbon monoxide \citep{perez2015, perez2018, teague2018}. However these methods to date have focused on inferring planetary mass companions, and not much can be said about more massive bodies. Companions of stellar mass produce large perturbations on the disc and open wide, deep cavities which can introduce fast radial flows as the outer disc material accretes onto the binary \citep{casassus2013, rosenfeld2014}. If inclined with respect to the outer disc, they can also produce warps and disc tearing \citep{facchini2013}, which will leave peculiar signatures in the kinematics. However these features can also be produced by planetary mass companions \citep[e.g. see][]{nealon2018, zhu2019}, so they are not necessarily a sign-post for circumbinary discs.

\citep{price2018} and \cite{calcino2019} computed circumbinary disc signatures in HD~142527 and IRS~48, respectively, and showed that large perturbations are introduced, particularly inside the cavity. \cite{price2018} showed that the fast radial flows seen in HD~142527 \citep{casassus2015} naturally occur due to the observed stellar companion. \cite{calcino2019} showed that asymmetries in the velocity map, as well as non-localised deviations in isovelocity curves of individual channel maps, can hint at the circumbinary nature of a disc. In this paper we expand on these findings by exploring observational signatures of circumbinary discs around intermediate mass ratio binary stars. Our aim is to derive kinematic criteria that signify the circumbinary nature of a disc in a quantitative fashion. We leave the application of these criteria to observations to the second paper in this series. The structure of this paper is as follows: We describe our modelling and synthetic observation methods in Section~\ref{sec:method}, and describe our resulting hydro simulations in Section~\ref{sec:res}. We introduce morphological and kinematic signatures robustly seen in the synthetic observations of our circumbinary discs in Sections~\ref{sec:co_em} and \ref{sec:kin_sig}. We derive and test kinematic criteria that quantify asymmetries in the velocity maps in Section 6. We discuss the applicability and caveats of our criteria in Section~\ref{sec:disc}, summarise our results in Section~\ref{sec:sum}.

\section{Methods} \label{sec:method}

\subsection{SPH Simulations}

\begin{table*}
    \centering
    \begin{adjustbox}{width=0.99\textwidth}
    \begin{tabular}{l|c|c|c|c|c|c|c|c|c|c|c|c}
    \hline
        Ref. & $q$    & $a$ (au) & $e$ & $i$ & $\omega$ & $M_\textrm{disc}$ (M$_\odot$) & $H/R_\textrm{ref}$ & $R_\textrm{ref}$ & $R_\textrm{in}$ (au) & $R_\textrm{out}$ (au) & $\alpha_{\textrm{SS}}$ & $N_\textrm{Orbits}$ \\
        \hline
        No Companion (NC)               & -    & -   & -   & -  & -  & 0.020  & 0.066 & 100   & 1     & 400 & $5\times 10^{-3}$   &  20 \\
        Planet (P)                      & $2.5\times 10^{-3}$  & 80  & 0.0 & 0.0 & 0.0 & 0.010  & 0.066 & 100   & 10    & 400 & $5\times 10^{-3}$   &  60 \\
        Multiple Planets (MP)           & $[2.5, 1.25]\times 10^{-3}$  & [75.6, 130]  & 0.0 & 0.0 & 0.0 & 0.010  & 0.066 & 100   & 10    & 400 & $5\times 10^{-3}$   &  60 \\
        Eccentric Planet (EP)           & $2.5\times 10^{-3}$  & 80  & 0.4 & 0.0 & 0.0 & 0.010  & 0.066 & 100   & 100    & 400 & $5\times 10^{-3}$   &  60 \\
        No Over-density (NOD)           & 0.25 & 40  & 0.0 & 0  & 0  & 0.010  & 0.066 &  100  & 63    & 400 & $5\times 10^{-3}$   &  1100 \\
        Over-density (OD)               & 0.2  & 30  & 0.0 & 0  & 0  & 0.005  & 0.05  &  45   & 45    & 120 & $1.5\times 10^{-3}$ &  500 \\
        Eccentric Companion (EC)        & 0.1  & 40  & 0.4 & 0  & 0  & 0.010  & 0.066 & 100   & 90    & 400 & $5\times 10^{-3}$   &  80 \\
        Light Inclined Companion (LIC)  & 0.15 & 40  & 0.5 & 30 & 0  & 0.010  & 0.066 & 100   & 90    & 400 & $5\times 10^{-3}$   &  20 \\
        Heavy Inclined Companion (HIC)  & 0.25 & 40  & 0.5 & 30 & 0  & 0.010  & 0.066 & 100   & 90    & 400 & $5\times 10^{-3}$   &  20 \\
        Polar Companion (PC)            & 0.2  & 40  & 0.5 & 90 & 90 & 0.010  & 0.066 & 100   & 90    & 400 & $5\times 10^{-3}$   &  60 \\
        Gravitationally Unstable (GI)   & -    & -   & -   & -  & -  & 0.75  & 0.05 & 100   & 10     & 400 & -   &  30 \\

        \hline
    \end{tabular}
    \end{adjustbox}
    \caption{A summary of the initial conditions of the models presented in this paper. Note that model OD is taken from \protect\cite{calcino2019}, but the disc parameters have been scaled. The duration of the simulations is shown in the final column and is measured in the number of orbits of the companion. For models NC and GI the number of orbits is defined at $R_\textrm{out}$, while for the multiple planets simulation it is the number of orbits of the outer planet.}
    \label{tab:ic}
\end{table*}

We simulated 11 circumbinary and circumstellar discs using the 3D smoothed particle hydrodynamics (SPH) code {\sc Phantom} \citep{phantom2018}. We did not include any dust component in our simulations since we are primarily concerned with the distribution and dynamics of the gas. In all simulations we used $N_\textrm{part} = 5\times 10^{6}$ SPH particles to model the gas disc. The central star and companion were modelled as sink particles \citep{bate1995}, which experience their mutual gravitational attraction, as well from the gas disc. Gas particles are free to accrete onto both sink particles provided they are within a specified accretion radius and are gravitationally bound.

Owing to the large parameter space of companion orbital parameters, we restricted our analysis to only a few orbital configurations.
We consider companions on both co-planar and inclined, as well as circular and eccentric orbits. We kept the parameters of the gas disc fixed where feasible. Specific orbital and disc parameters used in this study are listed in Table~\ref{tab:ic}, along with the reference names of each simulation. 

The gas discs in our simulations are initialised such that the surface density $\Sigma (R)\propto R^{-p}$ for $R_\textrm{in} < R < R_\textrm{out}$, where we set $p = 1$. The temperature profile of the disc is locally isothermal with $T(R) \propto R^{-2q_T}$, with $q_T=0.25$. The aspect ratio of the disc is set to $H/R_\textrm{ref}$ at $R_\textrm{ref}$, with specific values listed in Table~\ref{tab:ic}. The central sink particle is set to have a mass of 2 M$_\odot$, while the companion has a mass ratio of $q = M_{\textrm{C}} / M_{\textrm{P}}$, where the $q$ values for each simulation are listed in Table~\ref{tab:ic}. We use the SPH artificial viscosity $\alpha_{AV}$ to produce a \citet{shakura1973} alpha viscosity according to \citep{lodato2010}
\begin{equation}
    \alpha_{SS} \approx \frac{\alpha_{AV}}{10} \frac{ \left\langle h \right\rangle }{H},
\end{equation}
where $\left\langle h \right\rangle$ is the mean smoothing length around a cylindrical annulus and $H$ is the disc scale height. 
This prescription means that $\alpha_{SS}$ is a function of position since \citep{lodato2007}
\begin{equation}
    H = \frac{c_s}{\Omega} \propto R^{3/2-q_T}
\end{equation}
and 
\begin{equation}
    \left\langle h \right\rangle \propto \left( \frac{\Sigma}{H} \right)^{-1/3} \propto R ^{(p - q_T)/3+1/2}. 
\end{equation}
Our choice of $p$ and $q_T$ implies that $\left\langle h \right\rangle / H \propto R^{-1/2}$, and hence $\alpha_{SS}$ increases with decreasing radius. Our quoted values of $\alpha_{SS}$ are an average, which is obtained by finding the binned average of $\left\langle h \right\rangle / H$ as a function of $R$ and averaging this over all bins.
We use a value of  $\alpha_\textrm{SS} = 5\times 10^{-3}$ for all of our simulations except for the model containing a circumbinary over-dense lump (model OD, see Table~\ref{tab:ic}), where $\alpha_\textrm{SS} = 1.5\times 10^{-3}$. 

As the discs evolve the surface density decreases with time. However, this does not result in a substantial change in $\alpha_{SS}$ even in our longest duration simulation, No Over-density (NOD), which was evolved for 1,100 orbits of the companion. Our $\alpha_{SS}$ diverges most significantly from the initial value inside the cavity, where $\alpha_{SS}$ can reach $\sim$ 0.1. Despite such a large viscosity the radial velocity induced by accretion is still much smaller than the radial velocities induced by the binary companion. Hence the high viscosity does not have a significant effect on interpretation of the kinematics of our circumbinary discs. We discuss this further in Section \ref{sec:models}.

Both simulations with co-planar companions on circular orbits (models over-density, OD, and no over-density, NOD) initially form an over-dense feature orbiting the cavity edge at the Keplerian frequency.

It was shown in \cite{ragusa2020} that this over-density is generated during a phase of rapid growth in disc eccentricity. This eccentricity growth is thought to arise due to either the $(m,l) = (1, 1)$ outer circular Lindblad resonance or the $(m,l) = (3, 2)$ eccentric Lindblad resonance, which are located at
\begin{equation}
    R_{\rm L}=\left(\frac{m\pm 1}{l}\right)^{2/3}a_{\rm bin} \approx 1.59\ a_{\rm bin},
\end{equation}
where $a_{\rm bin}$ is the binary orbital separation. Only the outer resonances (i.e. the $m+1$) resonances lead to growth in eccentricity, while the inner ones ($m-1$) damp it. Both simulations are initialised with $R_\textrm{in}$ close to this location, and hence they both develop an over-density.

The feature in model NOD persists robustly for roughly 300 orbits of the companion, while in model OD the feature is seen well beyond 800 orbits. The reason why this feature dissipates in one model much earlier than the other is not fully understood \citep{ragusa2020}, but is likely related to a combination of the SPH resolution at the cavity edge as well as the viscosity. Since model OD is initialised with a disc extending to only 120 au (compared to 400 au in NOD), a higher resolution (and hence lower viscosity) is maintained. We show model OD at an earlier time evolution than model NOD since we are interested in how the over-dense feature changes the kinematic profile of the disc.

Of the three inclined models used in this study, two of them (models light inclined companion, LIC, and heavy inclined companion, HIC) are initialised with companions that are not in equilibrium with the disc. As such, the binary in these models strongly torques the disc which results in alignment of the disc and the binary. Previous literature suggests that such misalignments can be maintained as the disc undergoes oscillations around a stable configuration and may persist for thousands of binary orbits \citep{Martin&Lubow2017, Smallwood+2019, rabago2023}. Thus the inclusion of misaligned, unstable binaries is justified, as such objects are expected to exist \citep{Bate2018,Wurster2019}. 

Evolving these simulations for a similar duration as, for example, models OD and NOD, would lead to the discs becoming significantly misaligned from their initial orbits. We only evolve these simulations for 20 binary orbits so that their discs remain close to their initial inclination. This is long enough for a quasi-steady state to develop for the dynamic structure in and near the cavity, but not too long for the disc inclination to change substantially.

For all of our circumbinary disc models the orbital elements of the binary change less than 1\% compared with the initial values listed in Table~\ref{tab:ic}. For the planet (P) simulation, the semi-major axis reduced to 79.2 au and had negligible change in eccentricity. For the multiple planets simulation, the semi-major axes were reduced to $[74, 128]$ au and the eccentricity increased to $[0.026, 0.094]$. The eccentric planet (EP) simulation had in increase in semi-major axis to 82 au and a decrease in eccentricity to 0.33.

\subsection{Radiative Transfer Modelling and Synthetic Observations}\label{sec:rad}

We generated synthetic observations of our SPH simulations using the Monte Carlo radiative transfer code {\sc mcfost} \citep{pinte2006,pinte2009}. Since our simulations did not include the evolution of dust grains, the dust population was assumed to follow the gas in our radiative transfer calculations. The grains were set to have a power-law grain size distribution $dn/ds \propto s^{-3.5}$ for $0.03\mu$m $\leq s \leq 1$mm  with gas-to-dust ratio of 100. The gas mass from the simulations is adopted. The grains are assumed to be spherical, homogeneous, and composed of astronomical silicate \citep{weingartner2001}. 

We used $10^8$ Monte Carlo photon packets to compute the temperature and specific intensities at each wavelength. Images were then produced by ray-tracing the computed source function. We arbitrarily assume an inclination of $i = 30^{\circ}$, a position angle PA $= 270^{\circ}$, and a source distance of 100 pc. When generating CO isotopologue observations we assumed that $T_\textrm{gas} = T_\textrm{dust}$ and all molecules are at Local Themodynamical Equilibrium (LTE), along with constant abundance ratios across the disc relative to the gas mass. The ratios adopted were $^{12}$CO/H$_2 = 1\times 10^{-4}$, $^{13}$CO/H$_2 = 2\times 10^{-6}$, and C$^{18}$O/H$_2 = 1\times 10^{-7}$. These abundances are altered by photo-dissociation and CO freeze out ($T = 20$ K) following Appendix B of \cite{pinte2018b}. We assume that the primary star in every simulation has an effective temperature of $T_\textrm{eff} = 8000$ K and radius $R= 1.8 $ R$_\odot$, giving a blackbody luminosity of $\sim 12 $ L$_\odot$, typical for Herbig Ae/Be stars. The stellar properties for each companion are calculated from their final mass (almost identical to those listed in Table~\ref{tab:ic}) from the stellar tracks by \cite{siess2000} assuming an age of 3.5 Myr. For companions with planetary mass, their luminosity is adopted from \cite{Allard2001}. The final images are produced with a pixel resolution of $0.03\arcsec$.

Individual channels for the CO isotopologues are created at a separation of 50  ms$^{-1}$. We mimic the finite spectral resolution by linearly interpolating over 5 channels to produce 101 images between the first and last channel. 
These images were averaged after weighting by a Hann window function producing a width and separation of 250 ms$^{-1}$. The channels were then smoothed with a Gaussian beam assuming a beam size of $0.15\times 0.15$ arcseconds.\footnote{The code used to conduct these calculations, {\sc pymcfost}, is available at \url{https://github.com/cpinte/pymcfost}.} We choose this beam size as it is the standard beam size obtained from the MAPS survey \citep{mapsI}.

When adding white noise to our simulated observations, we assumed a specific noise levels of $F_\textrm{noise} = [1, 2.5, 5, 10]$ mJy. These noise levels correspond to an average peak signal-to-noise ratio of approximately $\textrm{SNR} = [170, 70, 35, 18]$ for the CO (3-2) line emission in a single channel, across all of the models. The noise levels we assume are readily achievable with a few hours of integration on source, however the signal-to-noise ratio of the lowest noise model is quite optimistic given we assume bright and hot central stars that produce brighter CO emission than would be seen around fainter stars. The $F_\textrm{noise} = 2.5$ mJy produces a channel signal-to-noise ratio closer to what has been obtained with previous ALMA observations \citep[e.g. the MAPS sample,][]{oberg2021}.

We leave a more detailed study of the different noise levels and how they affect our kinematic criteria derived in Section \ref{sec:kin_crit} in the Appendix.
The noise is generated using a random Gaussian with a mean of zero which we then convolve with a Gaussian beam.The convolved noise is then rescaled such that it has a final FWHM of $\textrm{F}_\textrm{noise}$. The noise is then added to the convolved observations to produce the final synthetic observations.

We used the code {\sc bettermoments} \citep{bettermoments2018} to generate moment maps of our synthetic observations and ALMA CO observations. 
We apply noise cuts when generating our moment maps. For the $\textrm{F}_\textrm{noise}=1$ mJy noise level, we apply a 5 RMS noise cut, while for the other noise levels the cut is 7 RMS. We used the first moment to generate our velocity maps, but also discuss and test other methods in the Appendix.

The source distance assumed, along with the general size of our discs in Table~\ref{tab:ic} and the adopted beam size imply that our discs are very well resolved. We assumed this to present a best case scenario of what kinematic signatures are and will be possible to observe with current generation interferometers such as ALMA. We did not take into account the image artefacts that can arise due to sparse $uv$-coverage. We test how changing the beam size and disc inclination affect our kinematic criteria in the Appendix. 

Some of the models listed in Table~\ref{tab:ic} contain binary configurations that lead to extremely depleted central cavities (e.g. models OD, NOD, and EC). Since the resolution of an SPH simulation is related to the mass of the gas at a specific location, some regions inside of the cavity are less resolved than others. In Appendix \ref{sec:res_study} we show that decreasing the SPH particle number does not significantly change our kinematic criteria. The main reason for this is that the less resolved portions of the disc do not produce a significant amount of CO flux compared with the higher density and better resolved portions. Furthermore, since we include the effects of photo-dissociation in our radiative transfer calculations, the ratio of CO in the low density regions is much lower than the prescribed ratios listed above, further reducing the observed CO flux. The addition of artificial noise to our simulated channel maps also ensures that no measurable level of flux is coming from the unresolved portions of the disc. This is evident in the velocity maps of Figure~\ref{fig:mom1_com}, where there is a lack of signal inside of the cavity of most simulations.

\section{Results} \label{sec:res}
\subsection{Hydrodynamical Models}\label{sec:hydro}

\begin{figure*}
    \centering
    \includegraphics[width=0.8\linewidth]{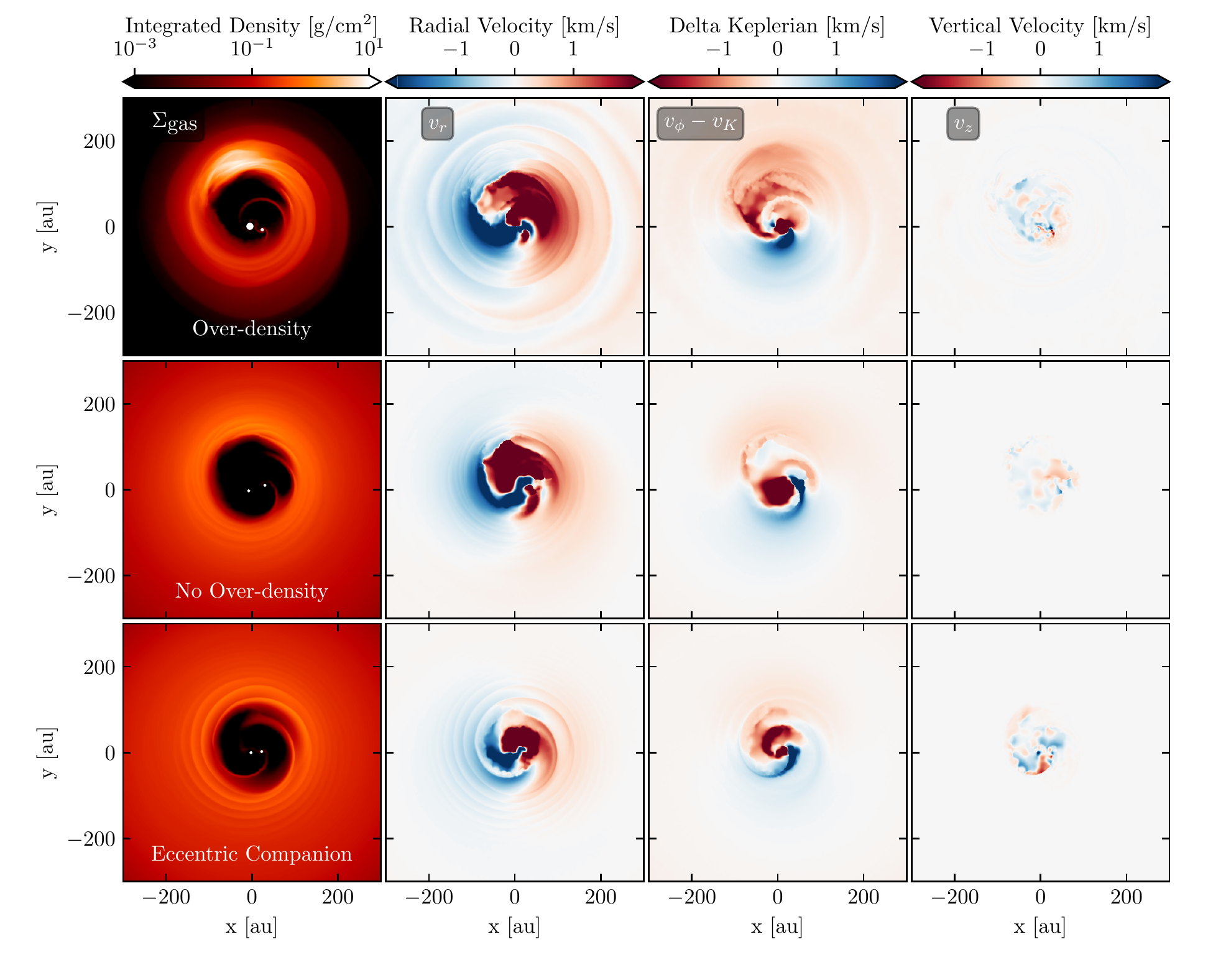}
    \caption{The surface density (\emph{left panel}), radial velocity (\emph{second from left}), deviation from Keplerian rotation (\emph{second from right}), and vertical velocity (\emph{right panel}) for a selection of our co-planar models listed in Table~\ref{tab:ic}. The white points in the left panel show the position and accretion radius of the sink particles. Both model OD and NOD have highly eccentric discs, which is seen in the velocity maps. In particular, gas motion is super-Keplerian at the pericentre of the eccentric disc, and sub-Keplerian at the periastron. The presence of an over-dense feature in model OD leads to the generation of spiral structure in the gas surface density and velocity perturbations, while only minor spiral structure is faintly seen outside of the cavity of model NOD. Thus we can distinguish between the spirals induced directly from the binary and spirals generated by the over-dense feature. Model EC also displays prominent spiral structure around the cavity in both surface density and velocity. In all co-planar models the velocity in the $z$-direction is negligible, and is dominated by noise (i.e. low particle resolution) in the very inner most regions of the cavity. }
    \label{fig:dens_vel_cop}
\end{figure*}

\begin{figure*}
    \centering
    \includegraphics[width=0.8\linewidth]{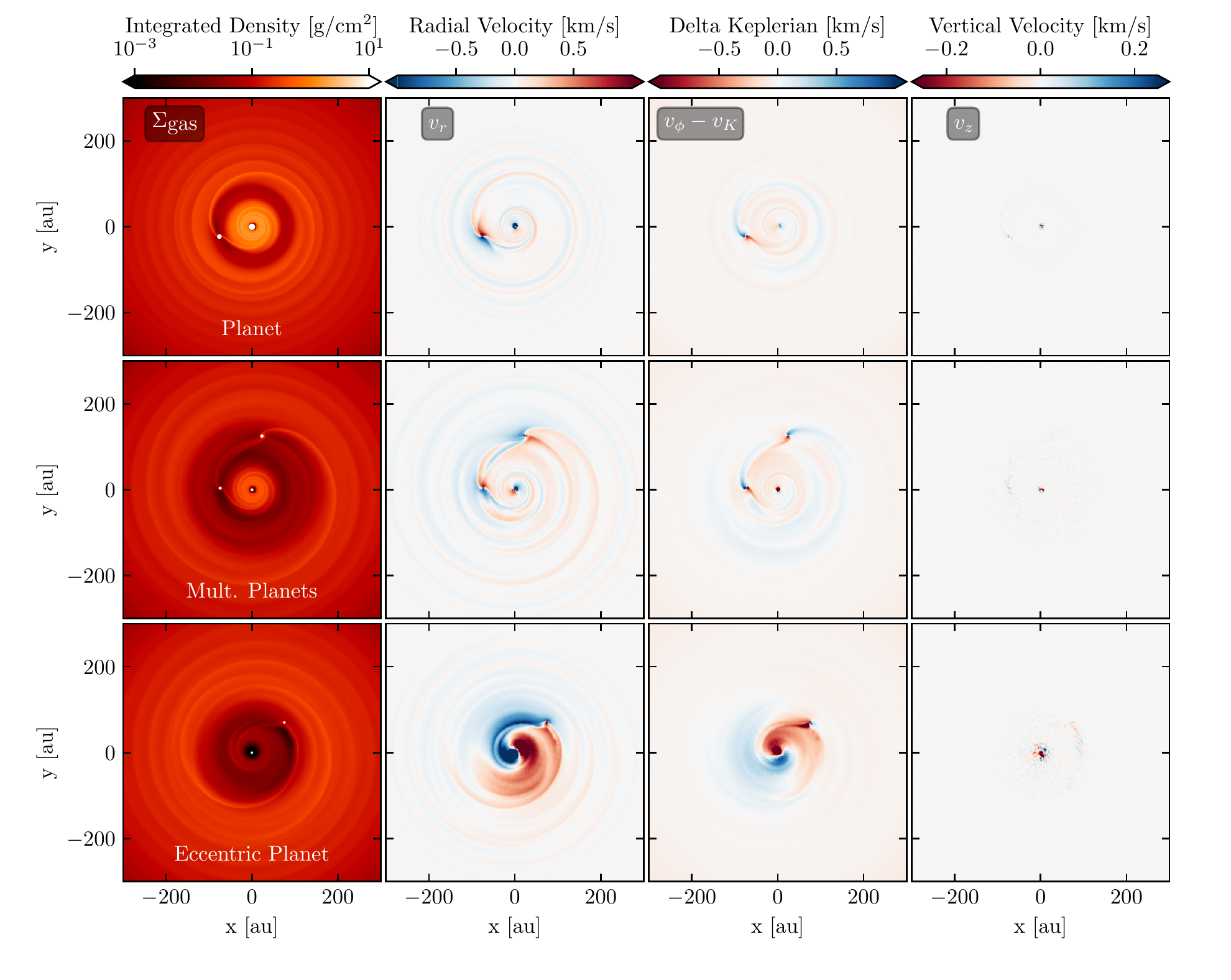}
    \caption{Same as Figure~\ref{fig:dens_vel_cop} but for our planet models. Perturbations in the velocity field are substantially lower than in the co-planar stellar mass companion models of Figure~\ref{fig:dens_vel_cop} for all models. The eccentric planet (EP) shows larger perturbations than the other planet models owing to the eccentricity of the companion which drives eccentric gas motion primarily inside of its' orbit.
    }
    \label{fig:dens_vel_plan}
\end{figure*}

\begin{figure*}
    \centering
    \includegraphics[width=0.8\linewidth]{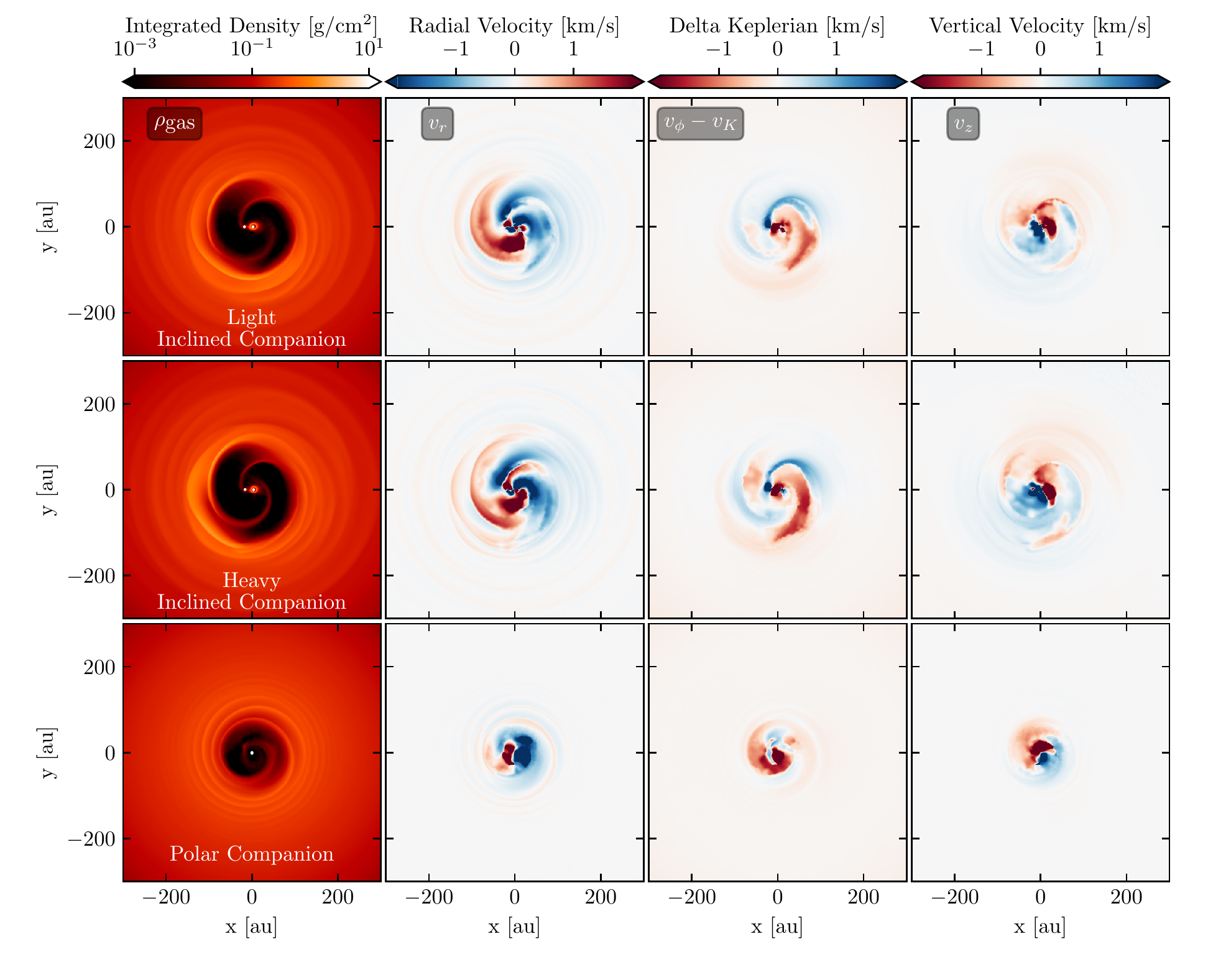}
    \caption{Same as Figure~\ref{fig:dens_vel_cop} but for our inclined models. Models LIC and HIC display abundant spiral structure inside and outside of the cavity. The spirals inside the cavity, and close to the cavity edge, are arising due to binary. In both model there is a non-negligible velocity component in the $z$-direction due to a slight warp of the disc. Model PC also contains spiral structure, though not as striking as the non-polar cases. Gas flowing inside the cavity is dragged away from the disc mid-plane by the companion, resulting in a large $v_z$. Substantial radial flows inside the cavity are also present in all models.}
    \label{fig:dens_vel_inc}
\end{figure*}

Figure~\ref{fig:dens_vel_cop} shows the surface density and velocity components for models OD, NOD, and EC, and models LIC, HIC, and PC in Figure~\ref{fig:dens_vel_inc}. The velocity components are the velocity in the radial direction, $v_r$, the deviation from Keplerian rotation assuming single point mass at the binary centre of mass, $\Delta V_\textrm{K}$, and the velocity in the vertical direction, $v_z$. All velocity components are measured from a thin slice about the mid-plane of the disc model. Keplerian velocity is computed using
\begin{equation}
    v_K = \left( \frac{G (M_P + M_C)}{r_\textrm{CM}} \right)^{1/2},
\end{equation}
where $r_\textrm{CM}$ is the radial location of the gas parcel with respect to the binary centre of mass.

\begin{figure*}
    \centering
    \includegraphics[width=0.8\linewidth]{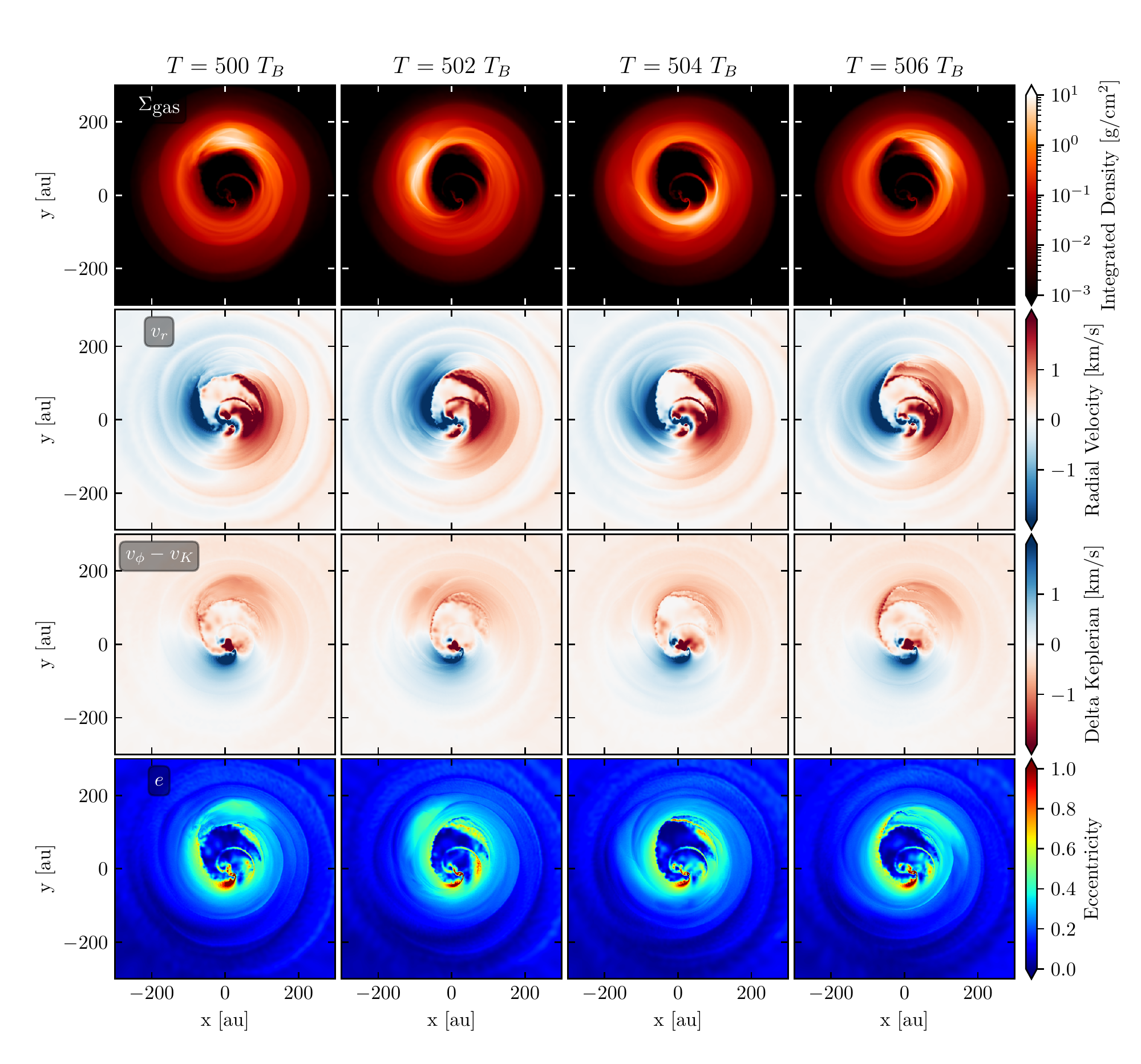}
    \caption{The time evolution of model OD with the eccentricity, $e$, of the gas particles included. The over-dense feature orbits once roughly every 7 binary orbits ($T_B$). As the over-dense feature orbits, outer density spirals trail it and propagate radially. These spirals perturb the velocity profile of the disc, and the separation of successive spirals is roughly determined by the orbital frequency of the over-dense feature. The azimuthal velocity is strongly perturbed in the radial direction through the over-density, where the velocity becomes increasingly sub-Keplerian with increasing radius. This is apparent in every snapshot presented, but appears stronger at apastron. This change in azimuthal velocity is mostly created by the change in eccentricity of the gas particles across the over-density.}
    \label{fig:dens_vel_irs}
\end{figure*}

\subsubsection{Co-planar models}\label{sec:cop_mod}

Starting with the co-planar models in Figure~\ref{fig:dens_vel_cop}, we observe the presence of an over-dense feature in model OD, while one is lacking in model NOD. Neglecting the presence of the over-dense feature for a moment, the morphology is similar between both models. Both discs are eccentric at the cavity edge, a feature which is seen in other studies of low eccentricity, co-planar binaries \citep{papaloizou2001, ragusa2017, hirsh2020, ragusa2020}. Model EC (eccentric companion) contains a co-planar companion with a modest eccentricity ($e=0.4$) which is lower in mass than the other two models. 

Non-circular Keplerian motion of the gas is apparent in the radial and azimuthal velocity components. These velocity profiles are consistent with our expectations of gas particles on an eccentric orbit; their azimuthal velocity component reaches a maximum at the pericentre of the disc, while it is at a minimum at the apocentre. The gas particles also have a large radial component to their velocity along the cavity edge due to their eccentric orbit. 

The presence of the over-dense feature changes both the density and velocity structure of the disc. Spiral arms emanating off the over-dense feature maintain a relatively high pitch angle as they propagate radially outwards. The over-dense feature also increases the accretion rate onto the sink particles in the cavity, as shown in previous studies \citep{farris2014, miranda2017}. The spirals are clearly visible in the radial and azimuthal velocity components. The radial component arises owing to the outward radial propagation of the spiral density waves \citep{rafikov2002,bollati2021}. 

To better explore these features, in Figure~\ref{fig:dens_vel_irs} we show multiple timesteps of model OD. Here we can clearly see the outwards radial propagation of the spiral density waves. Another feature of interest is the radially increasing deviation in azimuthal velocity seen across the over-dense feature. A change in velocity on the order of 500 ms$^{-1}$ occurs between the start and end of the over-dense feature in the radial direction. Although less obvious this is also evident in the other timesteps. 

At first impression one might assume this radial change in $v_\phi$ across the over-density is due to the gas pressure support. The change in velocity arising due to the gas pressure support can be derived from the Navier-Stokes equation assuming $v_r << v_\phi$ and the gas is in a circular orbit \citep[e.g.][]{pringle1981}
\begin{equation} \label{eq:rad_eq}
    v_\phi ^ 2 - v_K ^2 = \frac{c_s^2 r}{\rho} \frac{\partial \rho}{\partial r},
\end{equation}
where $c_s$ is the sound speed. If we take a slice along $x = 0$ in the left panel of Figure~\ref{fig:dens_vel_irs}, the change in velocity owing to the gas pressure support is not large enough to explain the observed change in $v_\phi$. Since the only other force present in our simulations is gravity, the change in velocity gradient must be arising due to the over-dense feature interacting with the time-varying gravitational potential. 

The cavity in all three circumbinary models is depleted by a factor of at least $10^4$, barring the occasional accretion stream entering the cavity which feeds the primary and secondary sink particles. This drop in density is consistent with the drops found in many transitional discs \citep{vandermarel2015, garg2021}. 

In comparison, the co-planar planet models shown in Figure~\ref{fig:dens_vel_plan} mostly show smaller velocity perturbations than the co-planar circumbinary models (note the change in the scale of the colourbar). Gas depletion co-located with the planets is much lower than in the circumbinary models. The eccentric planet (EP) model shows larger perturbations than the other two planet models due to the eccentricity of the planet, which causes eccentricity in the gas. Compared with the circumbinary models, model EP has a much lower gas depletion rate inside of the cavity.

\subsubsection{Inclined models}

Figure~\ref{fig:dens_vel_inc} shows the surface density and velocity components for our inclined models. Models LIC (light inclined companion) and HIC (heavy inclined companion) display prominent spiral structure both inside and outside of the cavity. The spirals inside the cavity are excited by a combinations of Lindblad resonances and accretion streams. These two models are similar to the model presented in \cite{poblete2020} which reproduced the spiral arms inside AB~Aurigae \citep{Tang2017}. We leave the interested reader to refer to Sections 3.1 and 3.2 of \cite{poblete2020} for a more complete description of the time evolution of the spirals. The spiral arms outside of the cavity are caused by a low amplitude over-dense feature orbiting the cavity similar to model OD. The velocity in the $z$-direction is non-zero due to the binary torque on the disc creating a warp. We do not present simulated observations of model LIC due to the similarities with model HIC. 

Model PC (polar companion) contains a companion on a polar orbit. This model has the lowest disc cavity radius of all the models presented, which is inline with both theoretical and numerical studies on eccentric and inclined binaries \citep{Miranda&Lai2015, hirsh2020}. We can also see that this particular orbital configuration is not as efficient at clearing gas inside the cavity, particularly compared with the co-planar models.

\section{Morphological Signatures} \label{sec:co_em}
\begin{figure*}
    \centering
    \includegraphics[width=0.8\linewidth]{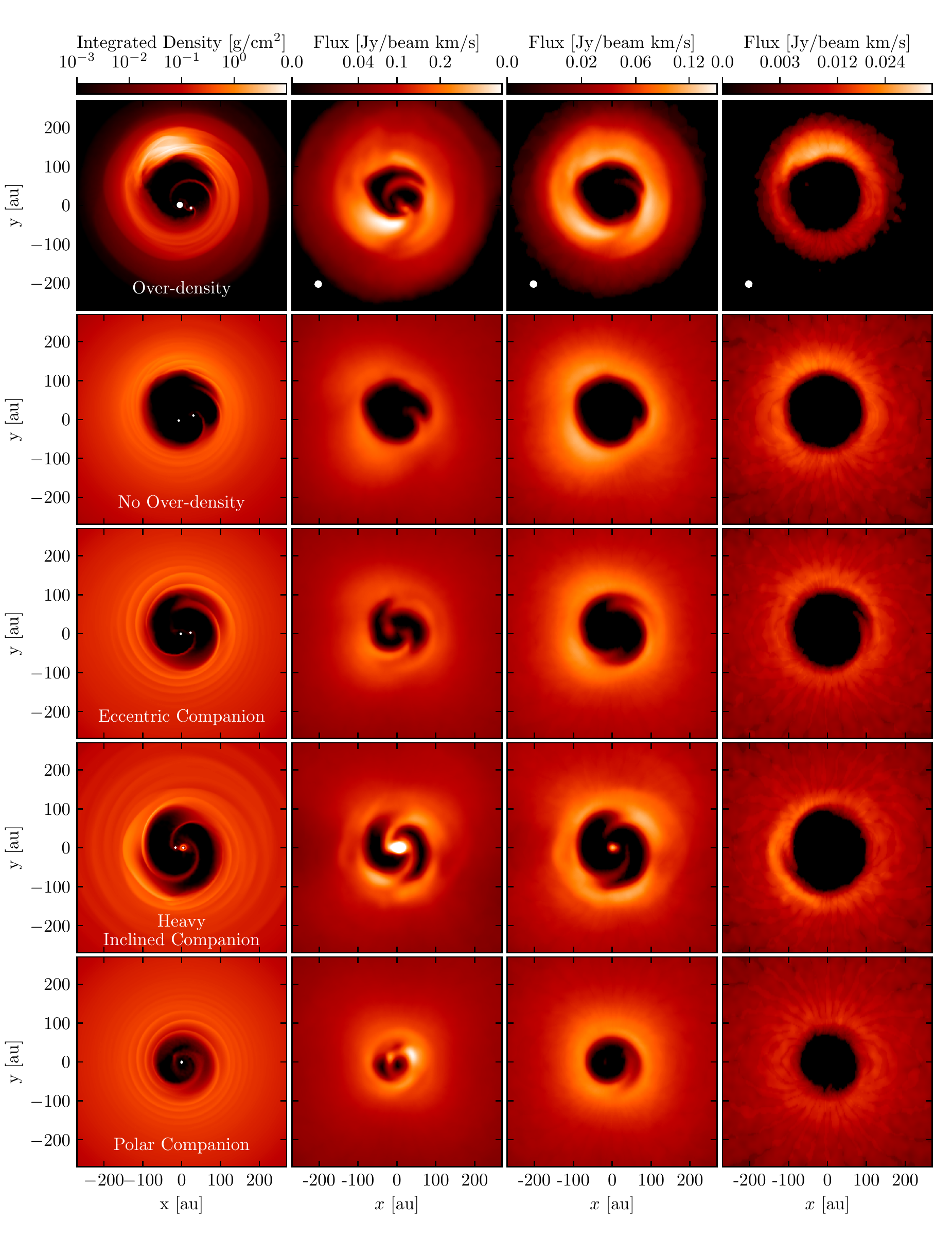}
    \caption{The integrated CO emission for several of the circumbinary models in Table~\ref{tab:ic} for the $\textrm{F}_\textrm{noise}=1$ mJy noise level. The left column shows the SPH integrated density, while the columns show deprojected $^{12}$CO (3-2), $^{13}$CO (3-2), and C$^{18}$O (3-2) integrated emission, in order. Due to the large depletion in gas in our circumbinary simulations, the cavity becomes progressively optically thin with less abundant isotopologues, but may remain optically thick in $^{12}$CO. The transition from an optically thick to optically thin cavity depends on the orbital parameters of the companion, our assumed initial gas mass, and the temperature profile of the disc. Spiral structure is observed both within and outside of the cavity.}
    \label{fig:mom0_com}
\end{figure*}

\begin{figure*}
    \centering
    \includegraphics[width=0.8\linewidth]{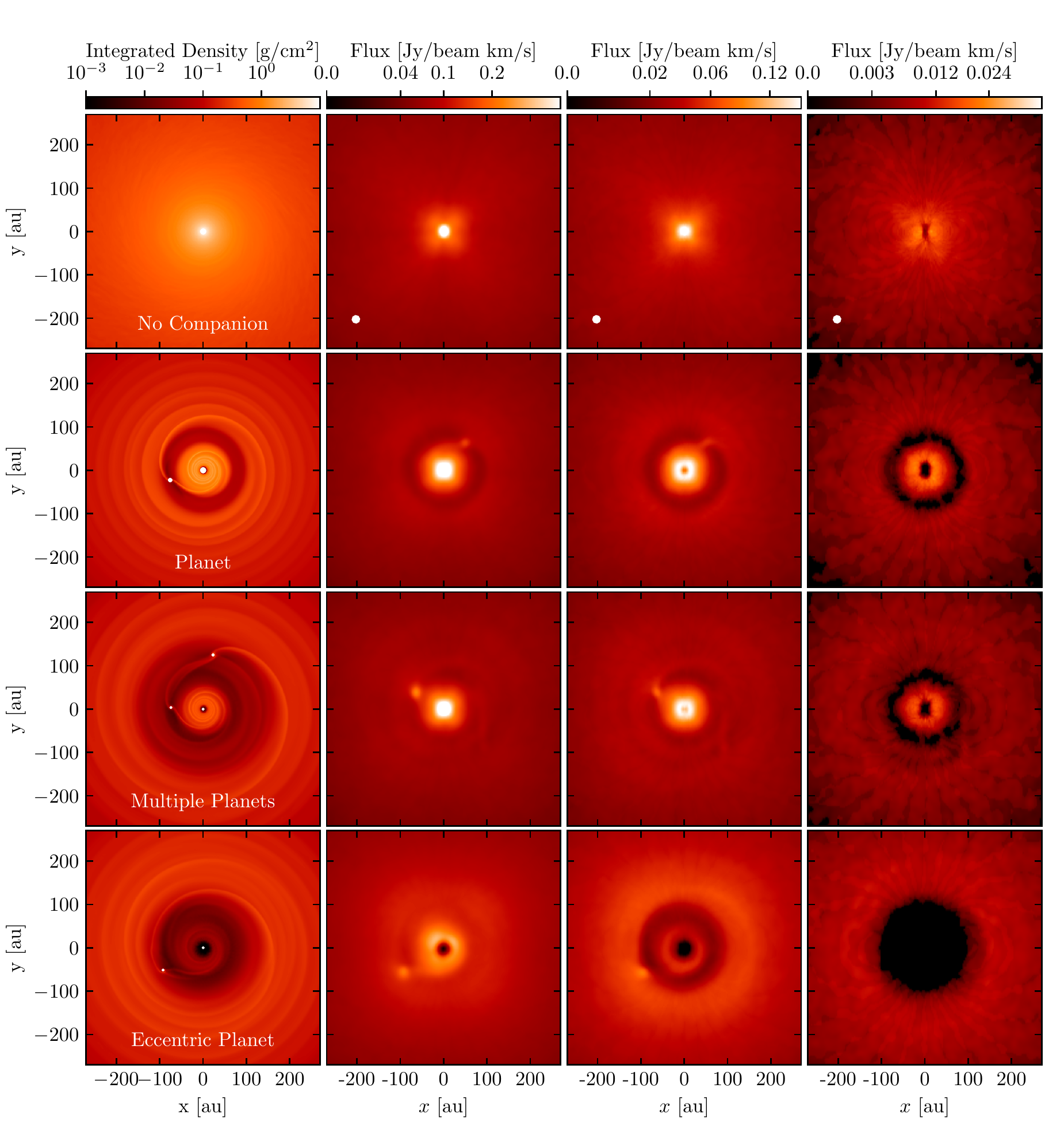}
    \caption{As in Figure~\ref{fig:mom0_com} but for the no companion and planet models. Aside from the eccentric planet model, no models show a cavity in any CO isotoplogue (the central small holes are artificial and created by the central sinks). }
    \label{fig:mom0_plan}
\end{figure*}

Figure~\ref{fig:mom0_com} shows our simulated CO emission for most of our circumbinary disc models (excluding model LIC), while Figure~\ref{fig:mom0_plan} shows CO emission for our no companion and planet models. The columns, from left to right, show the integrated density from the simulation, and the CO (3-2), $^{13}$CO (3-2), and C$^{18}$O (3-2) integrated emission.

\subsection{The Cavity}\label{sec:cavity}

The appearance of a cavity in CO isotopologues depends sensitively on the gas mass of the disc, the CO isotopologue abundances, and the temperature profile of the disc. It also depends on the nature of the binary. Inspecting the simulated CO observations in Figure~\ref{fig:mom0_com}, CO emission ranges from optically thin to optically thick. In general, the models with a co-planar companion are more efficient at clearing material in the cavity, and hence have a more optically thin cavity. Inclined models tend to allow more material into the cavity which in the case of model HIC leads to no cavity at all in $^{12}$CO. In all models the cavity is far more prominent in less abundant CO isotopologues, with C$^{18}$O most faithfully tracing the gas surface density. We also note that the cavity size increases as the CO isotopologue abundance decreases. 

When the cavity is eccentric and the main source of illumination is offset from the centre of the ellipse (as it is in a Keplerian orbit), the cavity wall is not uniformly illuminated. The cavity edge closest to the source of illumination is hotter than more distant regions, producing a temperature difference which can manifest as a brightness asymmetry. For example, in the over-density model the brightest region of the cavity edge is not the over-density, but the region closer to the source of illumination.

In comparison with the planet models in Figure~\ref{fig:mom0_com}, the CO emission in the cavity is much lower in the circumbinary disc models. A cavity is present in the eccentric planet (EP) model, however significantly more gas is present inside the cavity than in the circumbinary disc models causing only the less abundant C$^{18}$O emission to display a cavity. With this comparison we can confidently state that circumbinary discs will contain a cavity depleted in either CO or $^{13}$CO emission.

\subsection{Spirals Inside The Cavity} \label{sec:spiral_in}

Spiral structure is observed across the isotopologues inside the cavity, and are composed of spiral density waves excited by Lindblad resonances as well as accretion streams. The morphology of these spirals sensitively depends on the binary orbital parameters relative to the location of the disc. Highly inclined and/or eccentric companions tend to result in two prominent inner spirals that appear as a $m=2$ spiral mode. Our polar companion (PC) model produces a single prominent spiral inside the cavity. 
Our planet models in Figure~\ref{fig:mom0_plan} produces some faint spiral structure, however their contrast ratio is much lower than in the circumbinary models.

\begin{figure*}
    \centering
    \includegraphics[width=0.9\linewidth]{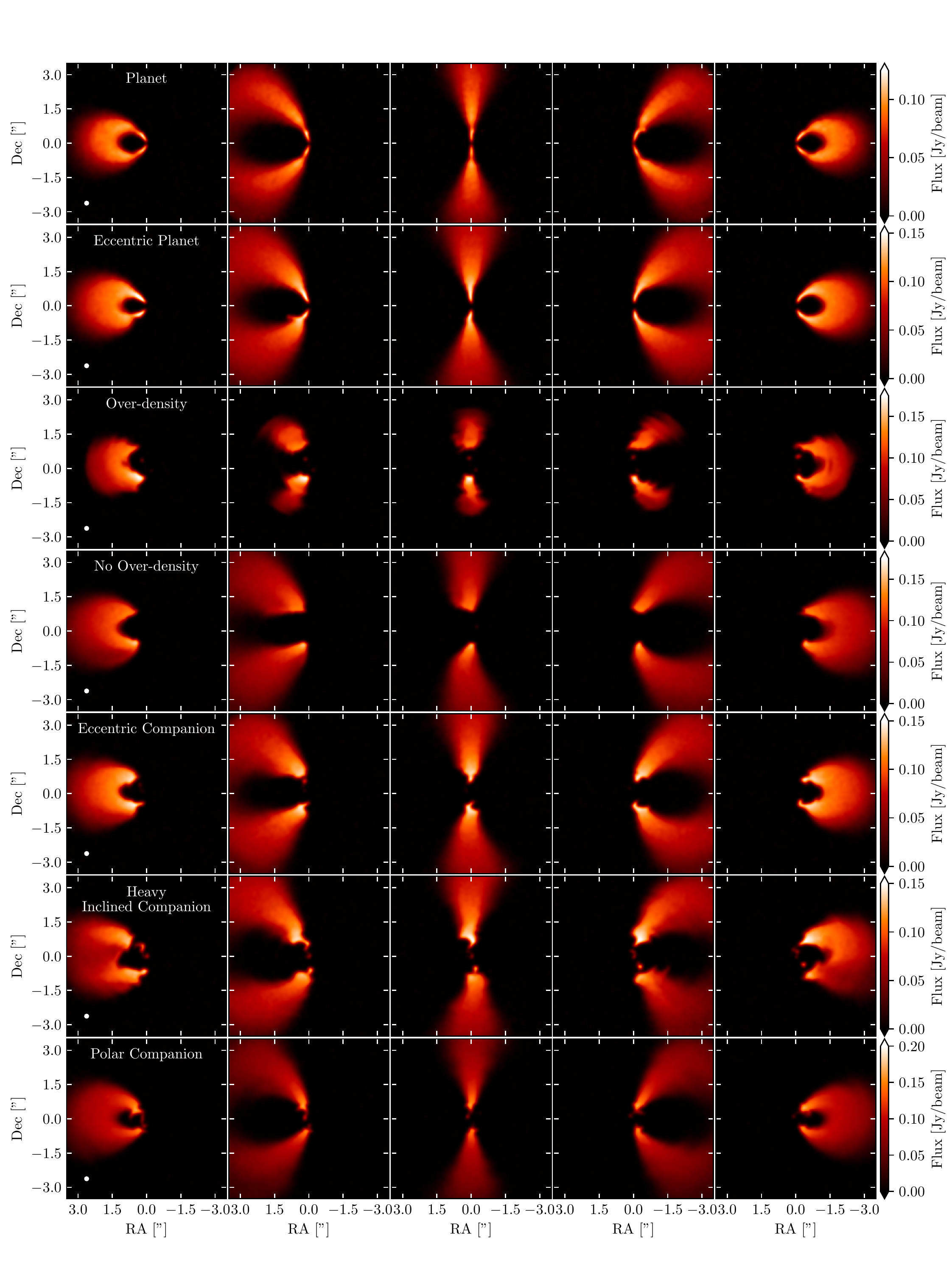}
    \caption{The isovelocity curves for several of the models in Table~\ref{tab:ic} for the $\textrm{F}_\textrm{noise}=1$ mJy noise level. In almost every case kinks or wiggles are seen across all of the channels in the circumbinary models. This is in contrary to planet-hosting discs, where the isovelocity curves are only strongly perturbed in the neighbourhood of the perturbing body \citep{pinte2018, pinte2019}. Models OD contains a much larger number of perturbations than model NOD (where very few are present), which are a direct result of the orbiting over-dense feature.}
    \label{fig:chan_com}
\end{figure*}

\subsection{Spirals Outside The Cavity} \label{sec:spiral_out}
Spirals outside of the cavity, if present, tend to be tightly wound and spatially co-located with the edge of the cavity. They are more evident in the models containing inclined and/or eccentric companions (i.e. EC, HIC, and PC) in surface density, but are not clearly visible in $^{12}$CO or $^{13}$CO integrated emission. Scattered light observations appear to be a better method of observing these spirals, which has been done in the circumbinary discs of GG~Tau~A \citep{Keppler2020} and HD~142527 \citep{fukagawa2006, rodigas2014,avenhaus2014, avenhaus2017}. These spirals arise due to Lindblad resonances between the secondary companion and the disc and dissipate as they propagate radially outward. The appearance of spirals in CO will be sensitive to the temperature profile of the disc. Our hydrodynamical models are assumed to be locally isothermal, and temperature gradients due to shocks and stellar radiation on the disc surface are not taken into account. These two effect could enhance the scale height at the location of the spirals, allowing them to intercept more stellar radiation than in our radiative transfer models and enhance their visibility in integrated emission.

Additional spiral structure outside of the cavity is seen in model OD that are a result of the orbiting over-dense feature. Contrary to spirals generated by the binary, these spiral structures are less tightly wound and emanate a substantial distance from the cavity. They are clearly seen in $^{12}$CO and $^{13}$CO. Faint spiral-like structures and also be seen in model HIC which contains a low amplitude over-dense feature orbiting the cavity. 

\begin{figure*}
    \centering
    \includegraphics[width=0.8\linewidth]{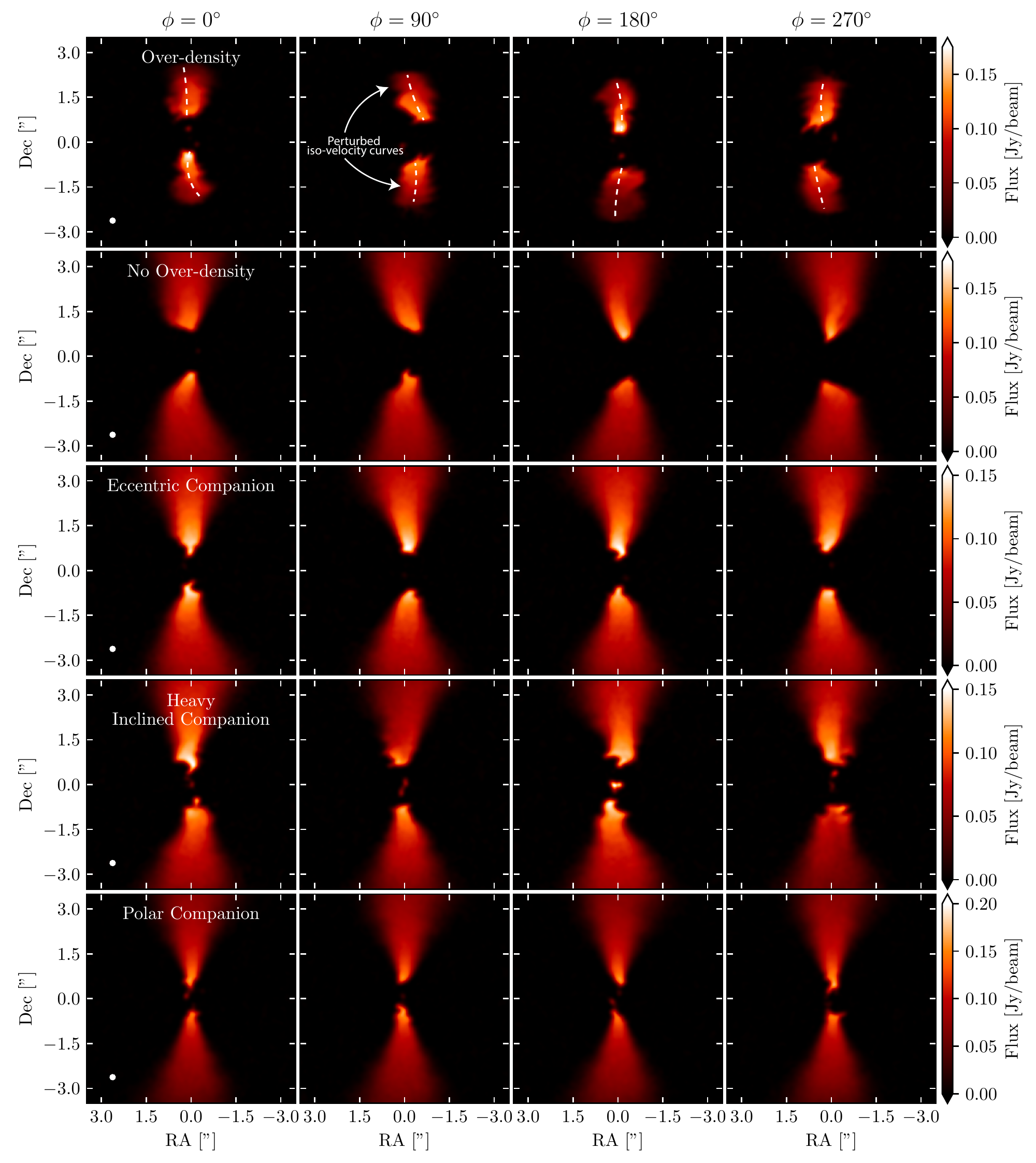}
    \caption{The $v_\textrm{los} = 0.0$ kms$^{-1}$ channels for the circumbinary disc models at various viewing angles $\phi$, using the $\textrm{F}_\textrm{noise}=1$ mJy noise level. The position angle of the disc is not changed between each column and is set such that for a Keplerian disc the isovelocity curves should point straight north and south. The direction of the isovelocity curves for our circumbinary discs is shifted compared to the expected orientation. This is particularly noticeable in the co-planar disc models where the disc eccentricity is high. We have annotated the perturbed iso-velocity curve for model OD with white dashed lines. For a Keplerian disc these lines should point along the North/South direction given the position angle of the disc. Since model OD has an eccentric disc, the iso-velocity curves for the $v_\textrm{los} = 0.0$ kms$^{-1}$ point away from this expected direction, and the columns show this is robust to the viewing angle. They also show that the velocity kinks in the vicinity of the cavity are robustly seen in most models regardless of the viewing angle.}
    \label{fig:chan_v0_com}
\end{figure*}

\begin{figure*}
    \centering
    \includegraphics[width=0.7\linewidth]{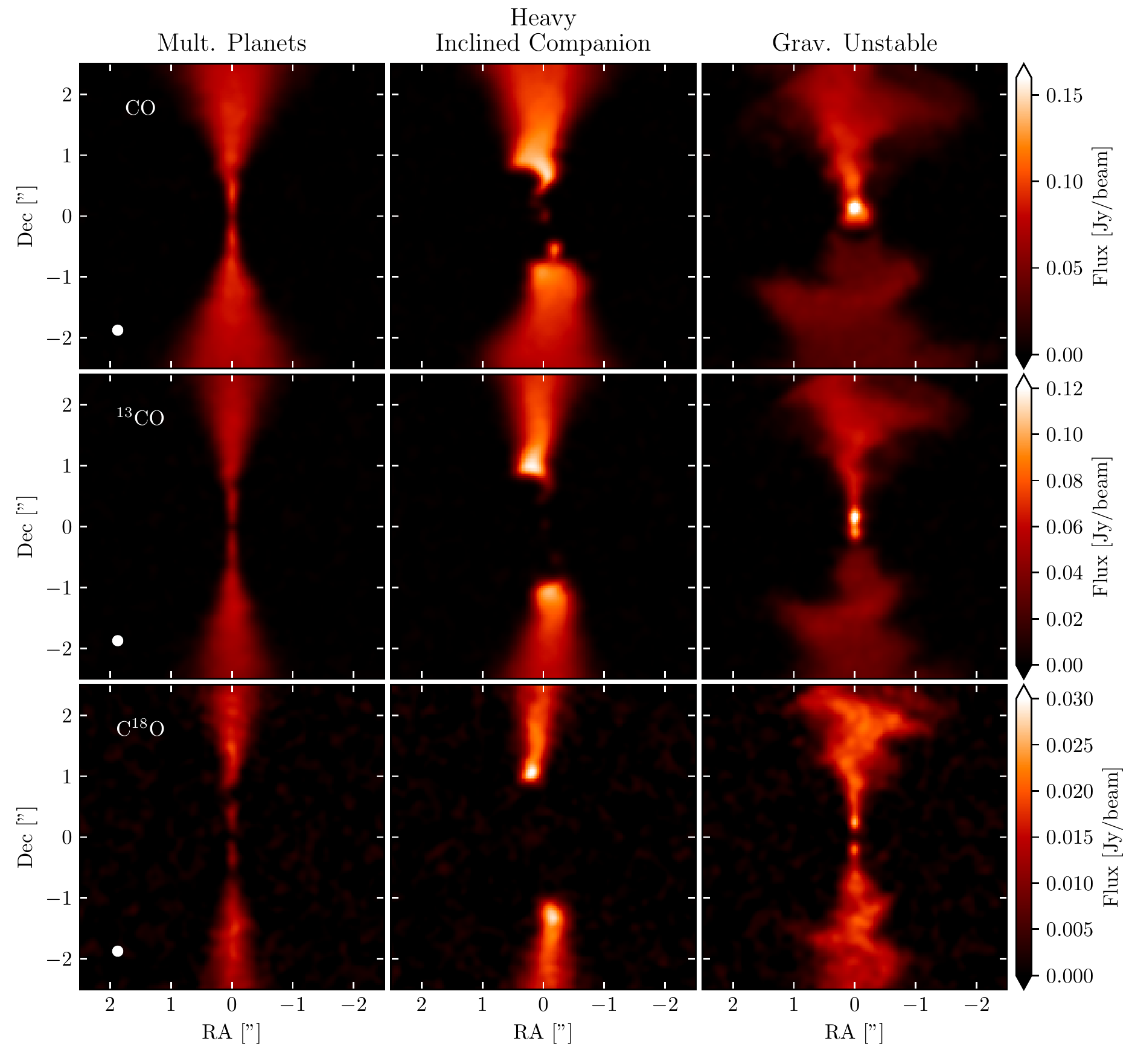}
    \caption{The $v_\textrm{los} = 0.0$ kms$^{-1}$ channels for the Multiple Planets, Heavy Inclined Companion, and Gravitational Instability models (columns) for each CO isotopologue (rows). Although all models show some degree of kinks, only the circumbinary model shows a kink in proximity of the cavity in CO, however the a depleted inner region does appear in the Multiple Planets.}
    \label{fig:chan_gi}
\end{figure*}

\section{Kinematic Signatures}\label{sec:kin_sig}
The kinematic profile of the disc is a valuable resource for determining the dynamic processes occurring inside of a protoplanetary disc. In the case of a circumbinary disc, the kinematics of the gas is heavily influenced by the interaction between the primary and companion. Therefore, searching for common kinematic signatures in numerical simulations of circumbinary discs can help shed light on the unknown dynamical processes occurring protoplanetary discs by observing the kinematic profiles.

\begin{figure*}
    \centering
    \includegraphics[width=\linewidth]{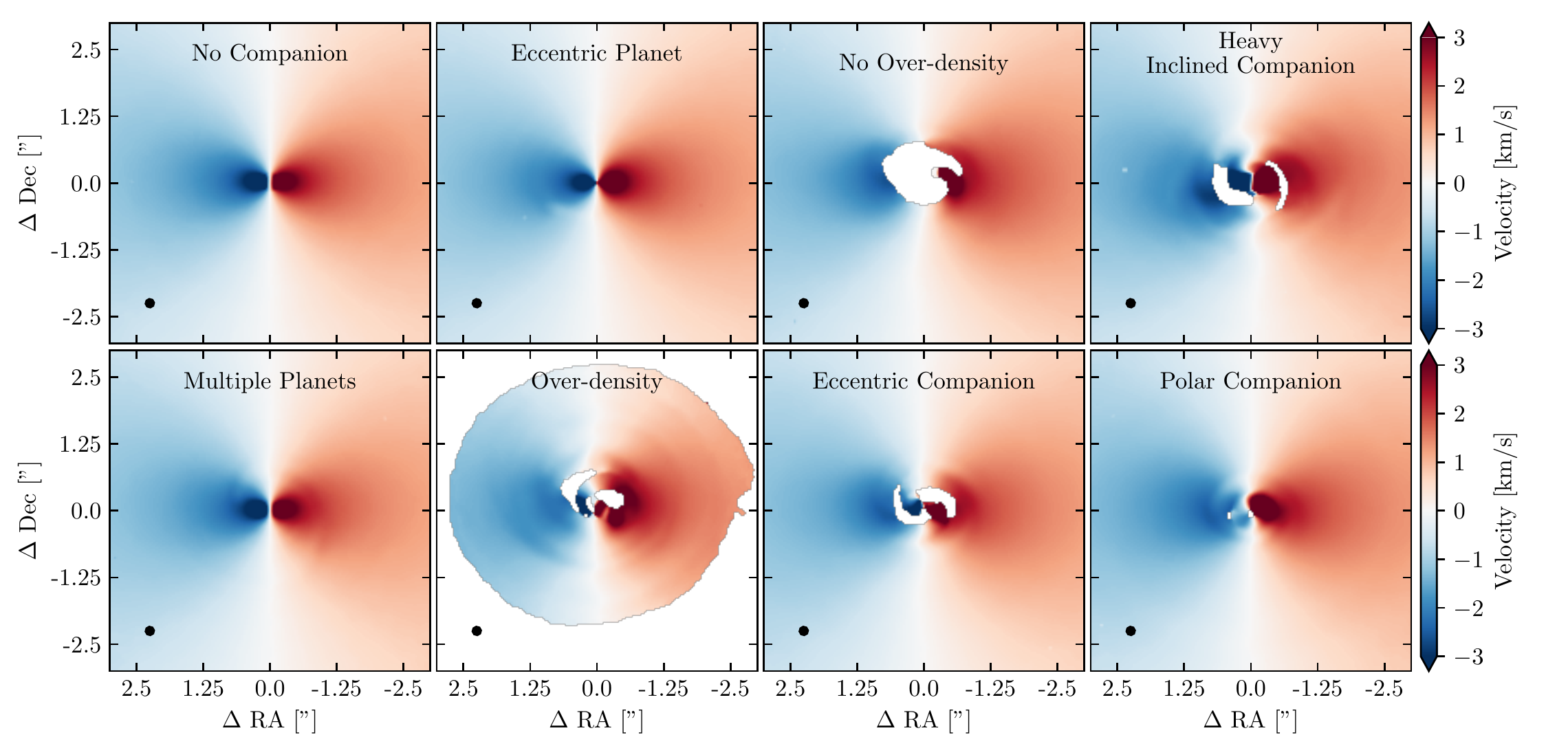}
    \caption{The Moment 1 velocity maps for all models (excluding LIC) using the channel maps presented in Figure~\ref{fig:chan_com} and the no companion model. The white sections in the centre arises due to a lack of signal inside the cavity of some of our models arises due to the large depletion of gas inside of the respective circumbinary disc model. We see that in all circumbinary disc models there are substantial deviations away from Keplerian rotation (top left panel).}
    \label{fig:mom1_com}
\end{figure*}

\subsection{Channel Maps}\label{sec:chans}

In Figure~\ref{fig:chan_com} we show the channel maps for most models. Starting with model OD (third row of Figure~\ref{fig:chan_com}), the individual channels are reduced in radial extent compared with other models owing to the smaller outer radius used in this model (see Table~\ref{tab:ic}). The channels are significantly perturbed from the channels of a circular Keplerian disc. For example, the presence of wiggles or `velocity kinks' \citep[see][for a definition]{calcino2022} across the iso-velocity curves are seen across all the channels shown. These kinks are a result of the perturbations in the velocity profile of the disc created by the over-dense feature, as shown in Figures \ref{fig:dens_vel_cop} and \ref{fig:dens_vel_irs}. We show the robustness of the appearance of wiggles in the channel maps in Figure~\ref{fig:chan_v0_com}, where we rotate each of the models in the azimuthal direction for the $v = 0.0$ km/s channel. Any model that shows wiggles at a particular viewing angle tends to show wiggles across multiple viewing angles. Thus the appearance of wiggles is robust to the viewing angle but depends on the orbital properties of the companion. 

We now compare the channels of model OD with model NOD. We begin by noting that azimuth angles $\phi = 0^{\circ}$ and $\phi = 180^{\circ}$ are orientated in such a way that the eccentricity vector of the disc is pointing toward and away from the observer, respectively. At orientations $\phi = 90^{\circ}$ and $\phi = 270^{\circ}$ the eccentricity vector is perpendicular to the line of sight. As the disc is rotated from $\phi = 0^{\circ}$ to $\phi = 90^{\circ}$, the portion of the isovelocity curve on the cavity edge stops pointing toward the projected centre of the disc. We have annotated the perturbation in the iso-velocity curves for model OD in Figure~\ref{fig:chan_v0_com}. Given our prescribed position angle when conducting the radiative transfer calculations (see Section \ref{sec:rad}), the $v=0$ km/s iso-velocity curves for a Keplerian disc would point north to south. However we can see that in model OD that this is not true for some orientations. This effect was described in \cite{calcino2019} and attributed to the eccentricity of the disc. We can better understand this phenomenon by comparing these channels to the velocity components in Figure~\ref{fig:dens_vel_cop}. When $\phi = 0^{\circ}$, the northern iso-velocity curve is tracing the CO along the most distance surface of the disc to the observer, which in this case traces the $x > 0$ side of the models in Figure~\ref{fig:dens_vel_cop}. The $v_r$ component on this side of the disc is strongly negative and is moving towards to the centre of mass, which is in the direction to the observer. Thus the emission close to the edge of the cavity appears spatially located where we would expect to see emission from blue-shifted material and not material with no motion with respect to the line-of-sight. There is a strong gradient in $v_r$ close to the cavity which is seen in the isovelocity curves as the tilt shown in Figure~\ref{fig:chan_v0_com}. This is also seen at $\phi = 270^{\circ}$.

Perturbations are also seen in models EC, HIC, and PC. Close to the cavity, these perturbations are a result of a combination of spiral arms and radial inflows into the cavity. The spiral arms outside of the cavity also appear in the channel maps, particularly in model HIC. With the exception of model OD, the spirals tend to be spatially located close to the cavity. 

In Figure~\ref{fig:chan_gi} we show the $v_\textrm{los} = 0.0$ kms$^{-1}$ channel for three models (MP, HIC, GI) and each CO isotopologue. Here it is seen that although all models contain kinks, the circumbinary model contains a large kink in proximity to the cavity. Although this could also be observed in a planet hosting model, it is not likely in a gravitationally unstable disc since as the gas density is so high that the CO should remain optically thick. 

\subsection{Velocity Maps}\label{sec:vm}

\begin{figure*}
    \centering
    \includegraphics[width=0.9\linewidth]{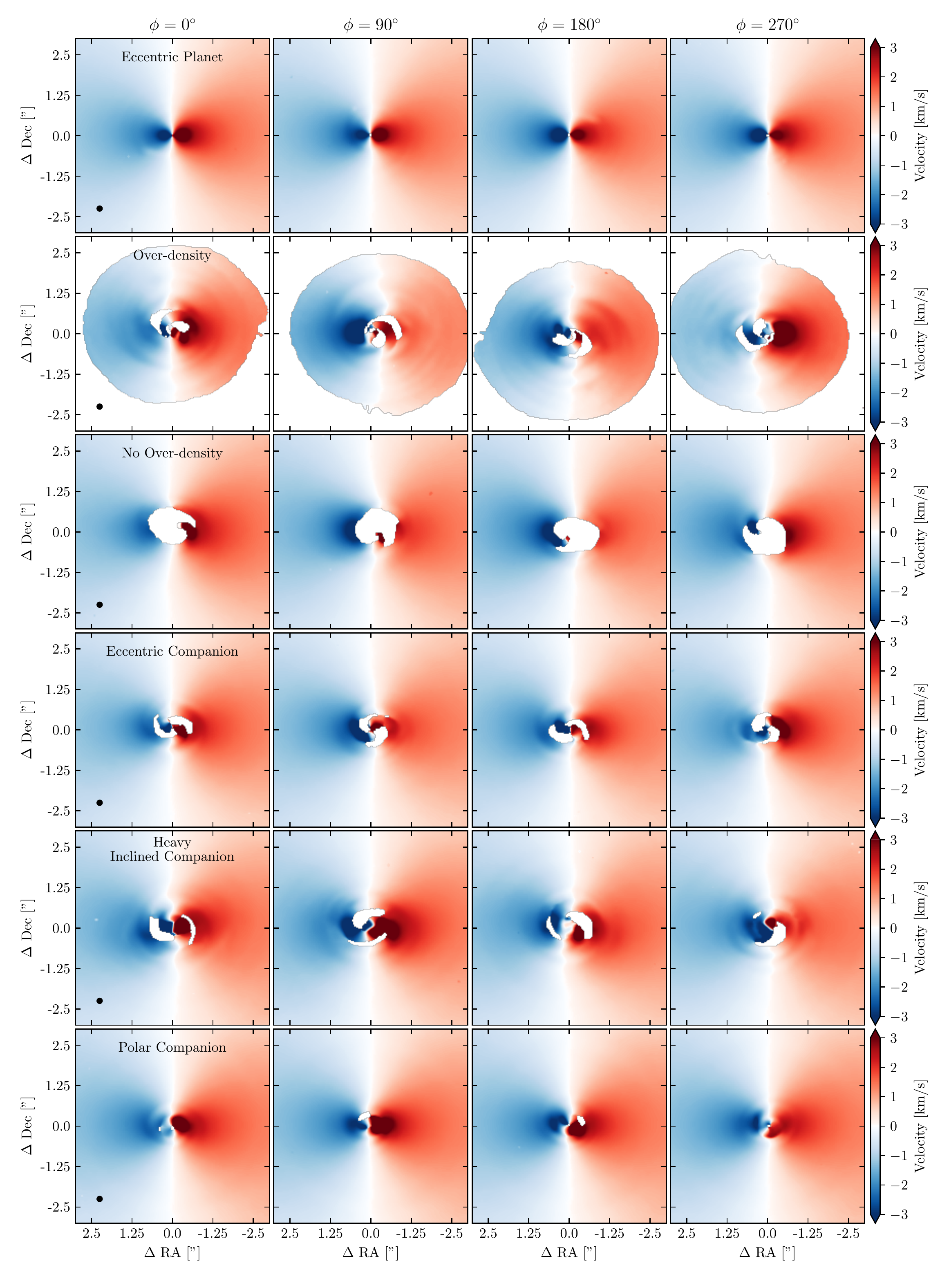}
    \caption{The moment 1 velocity maps for a selection of models across different orientations. In every circumbinary disc model a significant deviation from Keplerian rotation is seen in the central regions of the disc compared to model NC in Figure~\ref{fig:mom1_com}. }
    \label{fig:mom1_rot}
\end{figure*}

We present our velocity maps of most models in Figure~\ref{fig:mom1_com}. In some models (e.g. model no over-density) there is a lack of signal in CO emission which results in the white regions inside the cavity.

Comparing to model NC, significant deviations from Keplerian rotation are see in all circumbinary models, particularly in the regions close to and within the cavity (which has a projected radius of $\sim 1 " $). Inspecting Figures \ref{fig:dens_vel_cop} and \ref{fig:dens_vel_inc} (which have the models at the same azimuthal angle as the velocity maps), the majority of the deviation is likely due to the $\sim \pm 1$ km/s radial velocities inside the cavity. Asymmetries in the maximum velocity on each wing of the velocity map are quite common. For a disc with Keplerian rotation we expect $v_\textrm{max} \approx - v_\textrm{min}$, however in our circumbinary disc models the difference between maximum and minimum velocity can be as great as a factor of 2. In models OD and NOD the differences are most noticeable. In both of these cases the high velocity material in the redshifted (right) wing of the velocity map corresponds with the accretion streams being sent to the apastron of the cavity by the binary in Figure~\ref{fig:dens_vel_cop}. 

We test the robustness of these deviations to viewing angle in Figure~\ref{fig:mom1_rot}. When azimuth angle $\phi = 90^\circ$ and $\phi = 270^\circ$, models OD, NOD, EC, and HIC are orientated such that the eccentricity vector of their circumbinary discs is orientated tangential to the observer-disc line of sight. In these orientations the major deviations in the velocity map arise from the azimuthal velocity component of each disc, and is largely the result of each disc either being modestly (OD and NOD) or slightly (EC and HIC) eccentric. In orientations $\phi = 0^\circ$ and $\phi = 180^\circ$ most of the deviations are due to the radial velocity component.

\section{Kinematic Criteria}

Our models suggest that it is common for the velocity maps of circumbinary discs to display large deviations away from Keplerian rotation. One way to study these deviations is to subtract a best fit Keplerian disc, warped disc, or flared disc model \citep[e.g.][]{teague2018, teague2019, casassus2019, hall2020}. However this then makes the deviations model dependent, and several artefacts can present themselves after model subtraction \citep{teague2018, yen2020}. 
The deviations created by our circumbinary models tend to be so large that subtracting a rotation model is unnecessary to study them. We identify two ways these deviations are manifested in the velocity maps, defined as $V(x,y)$: substantial changes in velocity along the major-axis of the disc (which should follow a roughly $r^{-1/2}$ profile), and differences in the area enclosed in the map by a specific velocity. These form the basis of the two kinematic criteria we define below, and are displayed graphically in Figure~\ref{fig:v_com_fig}.

\subsection{Formulation}\label{sec:kin_crit}

\begin{figure*}
    \centering
    \includegraphics[width=0.7\linewidth]{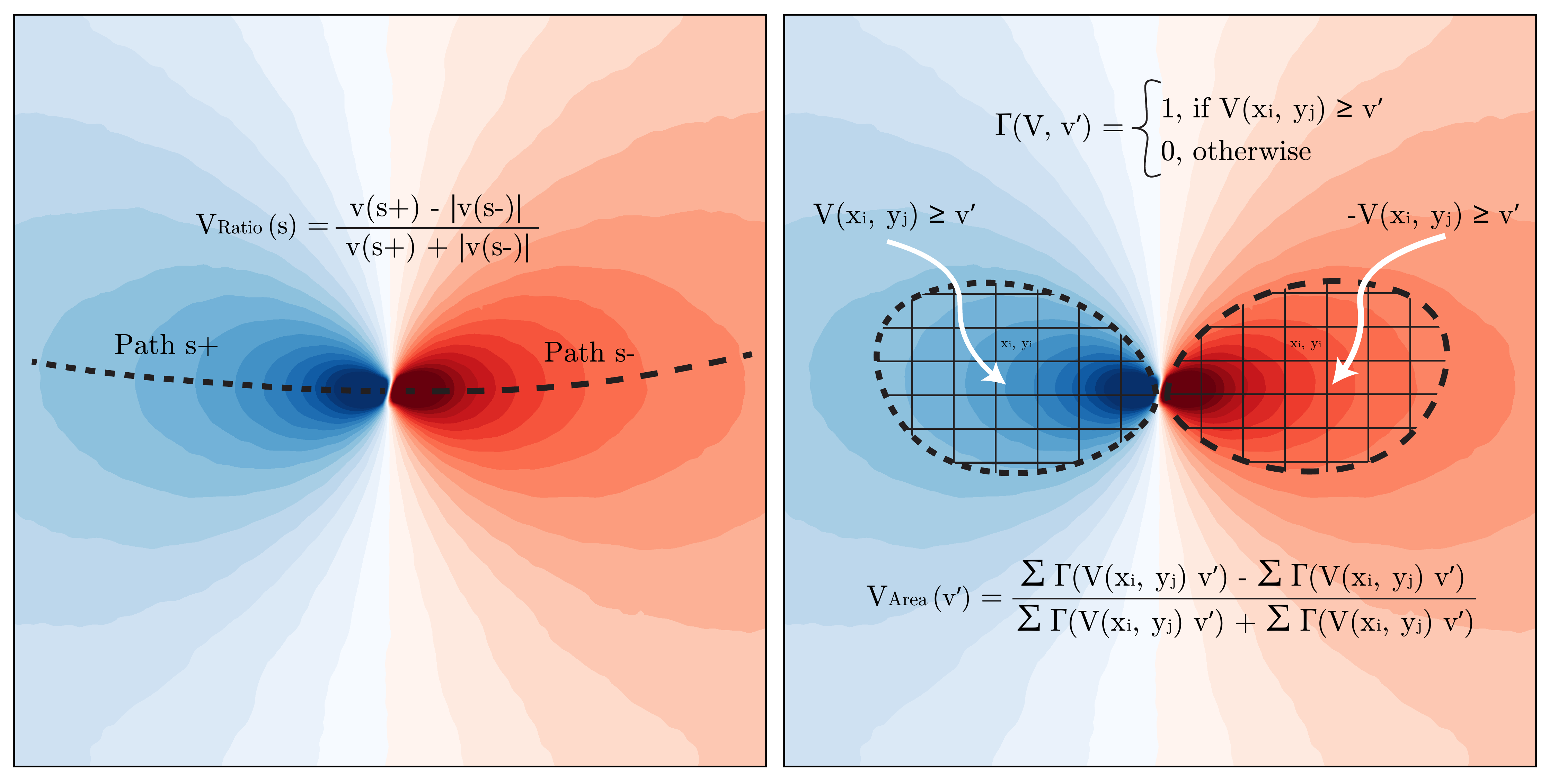}
    \caption{A graphical representation of the quantities $V_\textrm{ratio} (s)$ (left panel) and $V_\textrm{Area}$ (right panel). }
    \label{fig:v_com_fig}
\end{figure*}

Our first criteria is used to quantify significant deviations from axi-symmetric, circular Keplerian rotation as a function of position close to the semi-major axis of the disc
\begin{equation}
    V_{\textrm{Ratio}} (s) = \frac{v_\textrm{max} (s_+) - |v_\textrm{min} (s_-)|}{v_\textrm{max} (s_+) + |v_\textrm{min}(s_-)|},
\end{equation}
where $v_\textrm{max}(s_+)$ and $v_\textrm{min}(s_-)$ are the maximum and minimum velocities in $V(x, y)$ after subtraction of the systemic velocity of the system, and $s_+$ and $s_-$ are paths along the line $s$ in the positive and negative wings of the velocity map, respectively. Note that $v_\textrm{min}$ is a negative quantity so we take the absolute magnitude and by definition $V_{\textrm{Ratio}} (s)$ is nought for an unperturbed disc.

The path $s$ is defined as the radial positions close to the semi-major axis where the absolute velocity is the highest (see Figure~\ref{fig:v_com_fig}). This makes $V_{\textrm{ratio}} (s)$ more general than simply obtaining velocities along the semi-major axis, since disc flaring and warps can shift the maximum and minimum velocity a significant amount away from the semi-major axis \citep[for example, as seen in HD~163296][]{qi2015, isella2018}. The path $s$ is limited by $0 \leq |s| \leq \textrm{min}(\textrm{max}|s_+|, \textrm{max}|s_-|)$ where $s_+$ and $s_-$ are the paths on the blue and redshifted sides of the velocity map, respectively, and the maximum of these quantities is their greatest radial extent from their starting position close to the centre of $V(x,y)$. We also remove any points in $s_+$ and $s_-$ that are within one semi-major axis of the beam to reducing artificially inflated values of $V_\textrm{Ratio}(s)$ due to beam smearing. Thus we compare the velocity along the red and blueshifted sides of the disc. Small values of $s$ corresponds to regions close to the centre of the velocity map, while large values corresponds to the edge of the disc. There is no assumption on the centre of the disc or the binary centre of mass in the path $s$, however a central point can be defined by taking the point in the middle of the last points in $s_+$ and $s_-$. We describe our procedure for obtaining the path $s$ in Appendix \ref{sec:paths}. We verified that a more complicated technique of measuring the velocity difference of every pixel on each side of the velocity wing does not provide a better metric than our method explained above.

Our second criteria is related to the area of emission over a specific velocity threshold. In some of our models we find that although $V_\textrm{Ratio}$ may be small, the area enclosed by a specific absolute velocity can vary. To quantify this type of asymmetry we define the ratio

\begin{equation}\label{eq:va_def}
    V_\textrm{Area} (v) = \frac{n}{n + N_\textrm{beam}} \frac{\sum_{ij} \Gamma (\textrm{V}(x_i, y_j), v) - \sum_{ij} \Gamma (-\textrm{V}(x_i, y_j), v) }{\sum_{ij} \Gamma (\textrm{V}(x_i, y_j), v) + \sum_{ij} \Gamma (-\textrm{V}(x_i, y_j), v)}
\end{equation}
where $n$ is the number of pixels satisfying $|V(x_i, y_i)| \geq v $, $N_\textrm{beam}$ is the number of pixels in the beam, and
\begin{equation}
    \Gamma (\textrm{V}(x_i, y_j), v) = 
    \begin{cases}
        1,& \text{if } \textrm{V}(x_i, y_j) \geq v\\
        0,              & \text{otherwise}.
    \end{cases}
\end{equation}
Note that the denominator of equation \eqref{eq:va_def} is actually just $n$, so $V_\textrm{Area}(v)$ can be simplified as 
\begin{equation}\label{eq:va_def2}
    V_\textrm{Area} (v) = \frac{1}{n + N_\textrm{beam}}\left[ \sum_{ij} \Gamma (\textrm{V}(x_i, y_j), v) - \sum_{ij} \Gamma (-\textrm{V}(x_i, y_j), v)\right].
\end{equation}
The range of the velocity is limited to $0 \leq v \leq \textrm{max}(v_\textrm{max} - \Delta v, -v_\textrm{min} + \Delta v)$ and sample $v$ in steps $\Delta v$ equal to the spectral resolution of the observation. The addition of $\Delta v$ in the reduction of the range of $v$ ensures we do not sample the spectrally unresolved portions of the very inner disc, which can cause spurious values of $V_\textrm{Area}$. We weight the area ratio by $n/(n + N_\textrm{beam}$) so that unresolved portions of $V(x,y)$ do not dominate the quantity.

For a Keplerian disc $V_\textrm{Area}$ should be close to nought for any value of $v$.
This criteria also measures velocity asymmetries in the disc, however as opposed to $V_{\textrm{Ratio}} (s) $, it measures these asymmetries in terms of their spatial distribution.

\subsection{Weighting and Variance}
Protoplanetary discs come in wide range of sizes, masses, and environmental conditions. For our criteria to be robust more against these varying conditions, we weight our functions when computing the variance in them. As our results in Section \ref{sec:hydro} show, most perturbations in the kinematics are spatially coincident with the cavity. Hence when computing our criteria, we should down-weight the regions of the disc we do not expect to contain perturbations that arise due to the binary, and up-weight those that do. By doing so, we down-weight the regions over which the variance in outer edges of the disc which can be perturbed by outer companions, flybys, or infalling material. 

We find the weighted variance of the points in $V_\textrm{Ratio}$ by defining
\begin{equation}
    \sigma_{\textrm{Ratio}}^2 = \frac{1}{N_s}\sum_i w(s_i) \  V_{\textrm{Ratio}}(s_i)^2,
\end{equation}
where $N_s$ is the number of points in $s$, and $w(s_i)$ is a weighting function. For our purposes, we choose a cosine weighting function with
\begin{equation}
    w_B(s_i)  = 
    \begin{cases}
        1,& \text{if } s_i \leq 2\times r_\textrm{cavity} \\
        \cos ^2\left( \frac{\pi}{2} \frac{s_i - 2 r_\textrm{cavity}}{3 r_\textrm{cavity}} \right) & \text{if } 2\times r_\textrm{cavity} < s_i \leq 3\times r_\textrm{cavity} \\ 
        0,              & \text{if } s_i > 3\times r_\textrm{cavity},
    \end{cases}
\end{equation}
where $r_\textrm{cavity}$ is the radius of the cavity, which we choose as the peak of the gas surface density.\footnote{Often it is not possible to measure the peak of the gas surface density in observations. In place of this, the peak of the continuum offers a suitable replacement since dust grains concentrate at gas pressure maxima which coincides with the peak gas surface density \citep[e.g.][]{sierra2019}.} A limit of $3\times r_\textrm{cavity}$ is chosen based on the results of Section \ref{sec:hydro}. If line emission is not detected up to this radius, we truncate the weighting function by $r_\textrm{eff}$, which is the effective radius of the disc. We obtain $r_\textrm{eff}$ using $f_\nu (r_\textrm{eff})=x F_\nu$, where $x=0.9$, $f_\nu$ is the cumulative intensity profile, and $F_\nu = f_\nu (\infty)$ \citep[e.g. see][]{andrews2018b,Long_2019,long2022}. If the disc does not contain a cavity, then the weighting function $w_B(s)=0$, and thus the variance is also nought. This is desired since, as stated in Section \ref{sec:cavity}, we expect all circumbinary discs to contain a cavity. Later in this work we also use the flat weighting function
\begin{equation}
    w_F(s_i)  = 
    \begin{cases}
        1,& \text{if } s_i \leq r_\textrm{eff},\\
        0,              & \text{otherwise},
    \end{cases}
\end{equation}
for demonstrating the effect of the weighting function $w_B(s_i)$. 

We find the weighted variance in $V_\textrm{Area}$ by defining
\begin{equation}
    \sigma_\textrm{Area}^2 = \frac{1}{n_v}\sum_i V_\textrm{Area}(v_i)^2
\end{equation}
where $n_v$ is the number of sampled velocities. We slightly adjust the way $V_\textrm{Area}(v_i)$ is computed to apply our weighting. We replace $\sum_{ij} \Gamma (\textrm{V}(x_i, y_j), v) - \sum_{ij} \Gamma (-\textrm{V}(x_i, y_j), v) $ in equation \eqref{eq:va_def2} with $\sum_{ij} \Gamma_w (\textrm{V}(x_i, y_j), v) - \sum_{ij} \Gamma_w (-\textrm{V}(x_i, y_j), v) $, where $\Gamma_w (\textrm{V}(x_i, y_j), v)$

\begin{equation}
    \Gamma_w (\textrm{V}(x_i, y_j), v) = 
    \begin{cases}
        w(x_i, y_j),& \text{if } \textrm{V}(x_i, y_j) \geq v\\
        0,              & \text{otherwise}.
    \end{cases}
\end{equation}
where

\begin{equation}
    w(x_i, y_i)  = 
    \begin{cases}
        1,& \text{if } r_i \leq 2\times r_\textrm{cavity} \\
        \cos ^2\left( \frac{\pi}{2} \frac{r_i - 2r_\textrm{cavity}}{3r_\textrm{cavity}} \right) & \text{if } 2\times r_\textrm{cavity} < r_i \leq 3\times r_\textrm{cavity} \\ 
        0,              & \text{if } r_i > 3\times r_\textrm{cavity}.
    \end{cases}
\end{equation}
and $r_i = \sqrt{x_i^2 + y_i^2}$, where $x_i$ and $y_i$ are the deprojected disc coordinates. In this way asymmetries inside and close to the cavity are weighted more than asymmetries in the outer disc which do not arise due to the inner binary (see Section \ref{sec:osp}). Similar to our weighting procedure on $\sigma_{\textrm{Ratio}}^2$, we also compute the variance measurement $\sigma_\textrm{Area}^2$ with both our cosine weighting function and uniform weighting function. To easily distinguish between which weighting function as been used, we write $w_B \sigma_\textrm{Ratio}^2$ and $w_B \sigma_\textrm{Area}^2$ when the binary weighting function has been used, and simply $\sigma_\textrm{Ratio}^2$ and $\sigma_\textrm{Area}^2$ when a flat weighting function has been used.

\subsection{Testing the Kinematic Criteria}

\begin{figure}
    \centering
    \includegraphics[width=\linewidth]{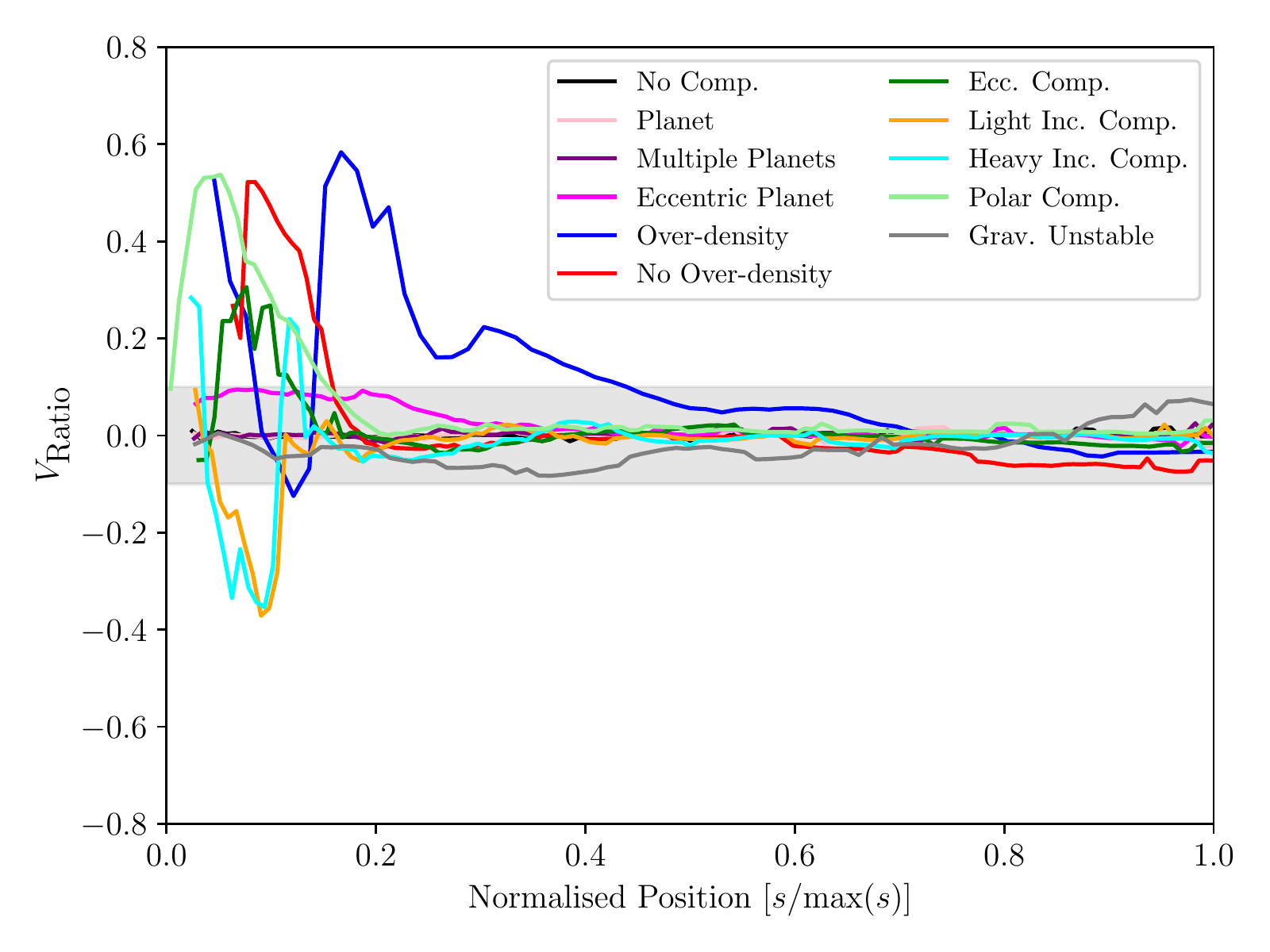}
    \caption{The velocity ratio $V_\textrm{Ratio}$ plotted as a function of normalised position, $s$/max($s$), for all of our disc models with $F_\textrm{noise} = 2.5 $ mJy} using a single viewing angle. Our no companion, planet model, and multiple planet models are indicated by the black, pink, and purple lines, respectively. These three models show much lower values for $V_\textrm{Ratio}$ compared with the circumbinary models, particularly when we approach the cavity, which is located roughly at $s/\textrm{max}(s)\sim 0.25$. These trends are essentially identical for different angles.
    \label{fig:vpv}
\end{figure}

For this section we show the results for the $F_\textrm{noise} = 2.5 $ mJy noise level since the resulting signal-to-noise level is readily achievable with ALMA. We plot $V_\textrm{Ratio}(s)$ in Figure~\ref{fig:vpv} for our disc models, but only include the azimuth angle $\phi = 0^{\circ}$ to avoid a cluttered appearance. We have normalised $s$ on the $x$-axis of the Figure. The position at nought on the $x$-axis is close to the centre of the velocity map. Since signal is lacking in this area for many of our models, the path $s$ can end on the cavity edge, most prominent in model NOD, where $V_\textrm{Ratio}(s)$ is no longer defined inside of $s\lesssim 0.1$.

In our circumbinary disc models the cavity radius is $\lesssim 100$ au, which is roughly one quarter of the total disc radii. In Figure~\ref{fig:vpv} this is where $V_\textrm{Ratio}$ starts deviate substantially from nought. The main exception is model OD (blue line), which has a smaller outer radii compared with the other models. 

In a planet-hosting discs, perturbations in the velocity field are expected to be mostly sub-sonic \citep[e.g. see][]{bollati2021, calcino2022}. Although planet masses approaching and exceeding the thermal mass produce super-sonic perturbations, it is reasonable to expect that these perturbations increase in amplitude with increasing companion mass \citep[e.g. see][]{dong2015b}. Thus we should also expect that $V_\textrm{Ratio}$ will increase with increasing companion mass. 
Using subsonic perturbations as an assumption, the maximum $V_\textrm{ratio}(s)$ at a particular value of $s$ for a planet hosting disc should be
\begin{equation}
    V_\textrm{Ratio, plan} \sim \frac{(v_K + c_s) - (v_K - c_s)}{(v_K + c_s) + (v_K - c_s)} = \frac{c_s}{v_K} = \frac{H}{r},
\end{equation}
where $c_s$ is the sound speed and $H$ is the scale height. For a typical protoplanetary disc $H/r \sim 0.1$, however in Figure~\ref{fig:vpv} we can see that all of our circumbinary disc models show a $V_\textrm{Ratio}$ a factor of few higher than this (shaded region) close to and within the cavity.

As expected, our no companion (NC), planet (P), and multiple planet (MP) models have a $V_\textrm{Ratio}$ much lower than the circumbinary models, and much lower than the theoretical maximum.
Our eccentric planet (EP) model shows inflated values of $V_\textrm{Ratio}$ compared with the other planet models owing to the eccentric gas motion induced by the planet, as discussed in Section \ref{sec:cop_mod}.

\begin{figure}
    \centering
    \includegraphics[width=\linewidth]{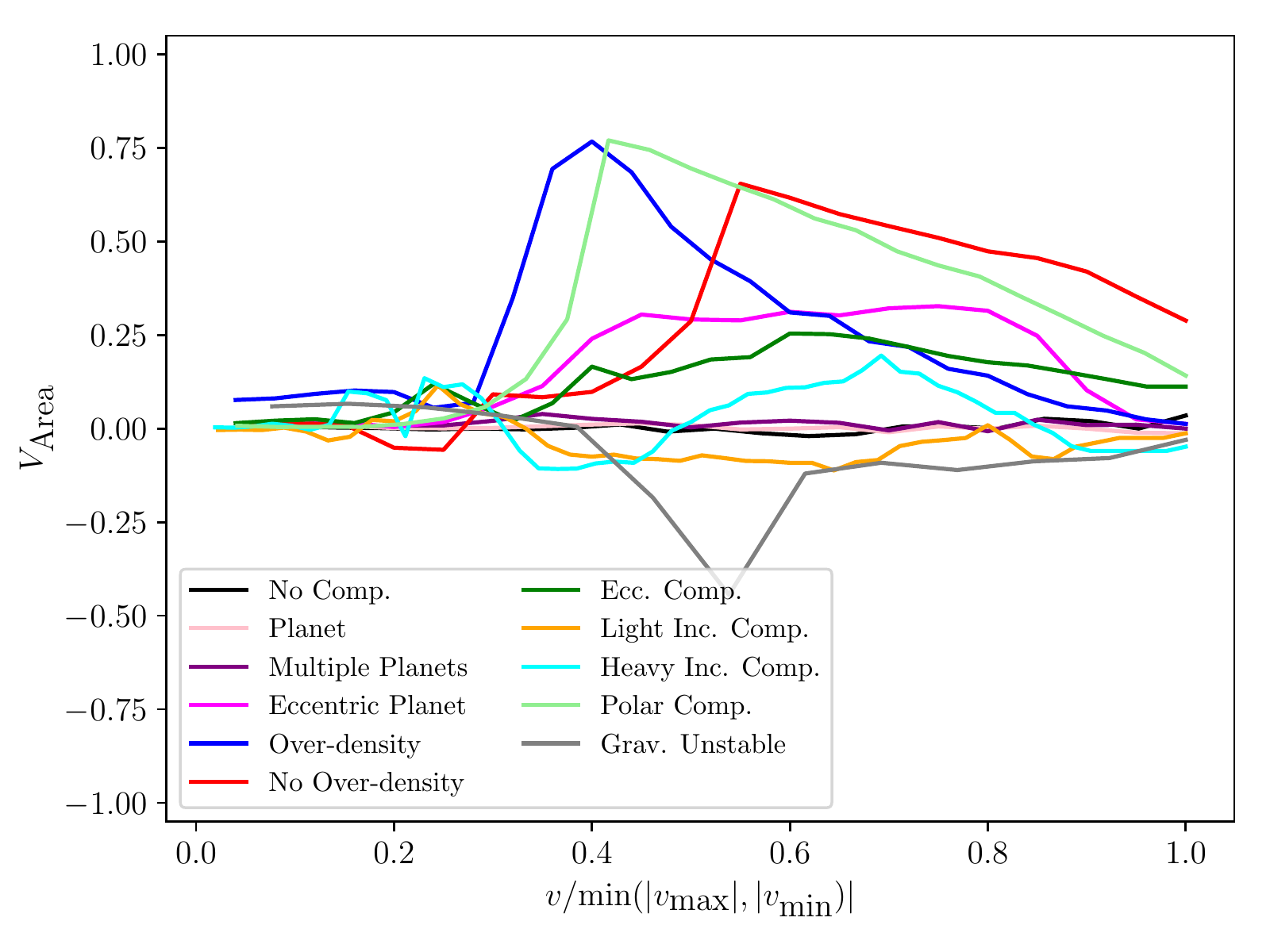}
    \caption{The parameter $V_\textrm{Area}$ plotted for each disc model with $F_\textrm{noise} = 2.5 $ mJy, including the no companion and planet models. We can see that $V_\textrm{Area}$ is much lower over the majority of the disc compared with the circumbinary disc models. Note that small values in velocity corresponds to large spatial scales.}
    \label{fig:va}
\end{figure}

In Figure~\ref{fig:va} we display our quantity $V_\textrm{Area}$ for a single azimuthal angle of all the models in Table~\ref{tab:ic}. As expected, $V_\textrm{Area}$ stays close to nought in models NC, P, and MP for all values of $v$, with slight fluctuations owing to the pixel resolution of the velocity map. For all of our circumbinary models presented $V_\textrm{Area}$ deviates substantially from zero. 

\begin{figure*}
    \centering
    \includegraphics[width=0.45\linewidth]{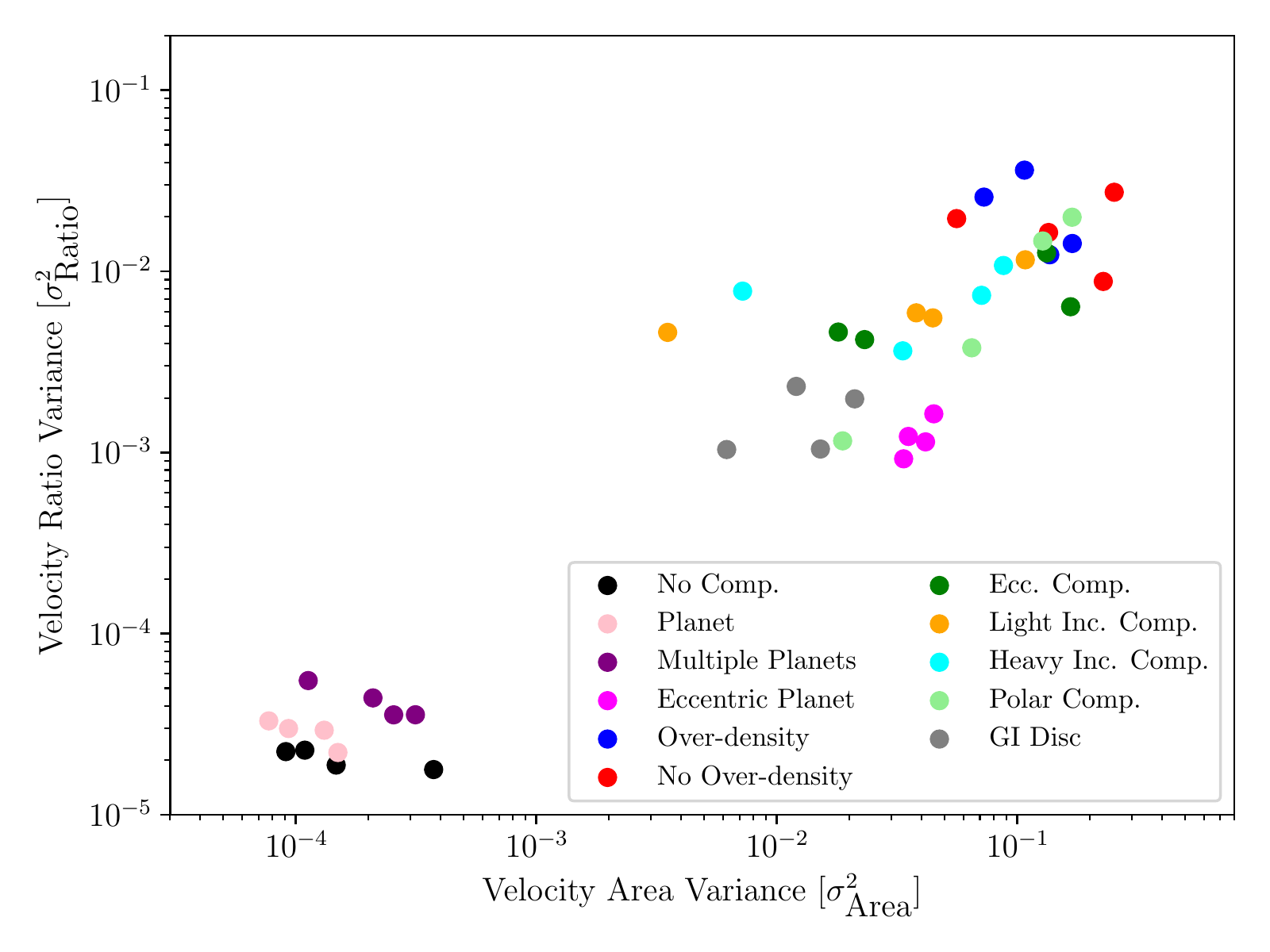}
    \includegraphics[width=0.45\linewidth]{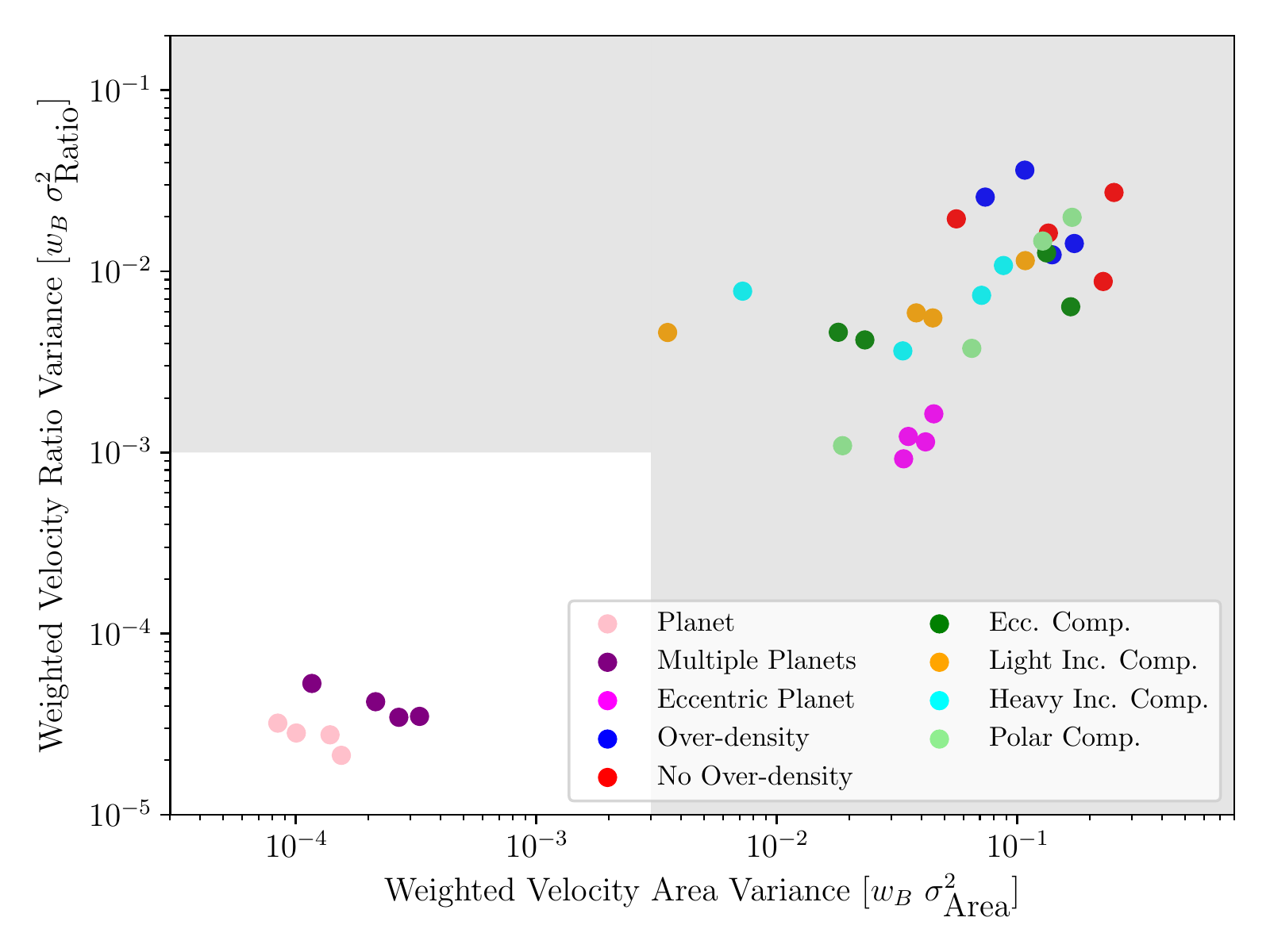}
  \caption{The variance in $V_\textrm{Ratio}$, $\sigma_\textrm{Ratio}$, versus the variance in $V_\textrm{Area}$, $\sigma_\textrm{Area}$ (left panel), and the weighted variance in $V_\textrm{Ratio}$, $w_B \sigma_\textrm{Ratio}$, versus the weighted variance in $V_\textrm{Area}$, $w_B \sigma_\textrm{Area}$ (right panel), for all of our disc models and $F_\textrm{noise} = 2.5 $ mJy. The four points for each model indicate each of the four viewing angles of our models tested. We see a correlation between the variance measurements with both weighting methods. The shaded region is derived empirically with $w_B\ \sigma_\textrm{Ratio}^2 > 0.003$ $w_B\ \sigma_\textrm{Area}^2 > 0.003$. All our circumbinary models satisfy the criteria in $w_B\ \sigma_\textrm{Area}^2$, while the models with substantial eccentricity (No Comp. and Over-density) mostly satisfy the criteria in $w_B\ \sigma_\textrm{Ratio}^2$. }
  \label{fig:sva_svpv}
\end{figure*}

We plot the values of $\sigma_\textrm{Ratio}^2$ versus $\sigma_\textrm{Area}^2$ for each of our models in the left panel of Figure~\ref{fig:sva_svpv}, along with their weighted counterparts in the right panel. Models NC, P, and MP all display low values of the variance quantities $\sigma_\textrm{Ratio}^2$ and $\sigma_\textrm{Area}^2$, while the other models show an elevated variance in either one or both of these quantities. This is expected since our analysis in Section \ref{sec:hydro} showed that these models display much larger perturbations in their velocity fields.

The variance $\sigma_\textrm{Area}^2$ in particular appears a robust indicator of perturbations as it is more than one order of magnitude larger for our binary discs than the planet hosting and no companion disc models. The exception to this is that our eccentric planet and gravitationally unstable model produces a large value of $\sigma_\textrm{Area}^2$ compared with the other models. 

Between the flat weighting and binary weighting functions, there appears to not be much difference in the variance measurements. The gravitationally unstable and the no companion models have weighted variances equal to zero since they do not contain cavities. However we can learn two things by comparing the affect each weighting function has. Firstly, since the planet hosting and no companion models have similar values for $\sigma_\textrm{Ratio}^2$ and $\sigma_\textrm{Area}^2$, this indicates that the planets do not have a significant affect on these quantities. Secondly, the weighted and unweighted variances are almost identical for the circumbinary and eccentric planet models, signalling that the perturbations causing the elevated variance measurements is originating from the cavity. This naturally raises the question for whether the weighting functions are necessary at all, however we argue they are since our simulations are quite idealistic. We assume the discs are evolving in isolation and do not consider any disc instabilities which would cause fluctuations in the velocity field, both of which will increase our variance measurements. Since a cavity is a theoretically expected \citep{artymowicz1994} and observationally supported \citep{casassus2013, dutrey2014} outcome of binary-disc interactions, the inclusion of a weighting function specifically targeting perturbations in and near the cavity is justified.

With this justification in mind, the shaded region in the right panel of Figure~\ref{fig:sva_svpv} encapsulates the area of the variance parameter space where we are more likely to see binary systems. The area is derived empirically with $w_B \sigma_\textrm{Ratio}^2 > 0.003$ and $w_B\sigma_\textrm{Area}^2 > 0.003$. It is robust to different noise levels, disc inclinations, and synthesised beam provided the cavity region is resolved by $\sim$5 beams (Appendix \ref{sec:robust}). Finally, there is a positive correlation between the quantities $\sigma_\textrm{Ratio}^2$ and $\sigma_\textrm{Area}^2$ which appears stronger with an increase in the disc inclination (Appendix \ref{sec:robust}).

\section{Discussion}\label{sec:disc}

The search for kinematic perturbations in protoplanetary discs is ramping up significantly, with several recently accepted proposals and one large program dedicated to this task.
While there now exists a theoretical framework for modelling planet induced kinks \citep{bollati2021, calcino2022}, and some work on the expected signatures of gravitationally unstable discs \citep{hall2020, terry2021}, few studies have explored kinematic perturbations in circumbinary discs.

\subsection{Kinematic and Morphological Criteria for Circumbinary Discs} \label{sec:criteria}
We now describe the three morphological features which together provide strong support for binarity.
In Section \ref{sec:co_em} we outlined several morphological features seen in integrated CO isotopologue observations. The presence of a cavity in $^{13}$CO and C$^{18}$O integrated emission is unanimous among our models, while a cavity may or may not be present in $^{12}$CO depending on the disc and companion properties. Thus our first indicator of a circumbinary disc is the presence of a cavity in either $^{13}$CO or C$^{18}$O integrated emission. Since a cavity should be present in essentially all circumbinary discs, our other morphological and kinematic criteria are determined in the context of a cavity hosting disc. 

Our second indicator from integrated CO isotopologue observations is the presence of spiral-like features in proximity to a cavity. Although not studied in the present work, spiral-like features could also be detected in scattered light observations, as they have been in the known circumbinary(-triple) discs HD~142527 and GG~Tau~A \citep{fukagawa2006, casassus2012, canovas2013, avenhaus2014, avenhaus2017, Keppler2020}. 

Our third indicator is non-localised velocity kinks in proximity of the cavity in the channel maps. We demonstrated in Section \ref{sec:chans} that kinks are seen on the edge of the disc cavity robustly in most models and viewing angles. We defined ``in proximity to the cavity'' to mean within $r \leq 1.5 \times r_\textrm{cavity}$. 

Our two final criteria are obtained from velocity maps of the disc. These are the kinematic criteria $w_B\  \sigma_{\textrm{Area}}^2$ and $w_B\ \sigma_\textrm{Ratio}^2$. We showed in Section \ref{sec:vm} that these criteria together, when above certain values, are indicators of binarity. Although more work should be done to differentiate between models, our present work allows us to summarise the following criteria that indicate binarity:

\begin{enumerate}
    \item Gas depleted central cavity
    \item Spiral arms in proximity to a cavity
    \item Non-localised wiggles in the channel maps in proximity to a cavity
    \item $w_B\ \sigma_\textrm{Ratio}^2 > 0.001$
    \item $w_B\ \sigma_\textrm{Area}^2 > 0.003$
\end{enumerate}
Table \ref{tab:com} summarises which criteria are met in the models we have tested. Our kinematic criteria essentially measure asymmetries in the velocity map. There are many ways that these asymmetries could arise, as discussed in Section~\ref{sec:osp}. Thus on their own, these criteria do not indicate binarity. However, in conjunction with the morphological criteria, our work indicates that a strong case for binarity can be made. 

From our short analysis in Appendix \ref{sec:robust}, either $^{12}$CO or $^{13}$CO emission can be used to measure the kinematic criteria provided that the peak SNR in an individual channel is greater than 50. The cavity region should be resolved with at least 5 beams in order to properly attain the kinematic criteria, however the kinematic criteria can still be met with lower resolution than this. The inclination of the disc is also important to consider, with very high and low inclinations presenting challenges. Low inclinations of $i \lesssim 5^{\circ}$ mean the projected azimuthal and radial perturbations are small, and hence $w_B\  \sigma_{\textrm{Area}}^2$ and $w_B\ \sigma_\textrm{Ratio}^2$ may not signify a binary even when one is present. For the higher inclinations, the main issue is that the outer disc surface can start to obscure the cavity region where most of the perturbations are expected. The CO emitting layer of the discs in this paper is somewhat lower than what is typically observed, so our criteria seem robust on our models with $i \lesssim 85^{\circ}$. However in observations this threshold might be somewhat lower, and whether the disc surface obscures the inner disc should be determined from integrated and peak intensity maps.

\begin{table*}
    \centering
    \begin{tabular}{l|c|c|c|c|c|}
    \hline
        Name      &  Cavity & Non-localised kink & CO Spirals & $w_B \ \sigma_\textrm{Ratio}^2 > 0.001$ & $w_B\ \sigma_\textrm{Area}^2 > 0.003 $ \\
        \hline
        No Companion (NC)      & \xmark\xmark\xmark\xmark  & \xmark\xmark\xmark\xmark & \xmark\xmark\xmark\xmark &  \xmark\xmark\xmark\xmark & \xmark\xmark\xmark\xmark  \\
        Planet (P)             & \xmark\xmark\xmark\xmark & \xmark\xmark\xmark\xmark  & \xmark\xmark\xmark\xmark &  \xmark\xmark\xmark\xmark & \xmark\xmark\xmark\xmark \\
        Multiple Planets (MP)             & \xmark\xmark\xmark\xmark & \xmark\xmark\xmark\xmark  & \xmark\xmark\xmark\xmark &  \xmark\xmark\xmark\xmark & \xmark\xmark\xmark\xmark \\
        Gravitationally Unstable (GI)             & \xmark\xmark\xmark\xmark & \xmark\xmark\xmark\xmark  & \xmark\xmark\xmark\xmark &  \xmark\xmark\xmark\xmark & \xmark\xmark\xmark\xmark \\
        Eccentric Planet (EP)  & \cmark\cmark\cmark\cmark & \xmark\xmark\xmark\xmark  & \cmark\cmark\cmark\cmark &  \cmark\cmark\cmark\xmark & \cmark\cmark\cmark\cmark \\
        No Over-density (NOD)  & \cmark\cmark\cmark\cmark & \xmark\cmark\cmark\xmark & \cmark\cmark\cmark\cmark &  \cmark\cmark\cmark\cmark & \cmark\cmark\cmark\cmark\\
        Over-density (OD)      & \cmark\cmark\cmark\cmark & \cmark\cmark\cmark\cmark  & \cmark\cmark\cmark\cmark &  \cmark\cmark\cmark\cmark & \cmark\cmark\cmark\cmark \\
        Eccentric Companion (EC)  & \cmark\cmark\cmark\cmark  & \cmark\cmark\cmark\cmark  & \cmark\cmark\cmark\cmark &  \cmark\cmark\cmark\cmark & \cmark\cmark\cmark\cmark \\
    Light Inclined Companion (LIC)  & \cmark\cmark\cmark\cmark  & \cmark\cmark\cmark\cmark  & \cmark\cmark\cmark\cmark &  \cmark\cmark\xmark\cmark & \cmark\cmark\cmark\cmark \\
        Heavy Inclined Companion (HIC) & \cmark\cmark\cmark\cmark & \cmark\cmark\cmark\cmark  & \cmark\cmark\cmark\cmark & \cmark\cmark\cmark\cmark & \cmark\cmark\cmark\cmark\\
        Polar Companion (PC)     & \cmark\cmark\cmark\cmark & \cmark\cmark\cmark\cmark  & \cmark\cmark\cmark\cmark & \cmark\cmark\cmark\cmark & \cmark\cmark\cmark\cmark \\
        \hline
    \end{tabular}
    \caption{A summary of which morphological and kinematic criteria are met in our models for the $F_\textrm{noise} = 2.5 $ mJy noise level. The four marks in each column are for each viewing angle in order from $\phi = 0^{\circ}$ to $\phi = 270^{\circ}$. Although spirals can be observed in planet hosting discs \citep[e.g. see][]{mentiplay2018}, they are not clearly visible in our planet model. Note that our non-localised kink criteria specifically refers to the immediate area in proximity of the cavity}.
    \label{tab:com}
\end{table*}

\subsection{Planet Signatures versus Circumbinary Signatures}
Several of our criteria that can be met in planet hosting discs. The appearance of gas and dust depleted cavities is both theoretically and observationally supported in planet hosting discs \citep{zhu2011, pinilla2012, keppler2018, long2018}. As dust grains collect into a ring inside a gas pressure maximum, their morphology can look indistinguishable from the dust ring around a circumbinary disc \citep[e.g. the case of GG~Tau~A,][]{dutrey2014}. Thus dust cavities are not exclusive to circumbinary discs. However such a feature is not present in all planet hosting discs \citep[for example, HD~163296][]{qi2015, isella2018}. Gas cavities, on the other hand, may be a more reliable indicator of a stellar companion inside of a cavity since a more massive companion can more efficiently clear material. 
Although this morphology can be observed in planet hosting discs, we can reasonably expect that a dust and at least a partially gas depleted cavity should exist in essentially \emph{all} circumbinary discs. 

Our second criteria is subject to some interpretation and it is possible that planet-hosting discs show non-localised kinks. For example, the spiral wake from an embedded planet in the disc around HD~163296 was reported by \cite{calcino2022}. These perturbations are small compared with those seen in circumbinary discs \cite{bollati2021}.
This is the reason we specify non-localised kinks in proximity to the cavity. Although a massive planet inside a cavity could produce non-localised kinks in proximity with the cavity, the large difference in mass between a planet and stellar companion will result in differences in the velocity kinks produced. A method used to derive the kink amplitude in circumbinary discs would be useful to compare with the kink amplitude generated by planets.

Spiral arms are also another feature which are expected to be seen in planet-hosting discs \citep{goldreich1979, goldreich1980, ogilvie2002, rafikov2002}. The brightness of planet-induced spiral arms in scattered light has been well studied using hydrodynamical simulations and radiative transfer \citep{dong2015b, zhu2015, fung2015, dong2016}. 
Planetary masses from as low as $\sim$ 0.5 M$_\textrm{J}$ could be enough to induce spiral arms that are observable in scattered light by current generation telescopes \citep{fung2015}. Thus, spiral arms, at least in scattered light, are also not a robust indicator of a circumbinary disc. However this may not be true in CO integrated emission and peak intensity. Several observational papers have found spirals in CO emission/peak intensity and have attributed them to planets \citep{Tang2017, boehler2018, phuong2020}. To our knowledge there are no works in the literature exploring the appearance of planet-induced spirals in CO isotopologues. Since the amplitude of companion-induced spirals correlates with the companion mass \citep{fung2015}, the appearance of planetary-induced versus binary induced spirals should differ, with the latter being more visible in CO isotopologue observations than the former \citep[e.g. see][]{mentiplay2018, poblete2020}. The only observational confirmation of this are the tentative and faint spirals in HD~163296 noted in the channels by \cite{calcino2022}, but not clearly seen in a model subtract peak intensity map \citep{teague2021}. Further investigation is needed to test this hypothesis. 

Our kinematic criteria are more robust to false positives however there are scenarios where we could see large values of $w_B\ \sigma_{\textrm{Area}}^2$ and $w_B\ \sigma_{\textrm{Ratio}}^2$. We have tested two such scenario in the present work in our eccentric planet and gravitionally unstable simulations. We found that the asymmetric flows introduced by the eccentric planet increase $\sigma_{\textrm{Area}}^2$ and $\sigma_{\textrm{Ratio}}^2$. Although eccentric planets have been proposed to explain the observed morphology of several discs in the literature \citep[e.g.][]{muley2019, calcino2020}, they likely do not make up a significant portion of the massive planet population in protoplanetary discs, whereas binary stars are common in the Universe. Additional simulations covering mass and orbital eccentricity are required to further explore their effects on our kinematic criteria.

\subsection{GI Wiggles versus Circumbinary Wiggles}

\cite{hall2020} showed that gravitationally unstable discs can produce significant deviations from Keplerian velocity. These perturbations can be detected in the iso-velocity curves (GI wiggles), or by subtracting the Keplerian rotation field from the velocity map of the disc. Further exploration of numerical simulations of gravitationally unstable discs by \cite{terry2021} found that the disc mass correlates with the wiggle amplitude in the channel maps. We found in the left panel of Figure~\ref{fig:sva_svpv} that our gravitationally unstable model produced relatively large values of $\sigma^2_\textrm{Ratio}$ and $\sigma^2_\textrm{Area}$ compared with the low perturbation models (i.e. NC, P, and MP). Although both gravitationally unstable discs and circumbinary discs display kinematic perturbations, there is a clear morphological difference between these discs in both the gas and dust distribution. Gravitationally unstable discs produce spiral arms that orbit at the Keplerian frequency and hence efficiently trap $\sim$mm sized dust grains \citep{rice2004, hall2020} which is observable with ALMA \citep{dipierro2014}. Thus GI discs should show evidence of dust trapping in spiral arms at mm wavelengths, while circumbinary discs are typically characterised by a cavity at mm wavelengths. Since binary induced spirals are not orbiting at the local Keplerian frequency outside the cavity, they should not trap dust. This allows for our kinematic criteria $w_B \sigma^2_\textrm{Ratio}$ and $w_B \sigma^2_\textrm{Ratio}$ to differentiate from these two classes, as seen in the right panel of Figure~\ref{fig:sva_svpv}.

It is plausible that very young systems can be both gravitationally unstable and contain a binary. One particular system where this might be the case is [BHB2007] 11, where $\sim$mm dust grains trace spiral arms and accretions streams around two young stars \citep{alves2019}. However for the more evolved Class II discs that display a central cavity, very few show dust associated with spiral arms \citep{vandermarel2021}, and hence GI is likely not significantly affecting these discs.

\subsection{Outside Perturbations and Other Applications}\label{sec:osp}

Perturbations arising due to phenomenon not originating from the host disc can occur, and may be responsible for some of the morphological features we have discussed. For example, the outer disc can be strongly perturbed by stellar flybys \citep{cuello2019, cuello2020, smallwood2023}, external companions \citep[e.g. as in HD~100453 and GG~Tau~A][]{white1999, benisty2017, gonzalez2020}, and post formation inflows/cloudlet capture \citep{dullemond2019, Kuffmeier2020, huang2020}. 

These outside influences will contribute to the kinematic and morphological criteria we derived in predictable ways that are distinguishable from a circumbinary disc. For example, an inflow or stellar flyby will affect $V_\textrm{Ratio} (s)$ and $V_\textrm{Area} (v)$ on the large spatial and low velocity scales. Hence in Figure~\ref{fig:vpv} we expect to see an increase in $V_\textrm{Ratio} (s)$ for large $s$, while in Figure~\ref{fig:va} we expect to see an increase in $V_\textrm{Area} (v)$ for small $v$. The same effect is also expected for bound gravitational bodies such as stellar and planetary companions. Our kinematic criteria could also prove useful in diagnosing protoplanetary discs influenced by these effects. 

Since our kinematic criteria are sensitive to any perturbations, care must be taken when interpreting their values in discs where there are clear and obvious outside perturbations that could also be spatially co-located with influences from an inner binary. Two examples of this are AB Aurigae and HD 100546, which were both proposed to be binary systems by \cite{poblete2020} and \cite{norfolk2021}, respectively. There is clear evidence of infalling material interacting with the disc around both systems \cite{dullemond2019, Kuffmeier2020}. Although this complicates the application of our kinematic criteria, their magnitude should correlate with the strength of any induced perturbation in the disc. For example, the degree to which infalling material is perturbing a disc. More massive and faster falling material will induce stronger perturbations in the outer disc which will correlate with $\sigma^2_\textrm{Ratio}$ and $\sigma^2_\textrm{Area}$. The same may also be true of gravitationally unstable discs, where more unstable discs display larger perturbations \citep{terry2021} and hence larger values of our kinematic criteria. This is also another justification for including a weighting function focused on the perturbations in and around a central cavity.

\subsection{Model Caveats}\label{sec:models}

Although not a topic of the present work, changes in the disc parameters such as scale height and viscosity could have some implications for the conclusions we draw. For example, changes in the disc viscosity can result in large changes in the circumbinary disc morphology \citep{rabago2023}. Substantial changes in the disc morphology are also seen in planet-hosting disc simulations \citep[e.g.][]{ataiee13a, zhang2018} where a lowering of disc aspect ratio and disc viscosity can result in an eccentric disc around Jupiter mass planets. \cite{zhang2018} produce a suite of simulations covering differing values of planet mass, disc scale height, and viscosity. They found that the velocity perturbations around the gap carved by the planet do depend on viscosity, with a lower viscosity producing larger amplitude perturbations that do not damp as quickly as in the higher viscosity simulations. However it has been shown that the amplitude of the velocity kink induced by planetary mass objects is not sensitive to the disc viscosity \citep{rabago2021}, rather the kink amplitude is dependent on the planet thermal mass which depends on the disc scale height \citep{bollati2021}. Since the gap structure and resulting velocity field depends on the planet mass, disc viscosity, and scale height \citep{fung2014}, there could be a degeneracy in our kinematic criteria between the companion mass and the disc properties. The uncertainty in the disc properties could blur the boundary between the planet-hosting and circumbinary discs.

However, we have reason to believe that even with introduced uncertainty in the disc properties, the correlation between companion mass and elevated values in our kinematic criteria will still hold.
The reason is simply that larger mass bodies will produce larger perturbations in the disc, regardless of what disc profile is chosen. More massive companions lead to more shocking in the disc, more depleted cavities, and more perturbations in the disc overall. In \cite{zhang2018}, the velocity perturbations around the planet induced gap only change by a factor of a few (provided the eccentricity in the disc is not excited) across a factor of two change in disc scale height and two orders of magnitude change in viscosity. This is in contrast with the roughly order of magnitude or more change in velocity perturbations seen between the planet-hosting and circumbinary disc models of Figures \ref{fig:dens_vel_cop}, \ref{fig:dens_vel_plan}, and \ref{fig:dens_vel_inc}. Therefore even substantial changes in disc properties still does not produce larger perturbations than transitioning from planetary to stellar companions.
Since our kinematic criteria are based on quantifying these perturbations, and more massive bodies lead to larger perturbations, they should produce a stronger signal in our criteria than less massive bodies.

\section{Summary}\label{sec:sum}

In this paper we have showcased some of the morphological features associated with circumbinary discs. We found that the presence of:
\begin{enumerate}
    \item a gas depleted cavity,
    \item spiral arms inside or outside of this cavity,
    \item and non-localised kinks in the channel maps,
\end{enumerate}
are robust indicators of binarity. We also found that the kinematics of circumbinary discs contain peculiar features and defined metrics to quantify these features in Section \ref{sec:vm}. These metrics quantify
\begin{enumerate}
    \item the ratio of maximum absolute velocity along a path close to the semi-major axis in each wing,
    \item and the ratio of the area of the disc enclosed by a specific absolute velocity in each wing.
\end{enumerate}
These kinematic and morphological metrics together provide robust indicators of binarity and can be used to infer the existence of a binary in cases where direct imaging remains challenging.

\section*{Acknowledgements}
JC acknowledges the support of LANL/LDRD program. HL acknowledges the support of NASA/ATP program and LANL/LDRD program. This research used resources provided by the Los Alamos National Laboratory Institutional Computing Program, which is supported by the U.S. Department of Energy National Nuclear Security Administration under Contract No. 89233218CNA000001. DJP and CP acknowledge funding from the Australian Research Council via FT130100034, DP180104235 and FT170100040. BJN. is supported by an Australian Government Research Training Program (RTP) Scholarship. VC acknowledges funding from the Australian Research Council via DP180104235 and from the Belgian F.R.S.-FNRS for financial support through a postdoctoral researcher fellowship. We used \texttt{plonk} for the figures in this paper \citep{Mentiplay2019}, which utilises visualisation routines developed for \texttt{splash} \citep{price2007}. 

\section*{Data Availability Statement}
The SPH code {\sc{phantom}} is available for use at \url{https://github.com/danieljprice/phantom}. The simulation setup files and dump files can be obtained through request to JC. {\sc{mcfost}} is available for use on a collaborative basis from CP. The code for computing the quantities $V_\textrm{Ratio}$, $V_\textrm{Area}$, $\sigma_\textrm{Area}^2$, and $\sigma_\textrm{Ratio}^2$, will be included in a future release of \texttt{eddy} \citep{teague2019}, which is available at \url{https://github.com/richteague/eddy}. 




\bibliographystyle{mnras}
\bibliography{paper} 



\appendix

\section{Obtaining the path $s$ for $V_\textrm{Ratio}(s)$}\label{sec:paths}

\begin{figure*}
    \centering
    \includegraphics[width=0.7\linewidth]{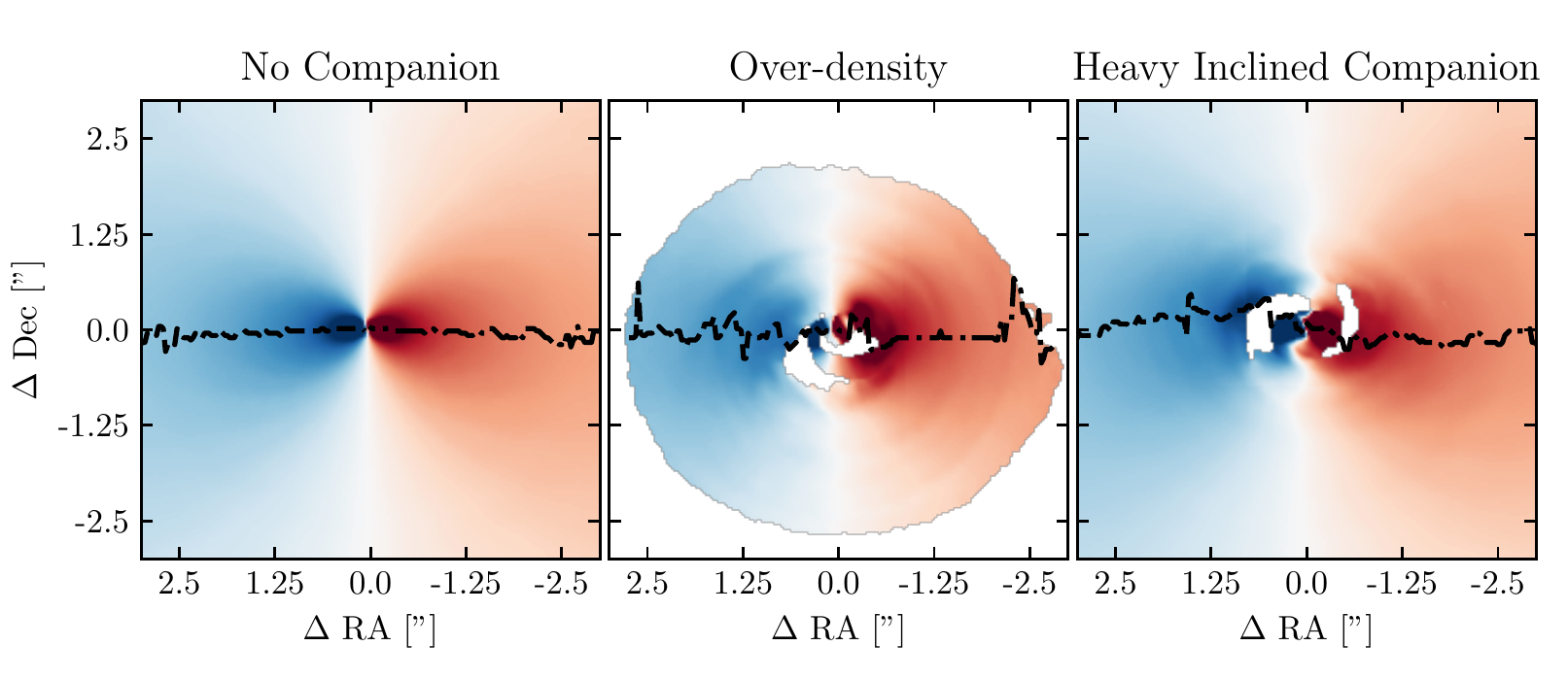}
    \caption{Examples of the paths $s_+$ (dashed line) and $s_-$ (dot-dash line) obtained by our method.}
    \label{fig:path_s}
\end{figure*}

To first obtain the path $s$ we rotate the velocity map $V(x, y)$ so that the semi-minor axis is parallel to the new $y$-axis, $y'$. For the column of velocity along a particular fixed value of $x'_i$, we select the maximum and minimum velocity along the slice $y'$. This is repeated for each $x'$ coordinate along the velocity map and results in two paths $s_-$ and $s_+$ which represent the minimum and maximum velocities in the velocity map. The paths are restricted by requiring that sudden drops in velocity close to the centre of $V(x', y')$ are cut when they drop below one third of the absolute maximum velocity of each wing. Figure~\ref{fig:path_s} shows how our method recovers the paths $s_+$ and $s_-$ in three of our velocity maps.


\section{Robustness of Kinematic Criteria to Velocity Maps}\label{sec:robust_v}

\begin{figure}
    \centering
    \includegraphics[width=\linewidth]{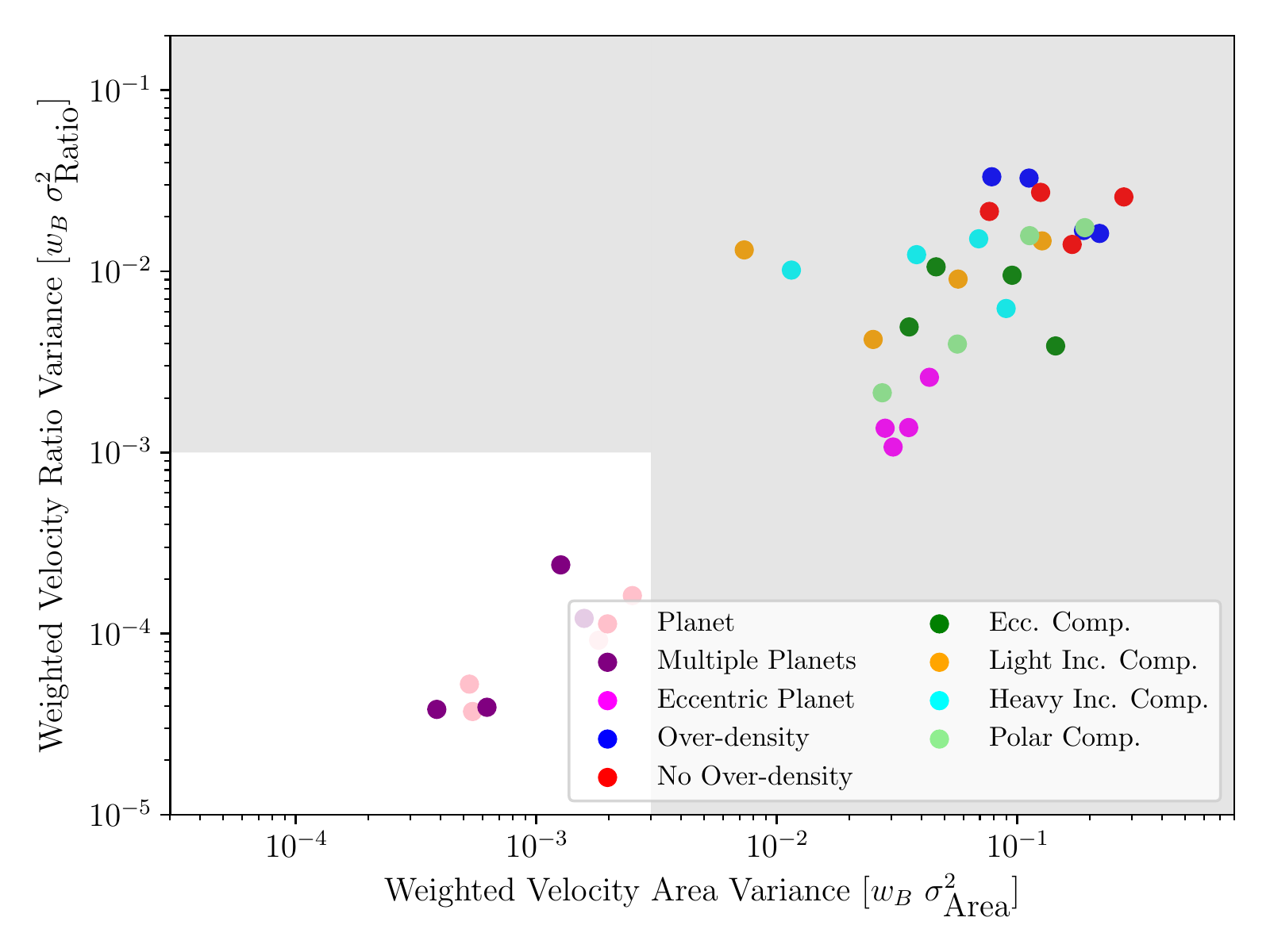}
  \caption{Same as Figure~\ref{fig:sva_svpv} but with the velocity map made using the quadratic method from \protect\cite{bettermoments2018}, along with a 5$\sigma$ cut made to the data.}
  \label{fig:quad}
\end{figure}

\begin{figure*}
    \centering
    \includegraphics[width=0.45\linewidth]{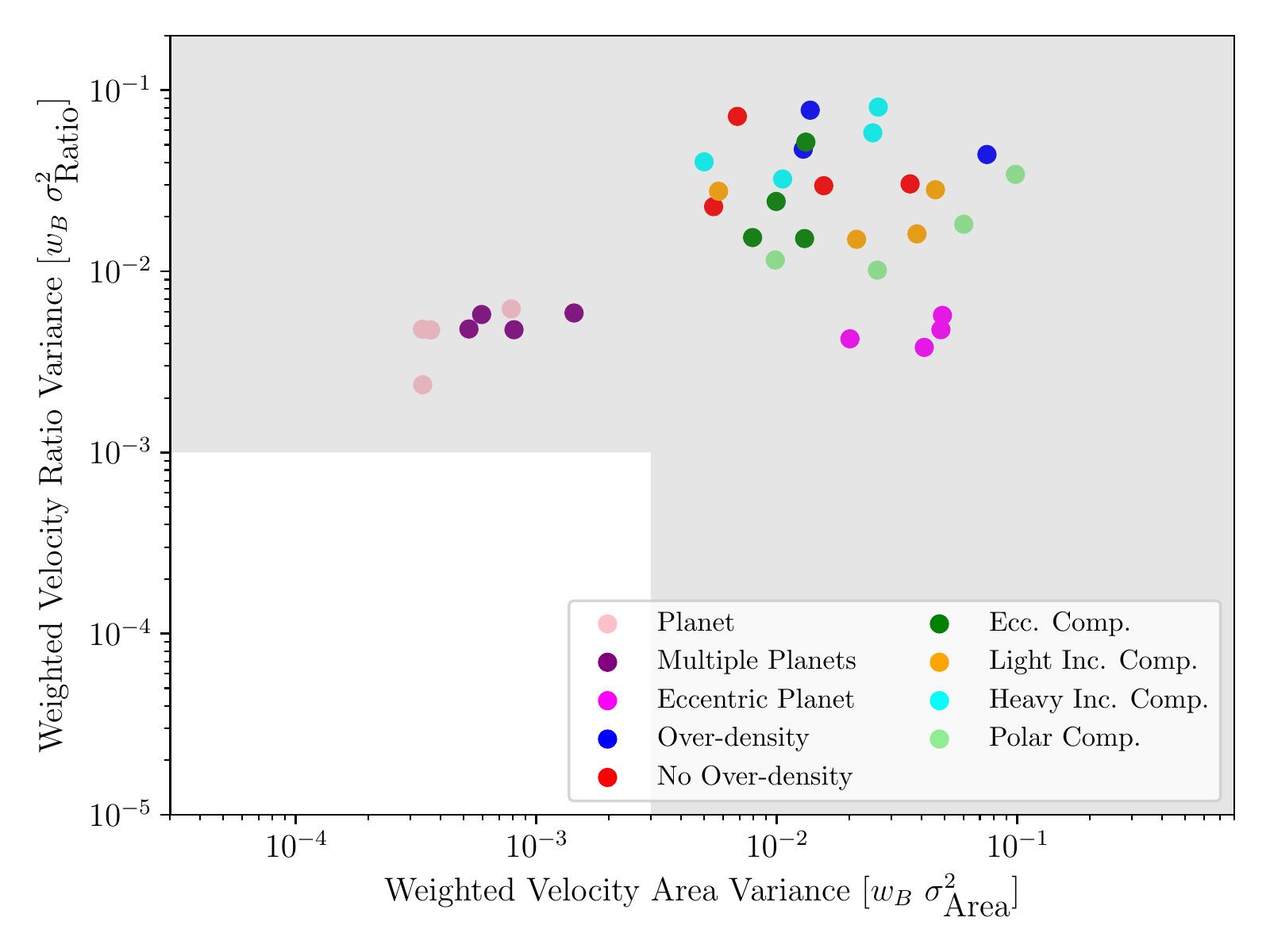}
    \includegraphics[width=0.45\linewidth]{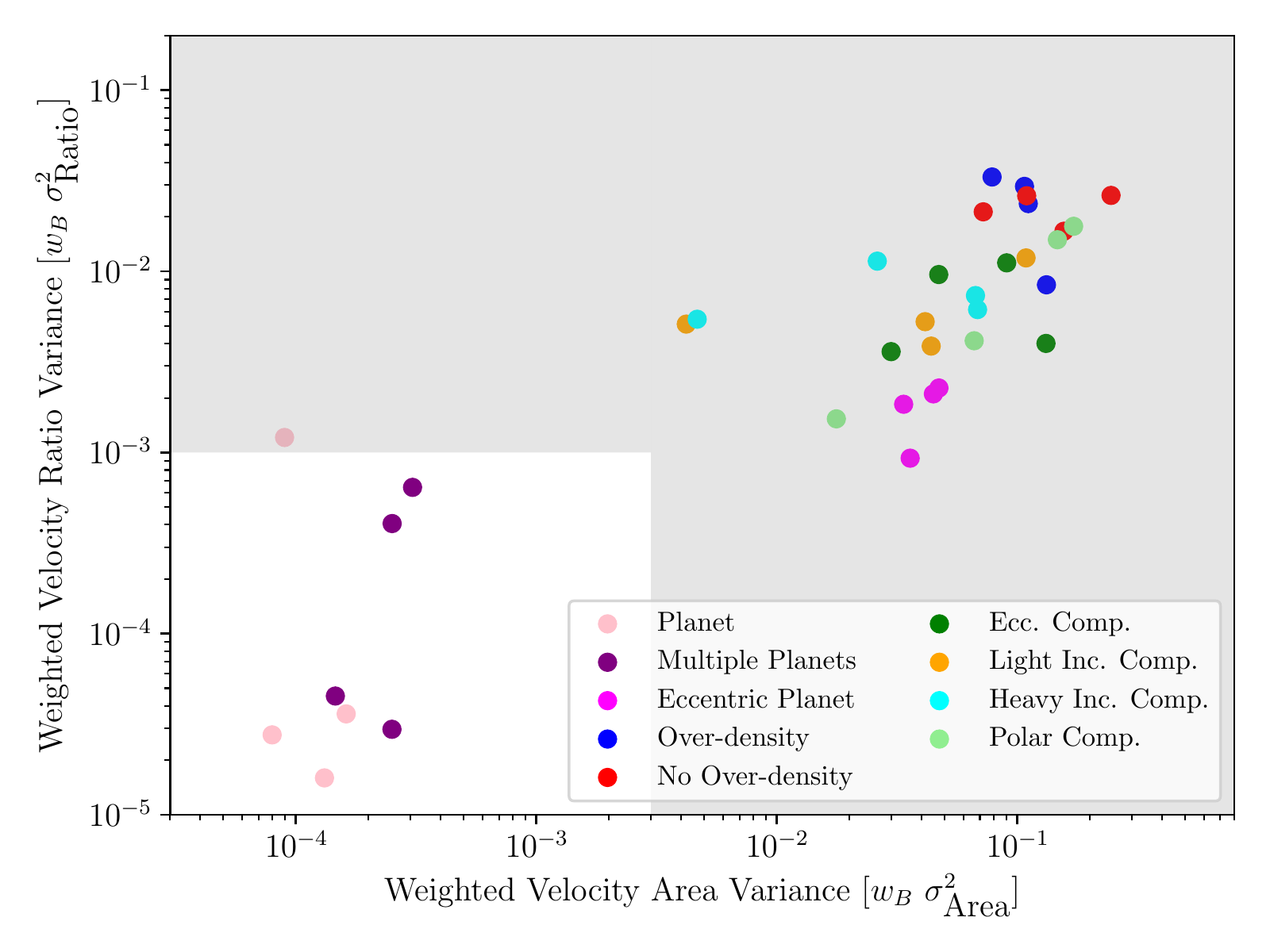}
  \caption{Same as Figure~\ref{fig:sva_svpv} but with a 3$\sigma$ cut (left panel) and 5$\sigma$ cut (right panel) applied to the simulated observations when making the velocity map.}
  \label{fig:noise_cut}
\end{figure*}

There are several methods commonly used when developing velocity maps. The most commonly used are the moment 1, peak velocity, and the quadratic method from \cite{bettermoments2018}. Along with these different methods, noise cuts and Keplerian masks are often applied to the data prior to generating the velocity maps.
Here we test how robust our criteria are to noise level cuts using the moment 1 method, and using the quadratic method. The peak velocity method obtains the velocity field by finding the peak velocity of each pixel in the map. This produces velocity maps where the velocity is sampled at the channel width, which leads to elevated values of $\sigma_\textrm{Ratio}$ and $\sigma_\textrm{Area}$. This is also somewhat present in velocity maps generated by the quadratic method, where $\sigma_\textrm{Area}$ is more affected than $\sigma_\textrm{Ratio}$, as seen in Figure \ref{fig:quad}. Although the values are elevated, they are still within our threshold for binarity. Thus using the quadratic method is suitable for creating the velocity maps that can be used with our kinematic criteria.

We test the moment 1 generated with cuts of 3 and 7 times the RMS noise level. The results presented in Figure \ref{fig:noise_cut} show that the kinematic criteria can show elevated values when the RMS noise cut is too low. This basically results from spurious velocity values in some pixels of the velocity map. A higher RMS noise cut corrects this. When applying the kinematic criteria on real observations care must be taken to ensure that there are no spurious pixels in the velocity map.

\section{Robustness of Kinematic Criteria to Observing Parameters}\label{sec:robust}

\begin{figure*}
    \centering
    \includegraphics[width=0.45\linewidth]{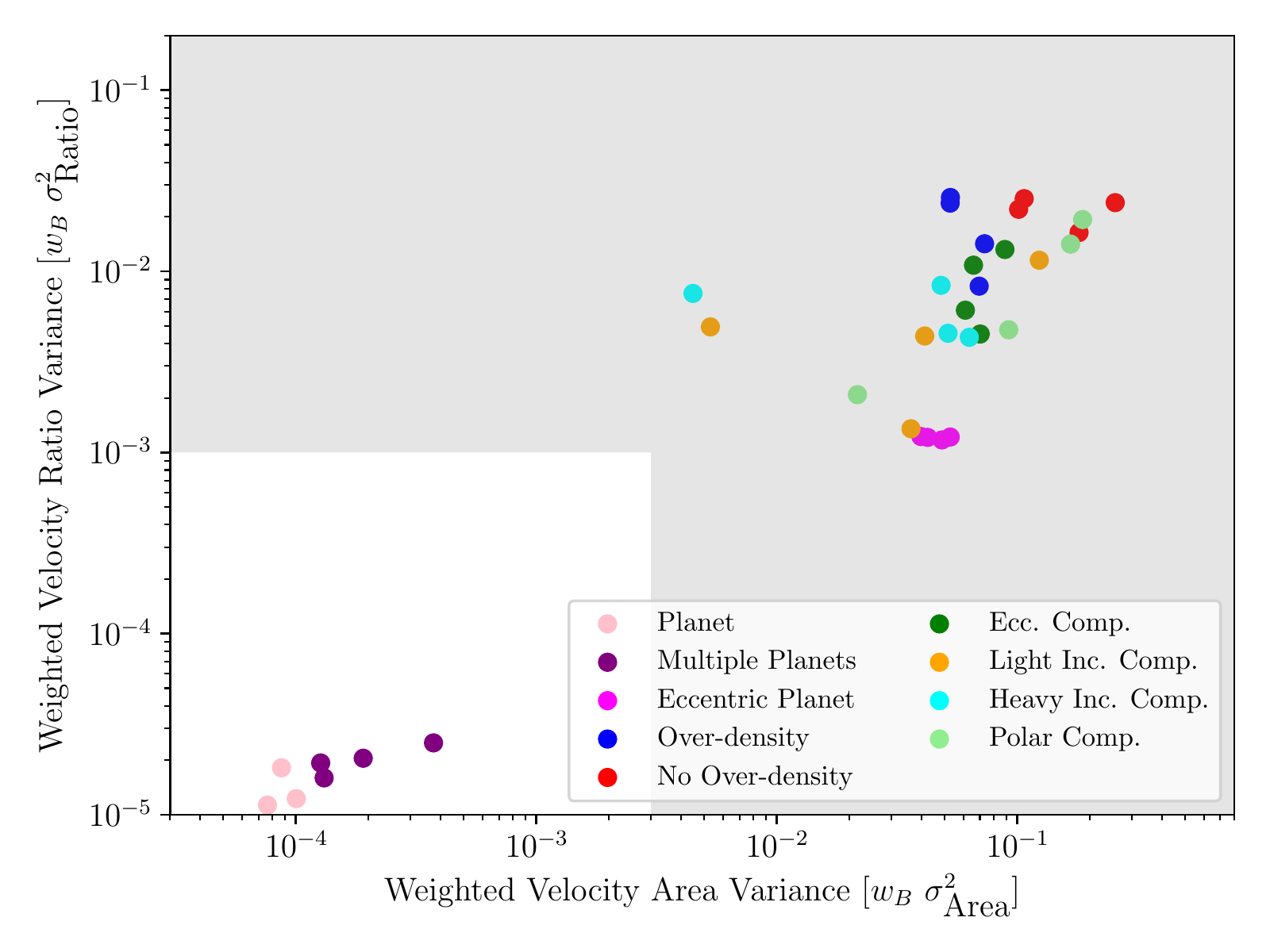}
    \includegraphics[width=0.45\linewidth]{sva_svpv_figure_mn_binary_weight_7sig.pdf} \\

    \includegraphics[width=0.45\linewidth]{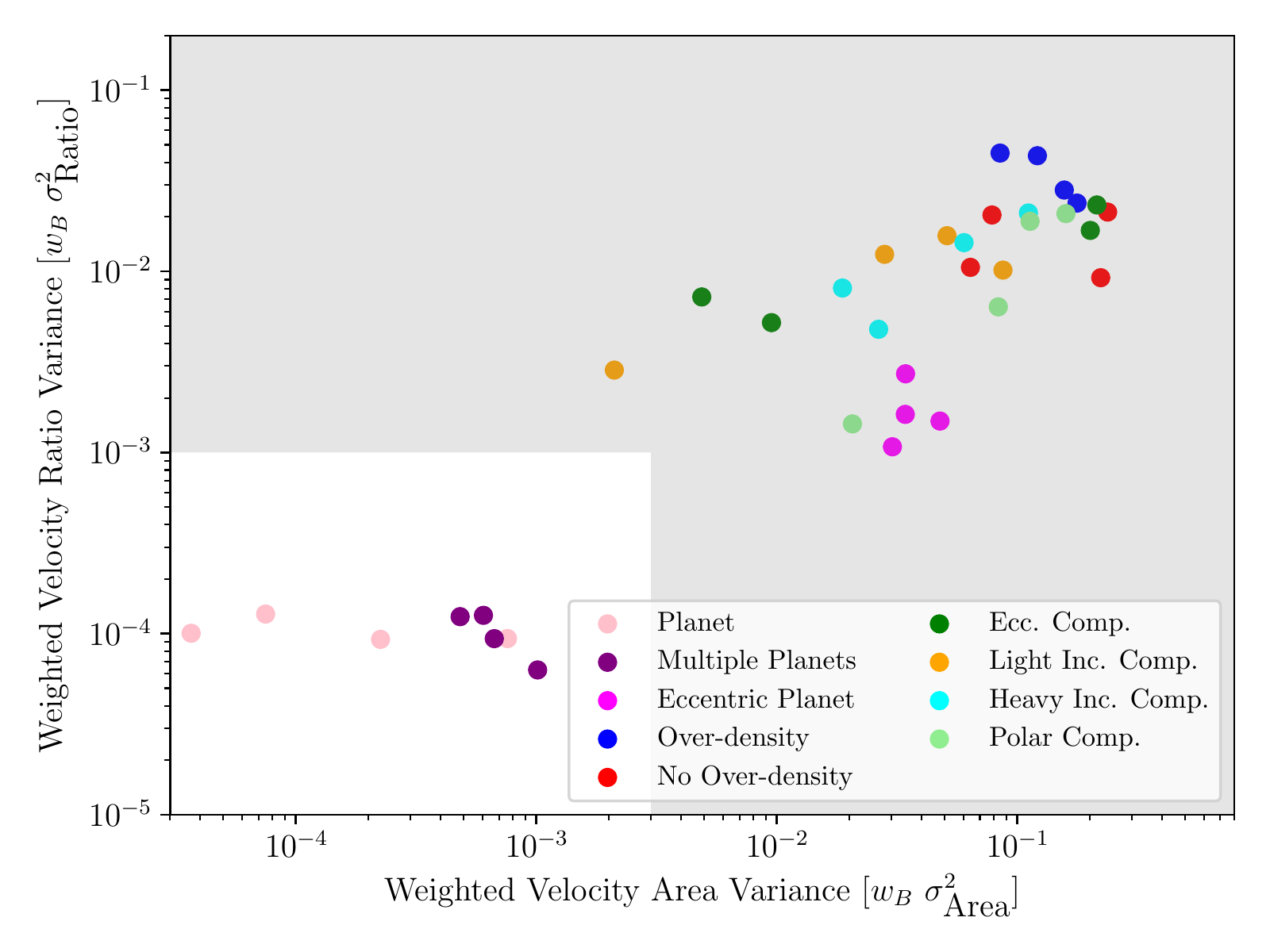}
    \includegraphics[width=0.45\linewidth]{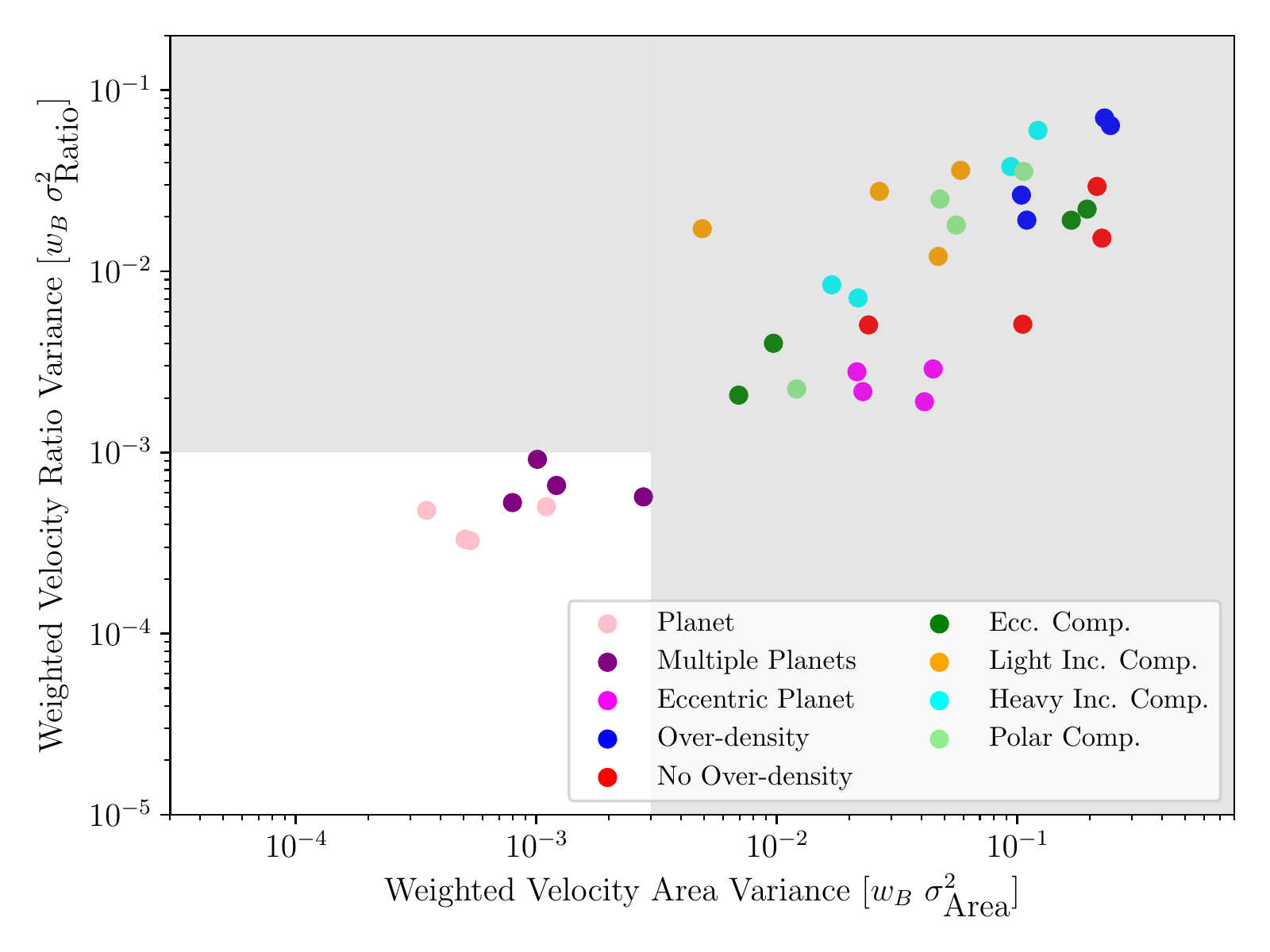}\\
  \caption{The variance $w_B\ \sigma_\textrm{Ratio}^2$ versus $w_B\ \sigma_\textrm{Area}^2$, for an RMS noise level of $1$ mJy/beam (top left), $2.5$ mJy/beam (top right), $5$ mJy/beam (bottom left), and $10$ mJy/beam (bottom right). }
  \label{fig:nl_com}
\end{figure*}

\begin{figure*}
    \centering
    \includegraphics[width=0.45\linewidth]{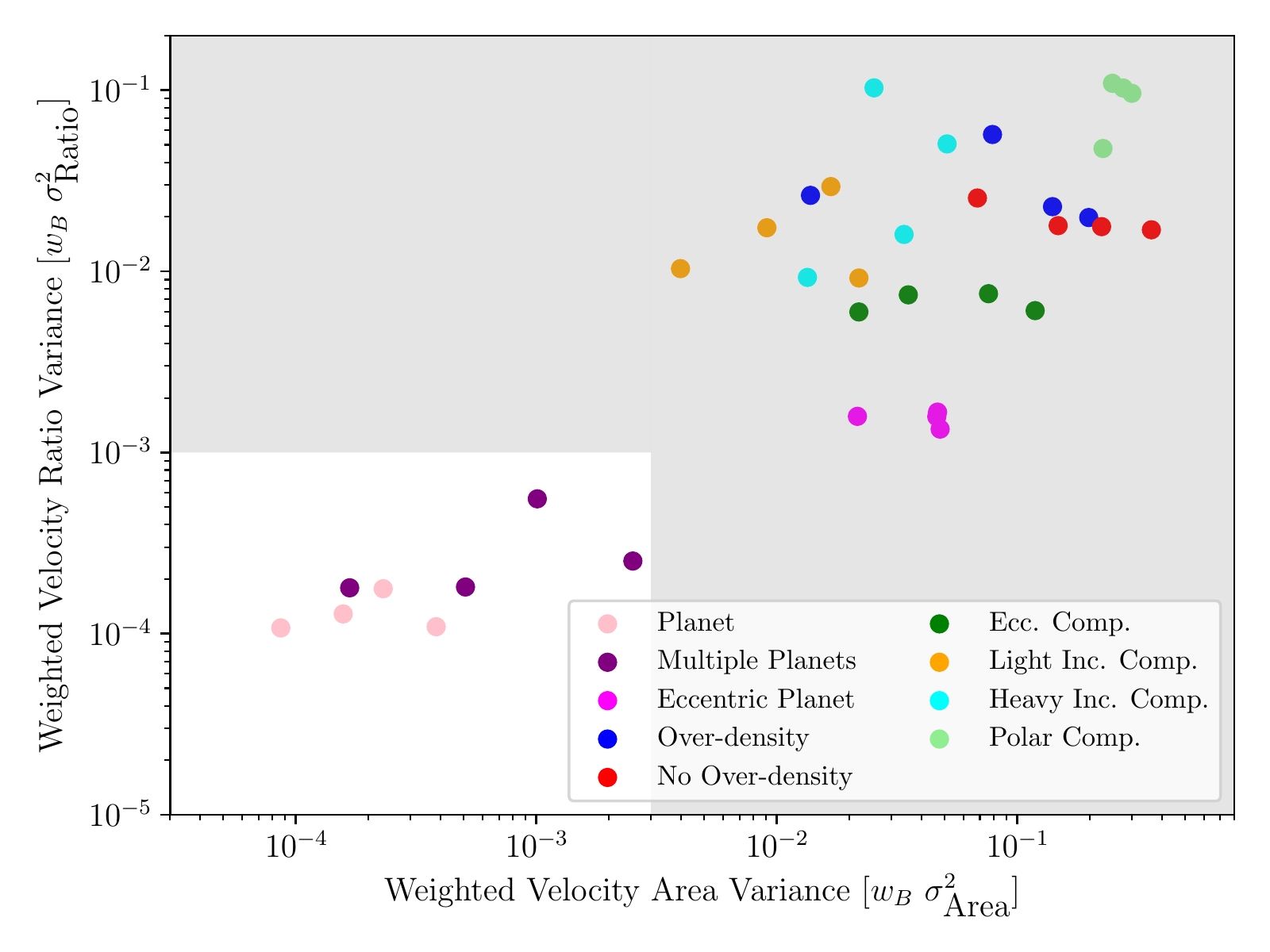}
    \includegraphics[width=0.45\linewidth]{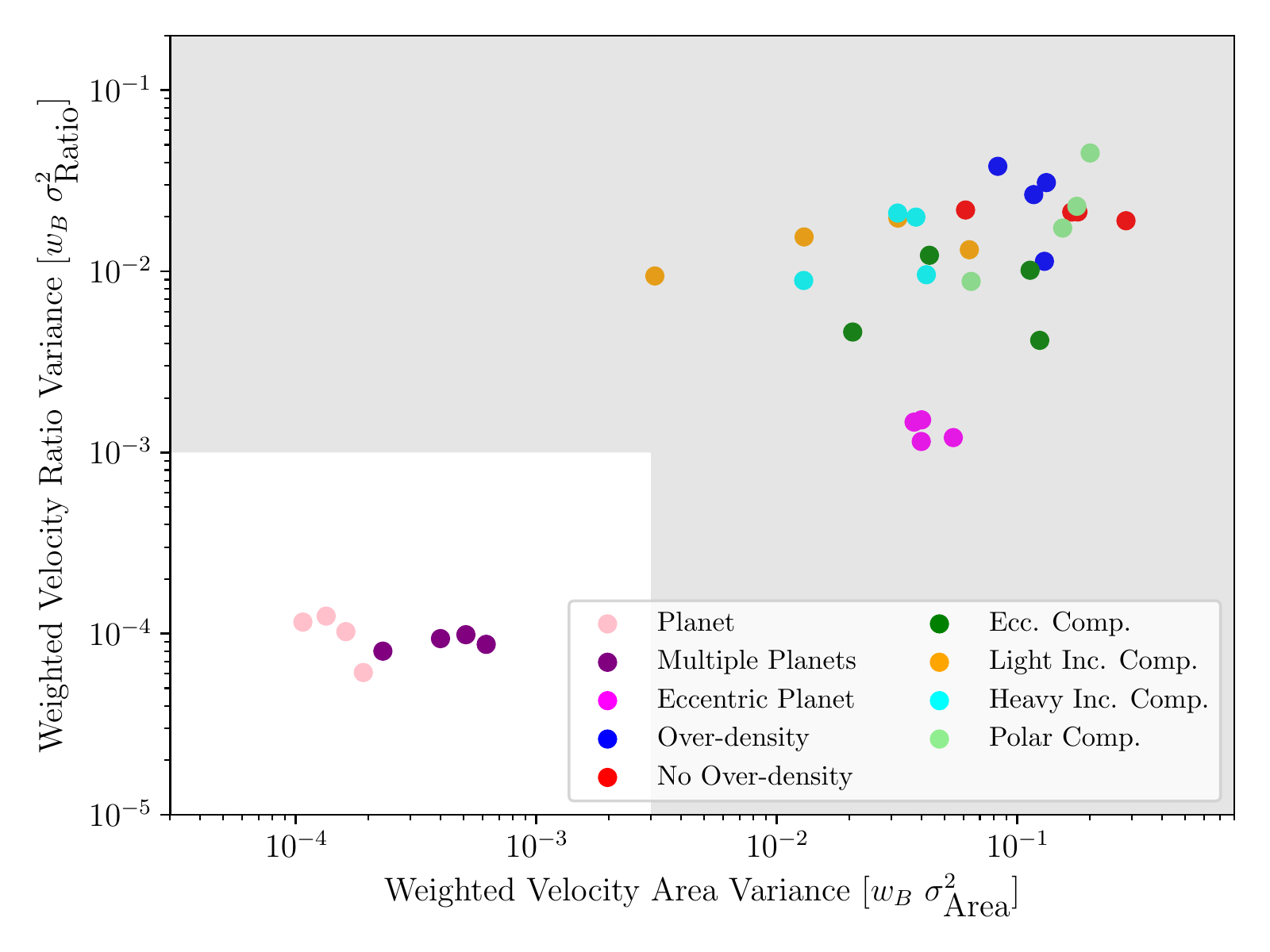} \\

    \includegraphics[width=0.45\linewidth]{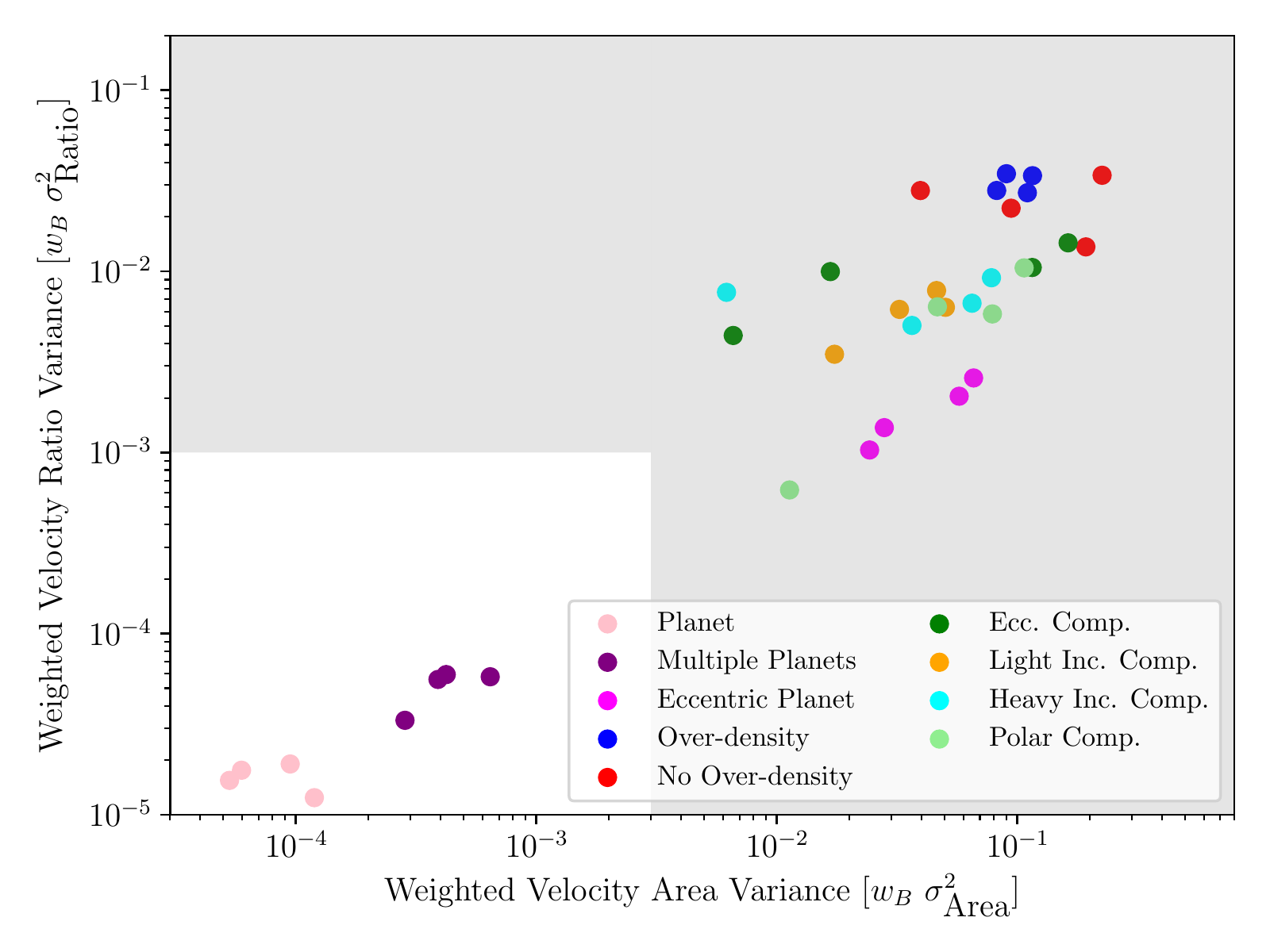}
    \includegraphics[width=0.45\linewidth]{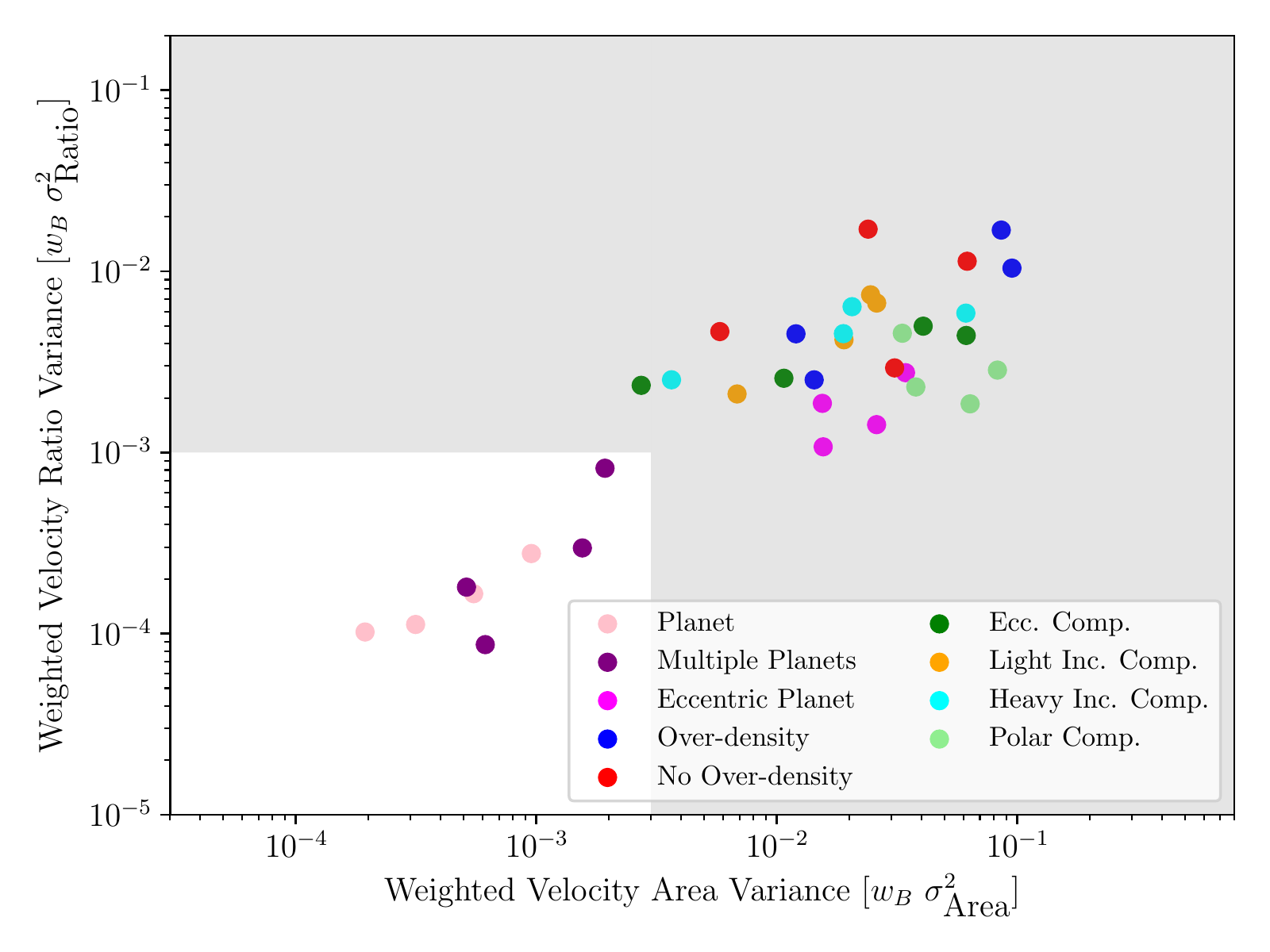}\\
  \caption{Same as Figure~\ref{fig:sva_svpv} but with inclinations of $i=5^{\circ}$ (top left), $i=15^{\circ}$ (top right), $i=60^{\circ}$ (bottom left) and $i=85^{\circ}$ (bottom right).}
  \label{fig:inc}
\end{figure*}

\begin{figure*}
    \centering
    \includegraphics[width=0.45\linewidth]{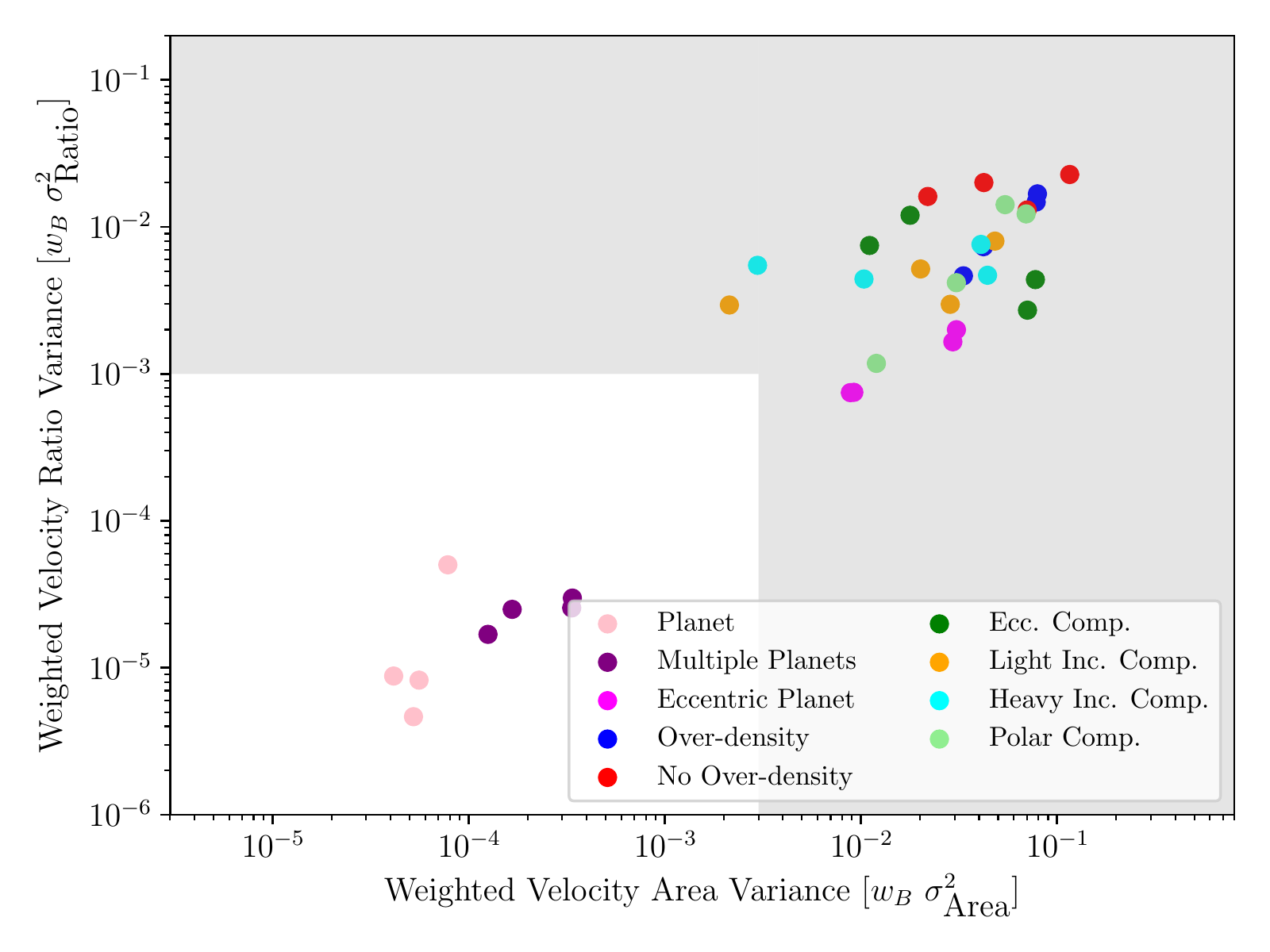}
    \includegraphics[width=0.45\linewidth]{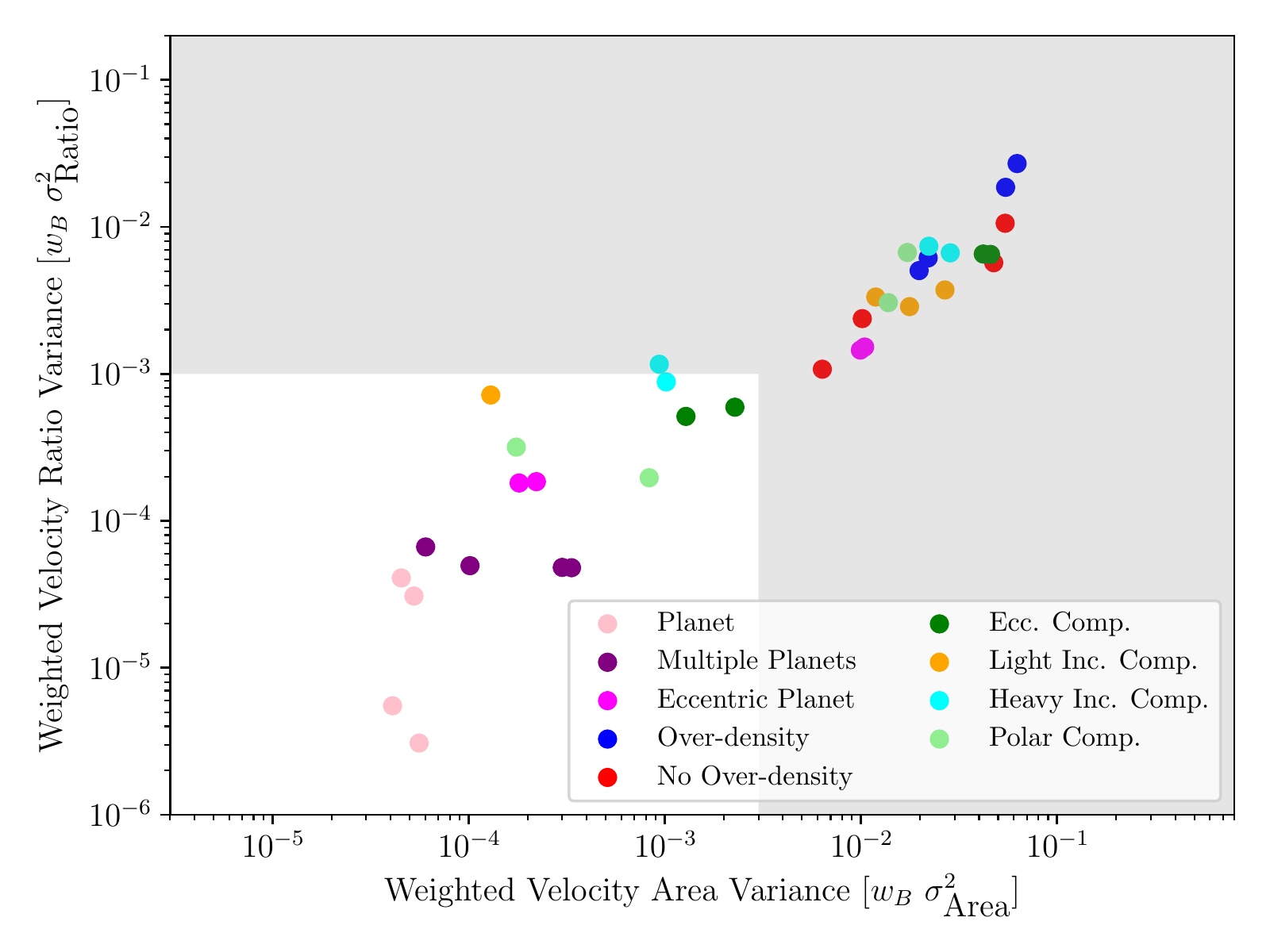}
  \caption{Same as Figure~\ref{fig:sva_svpv} but with a circular Gaussian beam of 0.3 arcseconds (left panel) and 0.6 arcseconds (right panel) applied to the simulated observations. Note that the axes limits have changed in these plots.}
  \label{fig:beam}
\end{figure*}

\begin{figure*}
    \centering
    \includegraphics[width=0.45\linewidth]{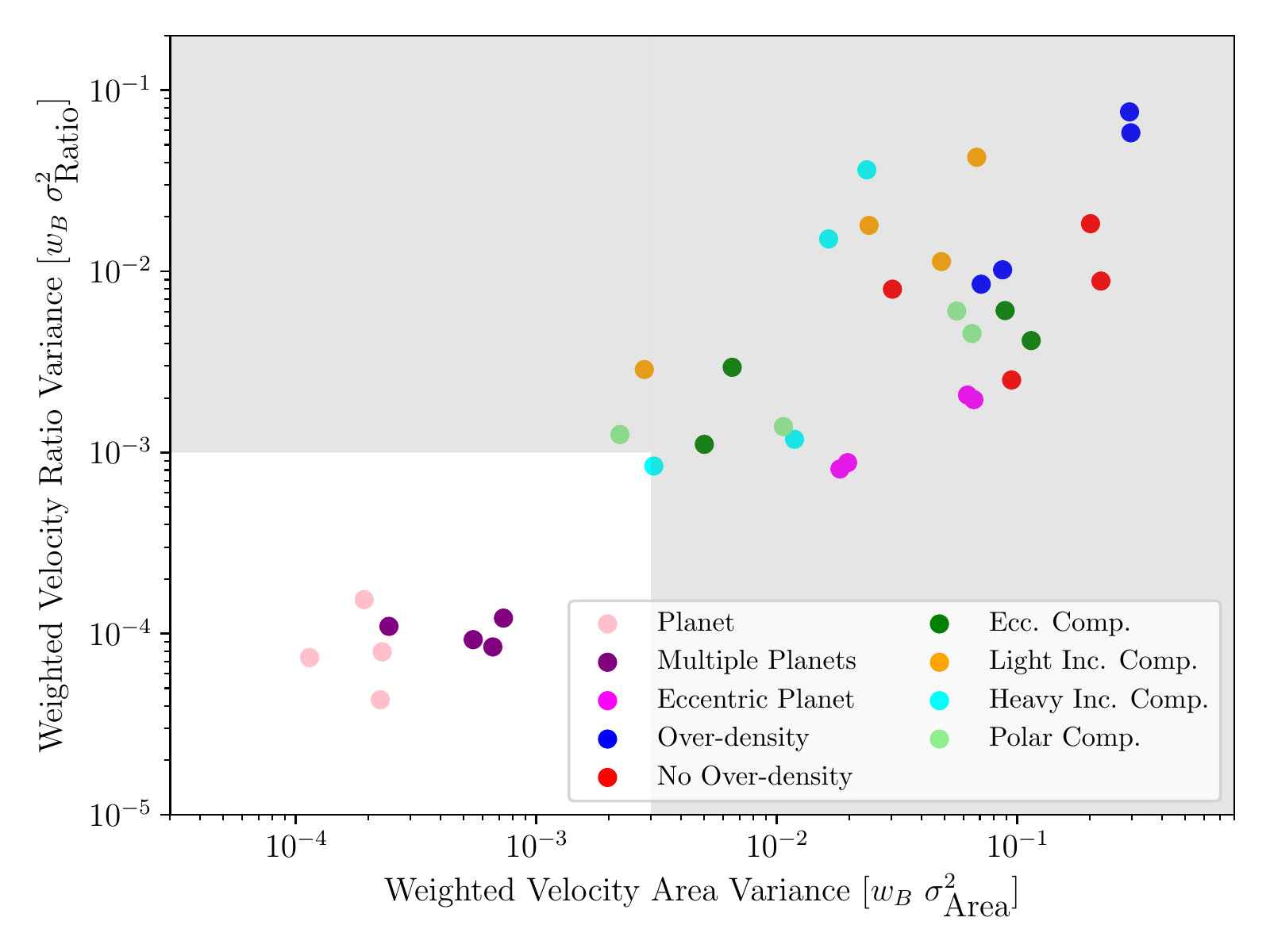}
    \includegraphics[width=0.45\linewidth]{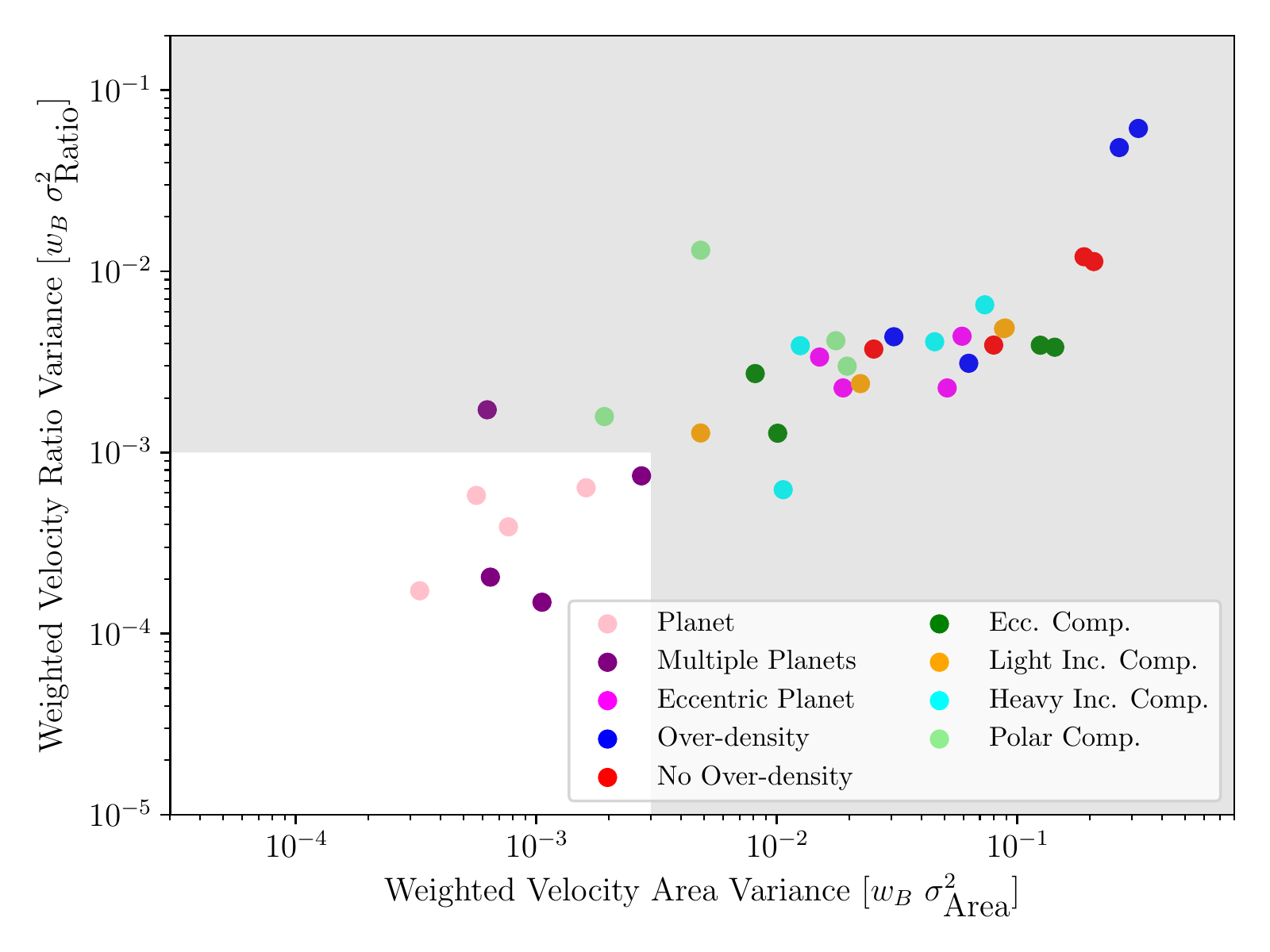}
  \caption{Same as Figure~\ref{fig:sva_svpv} but for the $^{13}$CO (3-2) line emission with noise levels $F_\textrm{noise} = 2.5$ mJy and $F_\textrm{noise} = 5.0$ mJy in the left and right panels, respectively.}
  \label{fig:13co_noise}
\end{figure*}

We test the robustness of our kinematic criteria in four ways: increasing the noise level of the observations (Figure~\ref{fig:nl_com}), changing the inclination of the observations (Figure~\ref{fig:inc}), increasing the beam size of the observations (Figure~\ref{fig:beam}), and changing the isotopologue used (Figure~\ref{fig:13co_noise}). 

Firstly, increasing the noise levels tends to remove the clear dichotomy with the circumbinary disc models and low-mass companion or no companion models. Figure~\ref{fig:nl_com} shows the binary weighted variance of $V_\textrm{Ratio}$ against the variance of $V_\textrm{Area}$ for an RMS noise level of $1$ mJy/beam (top left), $2.5$ mJy/beam (top right), $5$ mJy/beam (bottom left), and $10$ mJy/beam (bottom right). The noise levels correspond to an average peak signal-to-noise ratio of $\textrm{SNR} = [170, 70, 35, 18]$, respectively. The highest noise level results in elevated of our criteria and leads to false positives. It also leads to a reduction in the kinematic criteria of some circumbinary disc models. To be conservative, a peak SNR of at least 50 should be achieved in the observations for our criteria to be reliable.

Next we test the robustness of our kinematic criteria to changes in disc inclination by performing our analysis assuming inclinations of $i=5^{\circ}$, $i=15^{\circ}$, $i=60^{\circ}$, and $i=85^{\circ}$ in Figure~\ref{fig:inc}. We see that our kinematic criteria are mostly robust to the disc inclination, and that the correlation between $\sigma_\textrm{Ratio}^2$ and $\sigma_\textrm{Area}^2$ becomes tighter for discs seen at a higher inclination. Since both radial and azimuthal perturbations have a lower line-of-sight velocity at lower inclinations than higher ones, this trend is expected. Very low inclinations can start to reduce the projected line-of-sight velocity perturbations of the binary, and may result in false negatives. Very high inclinations also present challenges, as the inner disc region can be obstructed by the elevated CO emission from the outer disc. The planet models also start to show elevated values of $w_B\ \sigma_{\textrm{Ratio}}$ and $w_B\ \sigma_{\textrm{Area}}$.

We performed our analysis using larger synthetic beam sizes of 0.3 arcsecond and 0.6 arcseconds, shown in the left and right panels of Figure~\ref{fig:beam}, respectively. With a beam size of 0.3 arcseconds the cavity region is resolved with 3-8 beams depending on the model. At this resolution the kinematic criteria are still robust and mostly differentiate the circumbinary discs from the rest of the sample. However when the beam is increased to 0.6 arcseconds, and the cavity is only resolved with 2-4 beam sizes, many of the circumbinary disc models start to become less distinguishable from the no and low companion mass models. These results demonstrate that in order to obtain a robust estimate on our kinematic criteria, the observations should resolve the cavity with at least $\sim$5 beams. If the kinematic criteria are already met with lower resolution than this, our tests show that the criteria will still be met with higher resolution. Therefore, the resolution requirement is only needed if the kinematic criteria are not met in lower resolution data.

Finally we test our kinematic criteria with each CO isotopologue. Although the correlation between $\sigma_{\textrm{Ratio}}$ and $\sigma_{\textrm{Area}}$ still exists, none of the noise levels studied allow C$^{18}$O emission to reliably conform to our kinematic criteria, and we do not present their results here. C$^{18}$O line emission cannot be used to infer binarity from our kinematic criteria. $^{13}$CO line emission, on the other hand, can be used. However the risk of false positives is greater than for the $^{12}$CO (3-2) line. The peak signal-to-noise ratio of the $^{13}$CO (3-2) line in a single channel is $\textrm{SNR} = [122, 50, 26, 14]$ for the noise levels $F_\textrm{noise} = [1, 2.5, 5, 10]$ mJy. Figure \ref{fig:13co_noise} shows the kinematic criteria for the noise levels $F_\textrm{noise} = 2.5$ mJy and $F_\textrm{noise} = 5.0$ mJy in the left and right panels, respectively. When the noise level is $F_\textrm{noise} = 1.0$ mJy (not shown in the Figure), neither of the planet models show elevated values of $w_B\ \sigma_{\textrm{Ratio}}$ and $w_B\ \sigma_{\textrm{Area}}$. However there are false positives for the higher noise levels we test. Therefore, we require that if $^{13}$CO emission is used to meausre the kinematic criteria, a SNR greather than 50 is required.

\section{Robustness of Kinematic Criteria to Numerical Resolution}\label{sec:res_study}

\begin{figure*}
    \centering
    \includegraphics[width=0.6\linewidth]{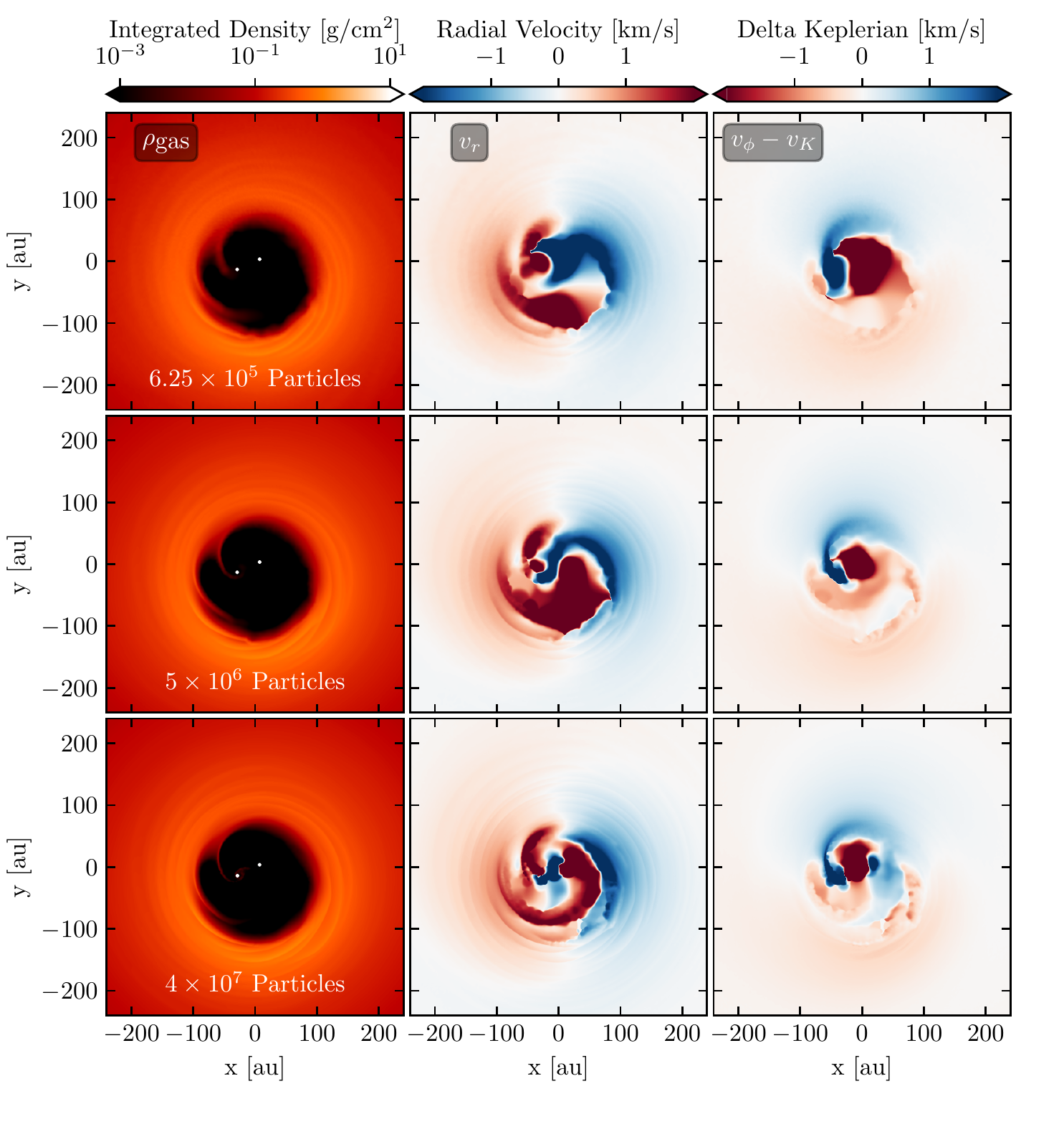}
    \caption{The surface density (first column), radial velocity (second column), and difference from Keplerian velocity (third column) for three resolutions of our no over-density model. }
    \label{fig:dens_vel_res_rt}
\end{figure*}

\begin{figure*}
\centering
    \includegraphics[width=0.7\linewidth]{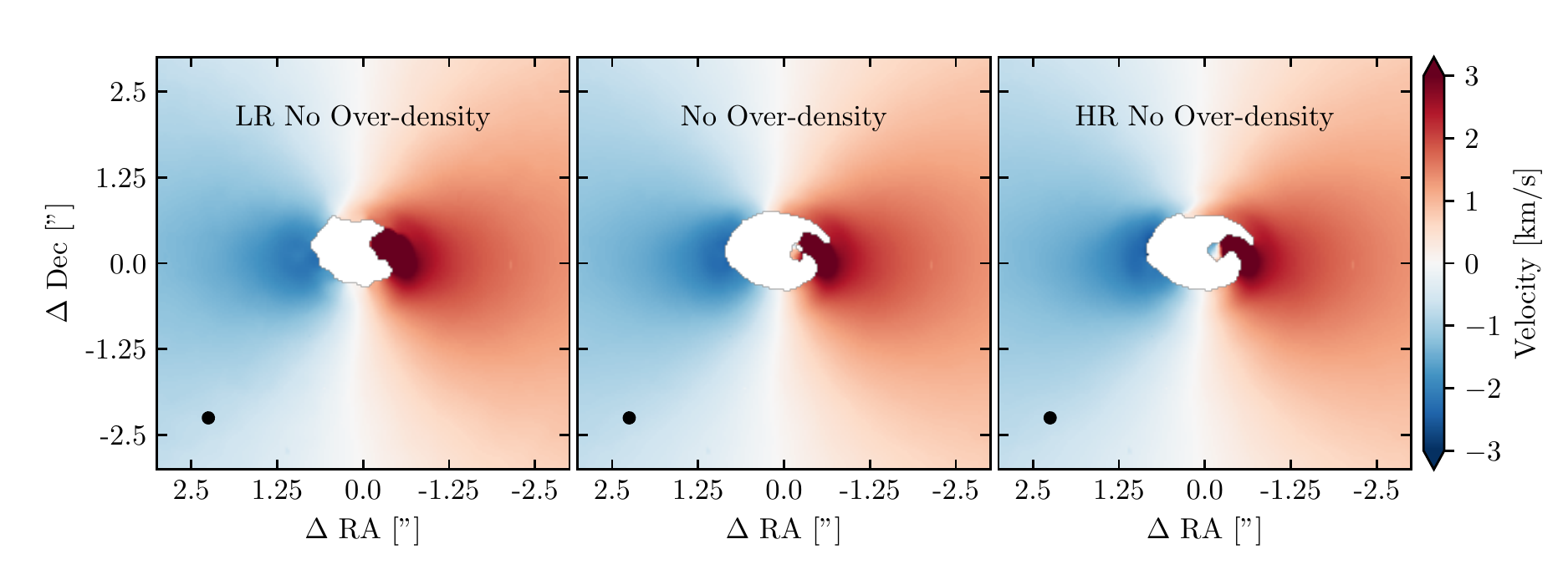} \\ 
    \centering
    \includegraphics[width=0.5\linewidth]{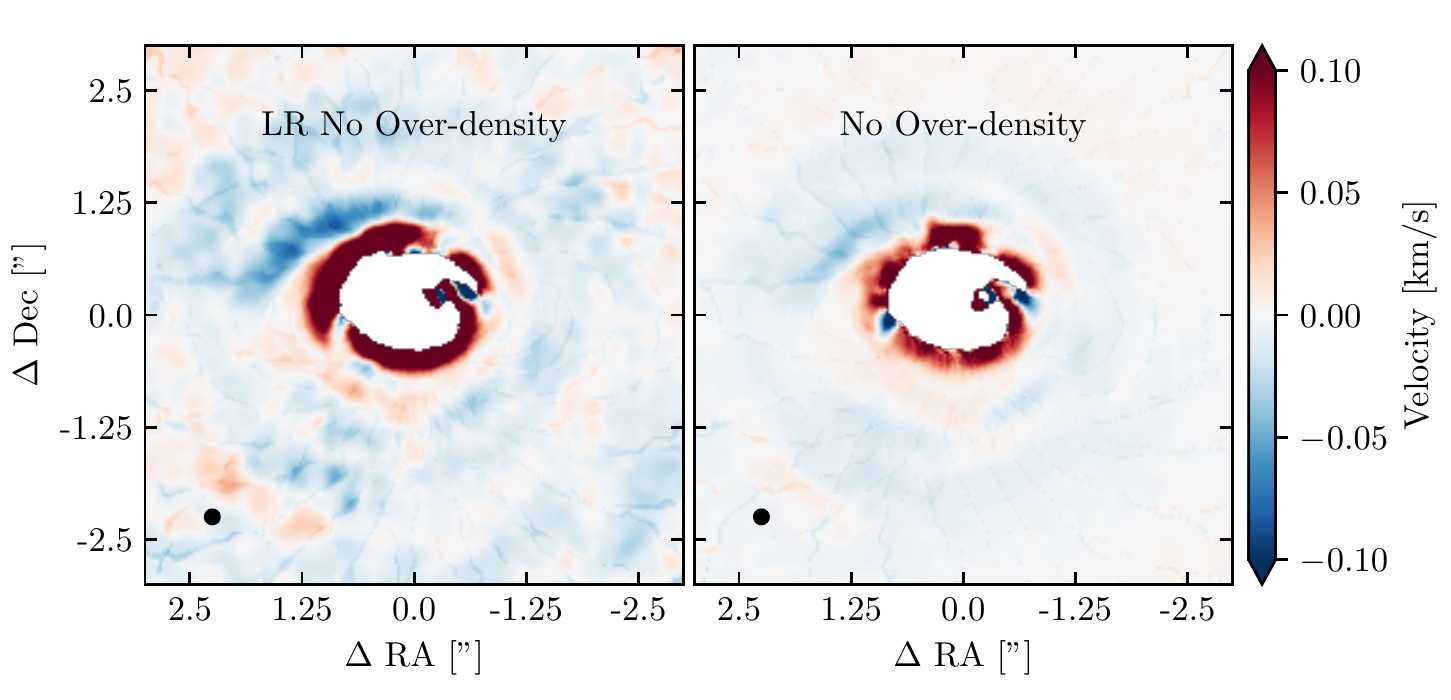}\\
  \caption{The moment 1 maps for the three resolution of our no over-density model shown in Figure \ref{fig:dens_vel_res_rt} (top row). We subtract the high resolution model moment 1 map (top row, right panel) from the other two models to produce the velocity residuals in the bottom panel. The differences in velocity between the models are not drastic enough to affect the conclusions of our paper.}
  \label{fig:mom1_res}
\end{figure*}

We explore how our results depend on the resolution of our SPH simulations by changing the SPH resolution in our no over-density model. We chose this simulation to conduct our resolution test since it has the lower density inside of the cavity. We reduced the number of SPH particles by a factor of 8 (which corresponds with a reduction in resolution by a factor of 2), and increase the particle number by a factor of 8. We then ran these adjusted resolution simulations for 20 orbits of the binary. We also run our original no over-density model for comparison. Figure~\ref{fig:dens_vel_res_rt} shows the disc surface density along with the radial and azimuthal deviations from Keplerian rotation. Although there is some difference in the velocity deviations inside the cavity, this low density region does not contribute any significant amount to the CO flux and hence is not observed in our synthetic observations. We demonstrate this by plotting the velocity map of each model (top row) and the difference between the low resolution and default resolution from the high resolution (bottom row) in Figure~\ref{fig:mom1_res}, for a fixed viewing angle. We ensured that the noise is the same in each model by fixing the random number generator seed. We see that there are difference greater than $\sim$100 m/s close to the cavity edge in our default simulation compared with the higher resolution run. However we do not expect perfect convergence since the binary in our model is not fixed, and how the gas torques the binary is sensitive to both the circumbinary and circumsingle discs around each component of the binary \citep{munoz2019}, which are not modelled in our simulations due to our sink properties. The default resolution used in our simulations is sufficient to obtain convergence on the properties of the cavity \citep{hirsh2020}.

\begin{figure*}
    \centering
    \includegraphics[width=0.5\linewidth]{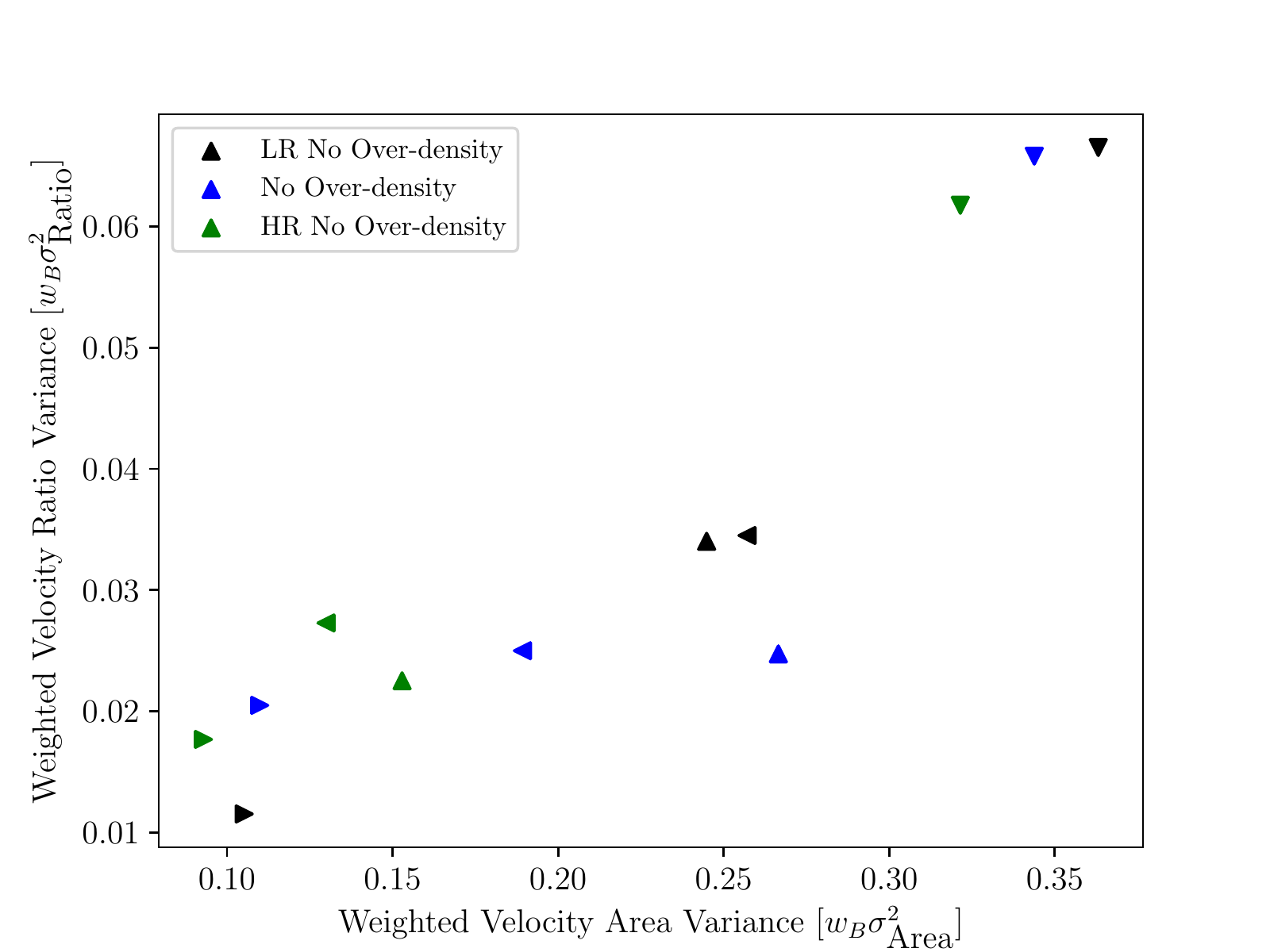}
    \caption{Our no over-density model (NOD) for three different resolutions (indicated by different colours) and four different viewing angles (indicated by the rotated triangles). Although the resolution can affect the measured variances, they still remain well above our threshold for binarity.}
    \label{fig:res_test}
\end{figure*}

We also measure our kinematic criteria on these simulations for the 4 different angles and the results are shown in Figure~\ref{fig:res_test}. Although there is a difference between each resolution, the viewing angle produces a much larger change in the criteria. The resolution of the simulation does not affect the conclusions we draw in this work.

\bsp	
\label{lastpage}
\end{document}